\documentclass[useAMS,usenatbib,a4paper]{mn2e}
\usepackage{amssymb,amsmath}
\usepackage{graphicx}
\usepackage{natbib}
\usepackage{journal_shortcuts}
\newcommand{\De}{\mathrm{d}}
\newcommand{\CW}{c_{\mathrm{w}}}
\newcommand{\CS}{c_{\mathrm{s}}}
\newcommand{\Sun}{_{\sun}}

\newcommand{\Pot}{_{\mathrm{pot}}}
\newcommand{\Kin}{_{\mathrm{kin}}}
\newcommand{\Turb}{_{\mathrm{turb}}}
\newcommand{\Wake}{_{\mathrm{wake}}}
\newcommand{\Heat}{_{\mathrm{heat}}}

\newcommand{\Perp}{_{\mathrm{perp}}}
\newcommand{\Par}{_{\mathrm{par}}}
\newcommand{\Gal}{_{\mathrm{gal}}}
\newcommand{\ICM}{_{\mathrm{ICM}}}
\newcommand{\ISM}{_{\mathrm{ISM}}}

\newcommand{\Ram}{_{\mathrm{ram}}}
\newcommand{\Drag}{_{\mathrm{D}}}
\newcommand{\DF}{_{\mathrm{DF}}}
\newcommand{\Disc}{_{\mathrm{disc}}}
\newcommand{\degree}{^o}

\newcommand{\K}{\,\textrm{K}}
\newcommand{\Kpc}{\,\textrm{kpc}}
\newcommand{\Mpc}{\,\textrm{Mpc}}

\newcommand{\CM}{\,\textrm{cm}}

\newcommand{\Myr}{\,\textrm{Myr}}
\newcommand{\Gyr}{\,\textrm{Gyr}}
\newcommand{\Kms}{\,\textrm{km}\,\textrm{s}^{-1}}
\newcommand{\Erg}{\,\textrm{erg}}
\newcommand{\ccm}{\,\textrm{cm}^{-3}}
\newcommand{\gccm}{\,\textrm{g}\,\textrm{cm}^{-3}}
\newcommand{\Presunit}{\,\textrm{erg}\,\textrm{cm}^{-3}}

\title[Ram pressure stripping induced galactic wakes]%
{Ram pressure stripping of disc galaxies orbiting in clusters. II. Galactic wakes}

\author[E. Roediger and M. Br\"uggen]%
{Elke Roediger%
\thanks{E-mail:
e.roediger@jacobs-university.de; m.brueggen@jacobs-university.de}
and  
Marcus Br\"uggen\footnotemark[1]%
\\
Jacobs University Bremen, P.O. Box 750\,561, 28725 Bremen,
Germany}

\begin{document}

\date{Accepted. Received; in original form }

\pagerange{\pageref{firstpage}--\pageref{lastpage}} \pubyear{2007}

\maketitle

\label{firstpage}

\begin{abstract}
We present 3D hydrodynamical simulations of ram pressure stripping of a disc
galaxy orbiting in a galaxy cluster. In this paper, we focus on the properties
of the galaxies' tails of stripped gas. The galactic wakes show a flaring
width, where the flaring angle depends on the gas disc's cross-section with
respect to the galaxy's direction of motion. The velocity in the wakes shows a
significant turbulent component of a few $100\Kms$. The stripped gas is
deposited in the cluster rather locally, i.e~ within $\sim 150\Kpc$ from where
it was stripped. We demonstrate that the most important quantity governing the
tail density, length and gas mass distribution along the orbit is the galaxy's
mass loss per orbital length. This in turn depends on the ram pressure as well
as the galaxy's orbital velocity.

For a sensitivity limit of $~\sim 10^{19}\CM^{-2}$ in projected gas density,
we find typical tail lengths of $40\Kpc$. Such long tails are seen even at
large distances (0.5 to $1\Mpc$) from the cluster centre.  At this sensitivity
limit, the tails show little flaring, but a width similar to the gas disc's
size.

Morphologically, we find good agreement with the HI tails observed in the
Virgo cluster by \citet{chung07}. However, the observed tails show a much
smaller velocity width than predicted from the simulation. The few known X-ray
and H$\alpha$ tails are generally much narrower and much straighter than the
tails in our simulations. Thus, additional physics like a viscous ICM,
the influence of cooling and tidal effects may be
needed to explain the details of the observations.

We discuss the hydrodynamical drag as a heat source for the ICM but conclude
that it is not likely to play an important role, especially not in stopping
cooling flows.
\end{abstract}


%
%
%
%
%
\section{Introduction}
%
\label{sec:intro}
\citet{gunn72} have been the first to propose the mechanism of ram pressure
stripping (RPS), i.e. the removal of a galaxy's gas disc due to its motion
through the intracluster medium (ICM). Since then, this process has been
studied theoretically and observationally. 

On the theoretical side, several groups have performed hydrodynamical
simulations, using either SPH or grid codes
(e.g. \citealt{abadi99,quilis00,schulz01,vollmer01a,marcolini03,roediger05,roediger06,roediger06wakes}). In
all these simulations, the model galaxy was exposed to a constant ICM wind in
order to isolate the ram pressure effect. These studies showed that the ram
pressure can remove a significant fraction of a galaxy's gas disc or can even
strip it completely. In many situations, the amount of gas lost can be
estimated by a simple analytical estimate based on \citet{gunn72} which
compares the ram pressure to the galaxy's gravitational restoring force.

However, when galaxies move through clusters, they do not experience constant
ram pressures, but the ram pressure varies along the orbit. \citet{vollmer01a}
were the first to simulate galaxies in a varying ram pressure. This group used
a sticky particle code to model the galactic gas disc and added the ram
pressure as an extra acceleration on all gas particles exposed to the
wind. This code was applied to several individual galaxies
(e.g.~\citealt{vollmer99,vollmer00,vollmer01,vollmer03,vollmer03a,vollmer04,vollmer05,vollmer06})
in order to disentangle their ram pressure histories. In these simulations,
the ram pressure was allowed to increase and decrease according to a given
temporal ram pressure profile but did not vary in direction. Besides this, the
sticky particle code cannot model hydrodynamical effects such as instabilities
that have been shown to play a role. Recently, \citet{roediger07} and
\citet{jachym07} have presented hydrodynamical ram pressure simulations of
galaxies moving through a galaxy cluster. Using an SPH code, \citet{jachym07}
have focused on rather compact clusters (similar to the Virgo cluster), so
they modelled the ICM-ISM interaction in the inner $140\Kpc$ of the cluster
only. Moreover, in their simulations, the galaxy always moves face-on on a
strict radial orbit. In these compact clusters, the ram pressure peaks are
rather short and ram pressure stripping becomes a distinct event. The ram
pressure peak can even be shorter than the stripping timescale, i.e. the
timescale needed to remove the gas from the galaxy's potential. In such cases,
the galaxy loses less gas than predicted. \citet{roediger07} (hereafter paper
I) have studied RPS in two clusters, one compact (though not as compact as
Virgo) and an extended one (similar to the Coma cluster). In this work, the
galaxies move on different orbits and and also with different
inclinations. Here, the ram pressure peaks are not short enough that the delay
in gas loss becomes important. As long as the galaxy does not move close to
edge-on or experiences significant continuous stripping, the analytical
estimate gives good predictions of the stripping
efficiency. \citet{tonnesen07} have performed a cosmological cluster
simulation where they can resolve ram pressure stripping with $\sim
3\Kpc$. They find that RPS is the main gas loss mechanism in clusters. They
stress that at a given cluster-centric radius, galaxies can experience a
variety of ram pressures e.g. due to ambient motions in the ICM. From an
analysis of cosmological simulations, \citet{brueggen07} find that the
majority of all cluster galaxies experience strong ram pressures within their
life-times.

Clearly, RPS provides an enrichment process for the ICM, which is known to
have a metallicity of about 1/3 solar. The metal enrichment of the ICM has
been modelled e.g.~by \citet{cora06,valdarnini03,tornatore04}. The effect of
RPS on the metal distribution in the ICM has been studied numerically by
\citet{domainko06}, which relies on the stripping criterion of \citet{gunn72}
to model the gas loss from the galaxies. Other enrichment processes are
galactic winds (\citealt{kapferer06}), outflows from AGN (\citealt{moll07}),
intracluster stars (\citealt{zaritsky04}). The recent generation of X-ray
observatories provides information about the distribution of metals in the
nearby clusters
(e.g.~\citealt{schmidt02,matsushita02,matsushita03,churazov03,degrandi04}),
which the models of metal enrichment have to explain. Recently, even the
enhanced metallicity in the tail of the ram pressure stripped elliptical
galaxy NGC~7619 in the Pegasus group (\citealt{kim07a}) has been reported.

Galaxies that experience ram pressure stripping are expected to have truncated
gas discs but undisturbed stellar discs. Such cases have been observed:
e.g.~NGC 4522 (\citealt{kenney99,kenney01,kenney04},\citealt{vollmer04a}), NGC
4548 (\citealt{vollmer99}) and NGC 4848 (\citealt{vollmer01}). Deep HI
observations also revealed long, one-sided tails for several galaxies
(\citealt{oosterloo05}, \citealt{chung07}). In addition to HI tails, there are
only very few galaxies that have X-ray (\citealt{wang04,sun05,sun06}) or and
H$\alpha$ tails (\citealt{gavazzi01,sun07,yagi07}).  However, the
interpretation of these observations as RPS tails is not
straightforward. E.g. the galaxy tails found by \citet{chung07} all belong to
Virgo spirals which are located at projected cluster-centric distances of 0.6
to $1\Mpc$. Given that ram pressure stripping is expected to be strongest near
cluster centres, this was a surprising result. The X-ray and H$\alpha$ tails
tend to be rather long (several 10 kpc), narrow ($< 10\Kpc$) and
straight. Moreover, especially the X-ray tails seem to be rare: \citet{sun07a}
have searched for additional cases in the Chandra and XMM data
of 62 galaxy clusters and did not find any. Also the HI tails seem to be less
common than expected: \citet{vollmer07} have carried out deep HI observations
in the vicinity of 5 RPS candidates in the Virgo cluster and did not find more
HI than was already known.

In order to understand the observations of galactic wakes, a better theoretical
understanding is needed. Here, we analyse the simulations
presented in paper I, i.e. hydrodynamical simulations of RPS of a
disc galaxy along orbits in galaxy clusters, with respect to galaxy wakes.
We focus on the following aspects: 
\begin{itemize}
\item Structure of galactic tails: width, length, substructure, velocity
  structure. 
\item Comparison to observations
\item Where in the cluster does the galaxy deposit stripped gas?
\item How much heating can RPS provide for the ICM?
\end{itemize}
%

\section{Method}
\label{sec:method}
Here we analyse the same simulations as in paper I.  We model the flight of a
disc galaxy through a galaxy cluster. The galaxy starts at a position $\sim 1$
to $2\Mpc$ (depending on orbit) from the cluster centre with a given initial
velocity. We use analytical potentials for the galaxy and the cluster, as this
reduces computational costs significantly. Given the high velocities of
cluster galaxies, the tidal effect on the galaxy is expected to be small.  The
work of Moore et al. (1996, 1998, 1999)\nocite{moore96,moore98,moore99}, \citet{mastropietro05a}
demonstrated that only harassment, i.e.  the cumulative effect of frequent
close high velocity encounters between cluster galaxies and the overall tidal
field of the cluster affects cluster galaxies seriously. 
Also less massive galaxies are affected more strongly by tidal
forces.  For group environments, there is evidence that the tidal forces also
influence the ram pressure stripping efficiency
(\citealt{mastropietro05b,mayer06}). We have, however, chosen a massive galaxy
(rotation velocity $200\Kms$) and follow only the first orbit in a cluster
environment. Thus we expect that tidal forces play a secondary role in our
case, although there may be some influence.
Similar to paper I,
\citet{jachym07} presented SPH simulations of RPS of a galaxy orbiting through
cluster. They modelled the ICM-ISM interaction only near the cluster centre,
but included the mutual gravity of all gaseous and non-gaseous galactic
particles as well as the gravitation of the cluster potential. They did not
find a significant influence on the stellar and DM components. Thus, we expect
that our treatment yields reasonable results. 

The orbit of the galaxy is determined by integrating the motion of a point
mass through the gravitational potential of the cluster.  In the course of the
simulation, the position of the galaxy potential is shifted along this orbit.

\subsection{Code} \label{sec:code}
The simulations were performed with the FLASH code (\citealt{fryxell00})
version 2.5, a multidimensional adaptive mesh refinement hydrodynamics code.
It solves the Riemann problem on a Cartesian grid using the
Piecewise-Parabolic Method (PPM).  The simulations presented here are
performed in 3D. The gas obeys the ideal equation of state with an
adiabatic index of $\gamma=5/3$.  The size of the simulation box is chosen
such that the galaxy's orbit during the simulation time (3 Gyr) fits into the
grid. Depending on the orbit, the size of the simulation box ranges between
$(2\Mpc)^3$ and $2\times 5\times 2 \Mpc^3$. All boundaries are reflecting.

The coarsest refinement level has a resolution of $\Delta x =62.5\Kpc$. For
most runs, we use 8 levels of refinement, i.e. the best resolution is $\Delta
x =0.5\Kpc$. In addition to the standard density and pressure gradient
criteria, our user-defined refinement criteria enforce maximal refinement on
the galactic gas disc and enforce stepwise de-refinement with increasing
distance to the galaxy. Details are described in paper I.  The important point
for the investigation of the wakes is that we limit the refinement outside
$50\Kpc$ around galaxy centre to $\Delta x =2\Kpc$ (6 refinement levels), and
we limit the refinement outside $150\Kpc$ around galaxy centre to $\Delta
x=15.6\Kpc$ (3 refinement levels). 
Together with the standard refinement
criteria, this leads to a typical resolution of $\Delta x =2\Kpc$ at $150\Kpc$
behind the galaxy. Given our resolution restriction, our analysis and
discussion of the wake properties is restricted to the first $150\Kpc$ behind
the galaxy.
 A discussion of the influence of the
resolution on our results is given in App.~\ref{sec:resolution}, 
we find our
results not to be sensitive to resolution.

The FLASH code offers the opportunity to advect mass scalars along with the
density.  In order to be able to identify the galactic gas after it is
stripped from the galaxy, we utilise a mass scalar, $f$, to contain
the fraction of galactic gas in each cell. Initially, this array has the value
1 in the region of the galactic disc and 0 elsewhere. As a result, every cell
where $f>0$ contains a certain amount of gas that has originally been inside
the galaxy. At each time-step, the quantity $f\rho$ gives the local density of
galactic gas. In the context of this paper, we will refer to this gas as
galactic gas or ISM even if it has left the galaxy.

\subsection{Model galaxy}
The galaxy model is the same as in \citet{roediger06} and paper I, i.e. a
massive spiral with a flat rotation curve at $200\Kms$. It consists of a dark
matter halo ($1.1\cdot 10^{11}M\Sun$ within $23\Kpc$), a stellar bulge
($10^{10}M\Sun$), a stellar disc ($10^{11}M\Sun$) and a gaseous disc ($5\cdot
10^{9}M\Sun$). All non-gaseous components just provide the galaxy's potential
and are not evolved during the simulation. For a description of the individual
components and a list of parameters please refer to RB06\nocite{roediger06}.

\subsection{Cluster model}
%
The cluster is set in hydrostatic equilibrium, where the density of the ICM
follows a $\beta$-profile,
%
\begin{equation}
\rho\ICM (r) = \rho\ICM{}_0 \left[ 1+\left( \frac{r}{R\ICM}  \right)^2  \right]^{-3/2\beta},
\end{equation}
while the ICM temperature is constant. Given the density and
 temperature distribution and thus pressure distribution throughout the
 cluster, the underlying gravitational acceleration due to the cluster
 potential is calculated from the hydrostatic equation. Here we
 present simulations for three clusters. The parameters are listed in
 Table~\ref{tab:ICM} along with two parameter sets for the Virgo cluster
 from the literature.
 \begin{table}
 \caption{ICM parameters: Core radius, $R\ICM$, beta-parameter, $\beta$,
   central ICM density, $\rho\ICM{}_0$, and ICM temperature, $T\ICM$, for
   clusters C1, C2 and C3. Also two parameter sets for the Virgo cluster are
   given: from \citet{matsumoto00} (M00) and \citet{vollmer01a} (V01)}
 \label{tab:ICM}
 \centering\begin{tabular}{lcccc}
 \hline
     &$R\ICM/\Kpc$  & $\beta$ & $\rho\ICM{}_0/(\gccm)$    & $T\ICM/\K$ \\
 \hline
 C1  & 50      & $0.5$   & $2\cdot 10^{-26}$ & $4.7\cdot 10^7$ \\
 C2 & $\cdot$ & $\cdot$  & $10^{-26}$        & $\cdot$ \\
 C3 & 386     & 0.705    &  $6.07\cdot 10^{-27}$  & $9.5\cdot 10^7$\\
Virgo: &&&&\\
M00& $14$ & $0.4$ & $3.69\cdot 10^{-26}$ & $2.6\cdot 10^7$  \\
V01& $13.4$ & $0.5$ & $4\cdot 10^{-26}$  &  \\
 \hline
 \end{tabular}
 \end{table}
%
%
 \begin{figure}
 \centering\resizebox{0.8\hsize}{!}{\includegraphics[angle=-90]{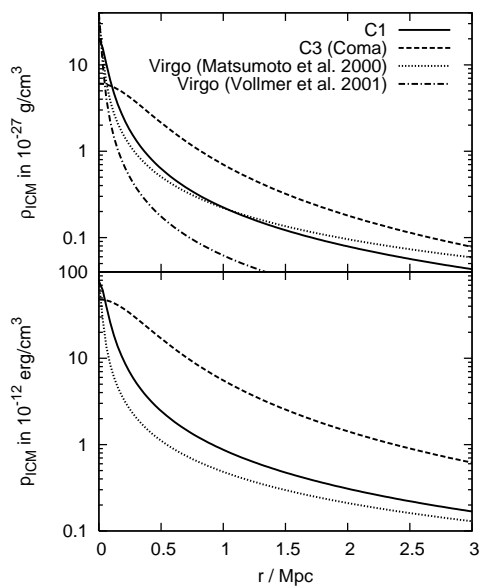}}
 \caption{Density and pressure profiles for model clusters C1 and C3. The
   profiles for cluster C2 are a factor of 2 lower than the ones of cluster
   C1. For comparison,
   two profiles for the Virgo cluster from the literature are shown.}
 \label{fig:cluster_profiles}
 \end{figure}
%
Figure~\ref{fig:cluster_profiles} shows the density and pressure profiles of
clusters C1 and C3.
The only difference between clusters C1 and C2 is a factor of 2 in 
density. These two clusters are compact, the ICM density is strongly peaked.
Thus, they are similar to the Virgo cluster, but not
as extreme as Virgo (see Fig.~\ref{fig:cluster_profiles}).  Cluster C3
resembles the Coma cluster (parameters see \citealt{mohr99}), which is very
extended.

\subsection{Galaxy orbit} \label{sec:orbits}
The orbit of the galaxy determines its ram pressure history.   We
aimed at constructing orbits with medium to high ram pressure peaks,
i.e. where the galaxy is expected to lose a significant fraction of its gas
or is even stripped completely. 
We concentrate on orbits of galaxies that could be regarded as
falling into the cluster for the first time, i.e. they start from a
sufficiently large distance from the cluster centre. Additionally, the
galaxies are bound to the cluster.

\begin{figure}
 \centering\resizebox{\hsize}{!}{\includegraphics[angle=-90]{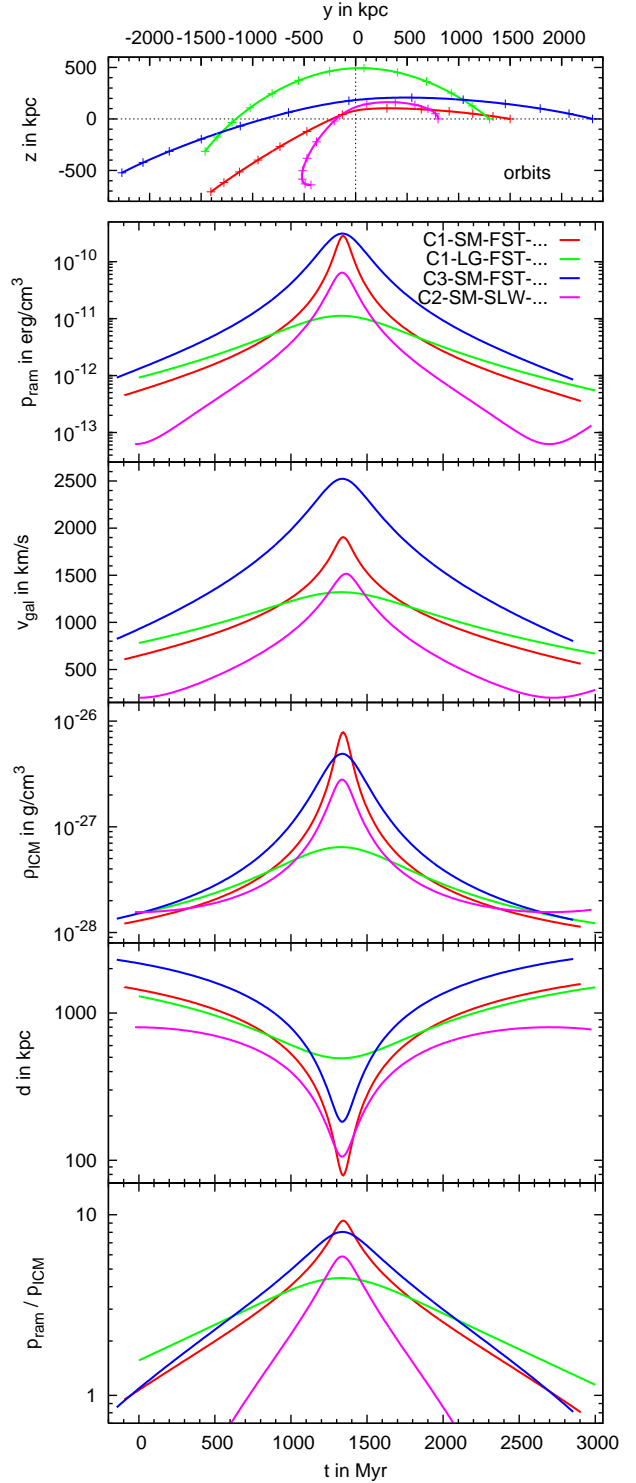}}
 \caption{Summary of galaxy orbits for our simulations. The top panel shows the orbits, which are in the
   $y$-$z$-plane. Crosses mark the position of the galaxy in intervals of
   $250\Myr$. The next panels show the temporal evolution of  ram pressure,
   $p\Ram$, galaxy velocity, $v\Gal$, ICM density along orbit, $\rho\ICM$,
   distance to cluster centre, $d$, and ratio between ram pressure and local
   ICM pressure, $p\Ram/p\ICM$. For an explanation of the labels see text.}
 \label{fig:orbits_profiles}
 \end{figure}

\begin{figure}
 \centering\resizebox{\hsize}{!}{\includegraphics[angle=0]{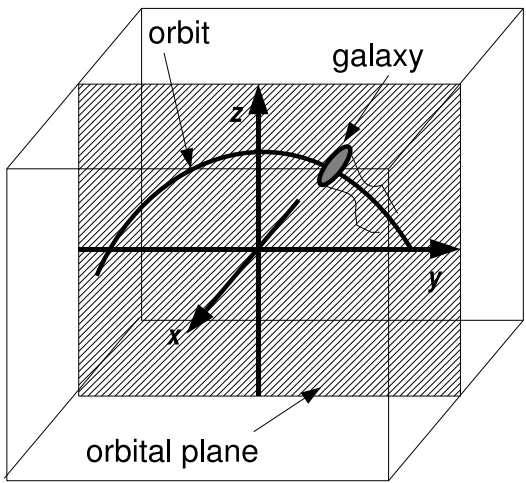}}
 \caption{Sketch showing the position of the orbital plane in the simulation
 box. Throughout this work, we use two lines-of-sight (LOS) to view the
 galaxy: along the $x$-axis (towards negative direction), which means the LOS
 is perpendicular to the orbital plane; and along the $z$-axis (towards
 negative direction), which means the LOS is in the orbital plane.}
 \label{fig:sketch}
 \end{figure}

Here we present simulations for four orbits that are summarised in
Fig.~\ref{fig:orbits_profiles}.
%
%
Three of these orbits are highly radial, whereas the fourth has a large impact
parameter.  The labels of the runs represent the cluster (C+number), small or
large impact parameter (SM or LG), fast or slow galaxy (FST or SLW),
inclination (F for near face-on, M for medium, E for near edge-on; FE for
first near face-on but then near edge-on, etc.).  In all cases, the galaxy
orbits in the $y$-$z$-plane. Figure~\ref{fig:sketch} gives a sketch of the
orbital plane and galaxy orbit in the simulation box.
%
%
Table~\ref{tab:runs} lists the simulation runs
and the amount of lost gas according to paper I.
%
\begin{table*}
\caption{List of runs. Run labels code the cluster (C+number), small or large
  impact parameter (SM or LG), fast or slow galaxy (FST or SLW), inclination
  (F for near face-on, M for medium, E for near edge-on; FE for first near
  face-on but then near edge-on, etc.). Initial galaxy coordinates $x\Gal{}_0$
  and $z\Gal{}_0$ are always zero, $y\Gal{}_0$ is given in the second
  column. The third column lists the initial galaxy velocity. The fourth and
  fifth columns list the impact parameter and velocity, respectively. The
  inclination listed in the sixth column is the angle between the galaxy's
  rotation axis and the $y$-axis. The seventh column states the fraction of lost
  gas according to paper I.}
\label{tab:runs}
\centering\begin{tabular}{lrrrrrr}
\hline
label        & $y\Gal{}_0$ & initial     & impact    & impact   & incli-       & fraction \\
             &  in         &  velocity   & parameter & velocity & nation       & of lost\\
             &   kpc       &  in $\Kms$  & in kpc    & in $\Kms$& in $\degree$ &  gas\\
\hline
C1-SM-FST-MF  &  1500     & $(0, -600,100)$ &   79    & 1904  & $-30$ & 100\%\\
C1-SM-FST-E   &  $\cdot$  & $\cdot$         & $\cdot$ & $\cdot$ & $80$  & 100\% \\
\hline
C1-LG-FST-MF  &  1300     & $(0, -600,500)$ &   492   & 1320    & $-30$ & 85\% \\
C1-LG-FST-EF  &  $\cdot$  & $\cdot$         & $\cdot$ & $\cdot$ & $-60$ & 72\% \\
C1-LG-FST-FE  &  $\cdot$  & $\cdot$         & $\cdot$ & $\cdot$ & $45$  & 77\%\\
\hline
C2-SM-SLW-FME &  800      & $(0, 0,200)$    &  105    & 1516    & $30$  & 86\%\\
C2-SM-SLW-EMF &  $\cdot$  & $\cdot$         & $\cdot$ & $\cdot$ & $-60$ & 90\%\\
C2-SM-SLW-MFM &  $\cdot$  & $\cdot$         & $\cdot$ & $\cdot$ & $-20$ & 95\%\\
\hline
C3-SM-FST-MF  & 2300      & (0,-800,200)    &  182    & 2524    & 30    & 100\%\\
\hline
\end{tabular}
\end{table*}
%


\section{Results}
%
\label{sec:results}
As the galaxy moves through the cluster, the ram pressure stripped gas forms a
tail behind the galaxy.  Some slices in the orbital plane showing
the colour-coded gas density can be found in paper I and in
Fig.~\ref{fig:res_flow}. ISM densities in the tail are around $10^{-26}\gccm$
close to the galaxy ($\sim 20\Kpc$ distance from galaxy centre) and
decrease to a few $10^{-28}\gccm$ at larger distances ($50\Kpc$).

\subsection{Projected densities} \label{sec:projections}
\begin{figure}
\centering\resizebox{0.49\hsize}{!}{\includegraphics[angle=0,width=0.32\textwidth]{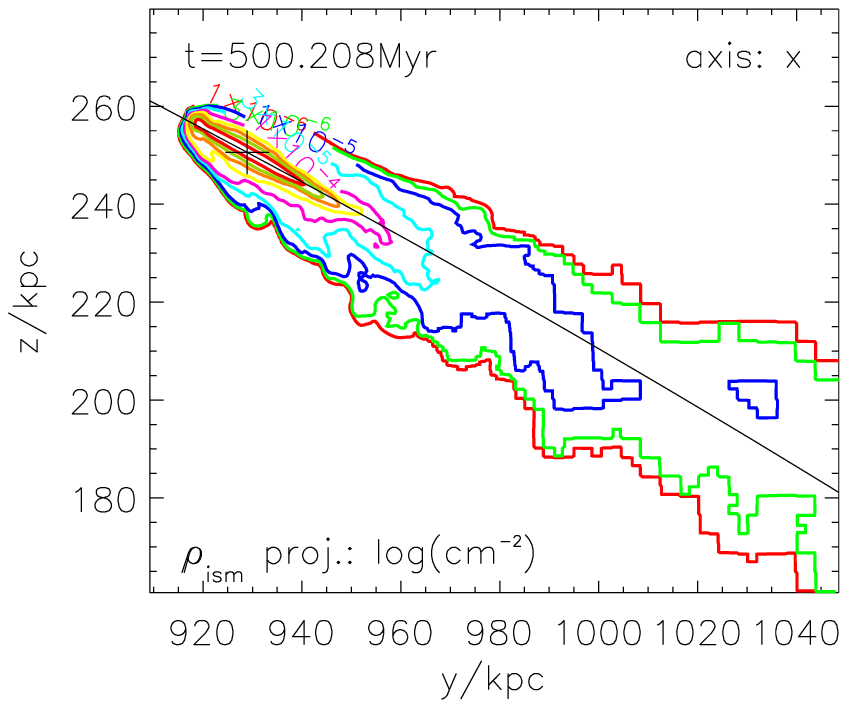}}
\centering\resizebox{0.49\hsize}{!}{\includegraphics[angle=0,width=0.32\textwidth]{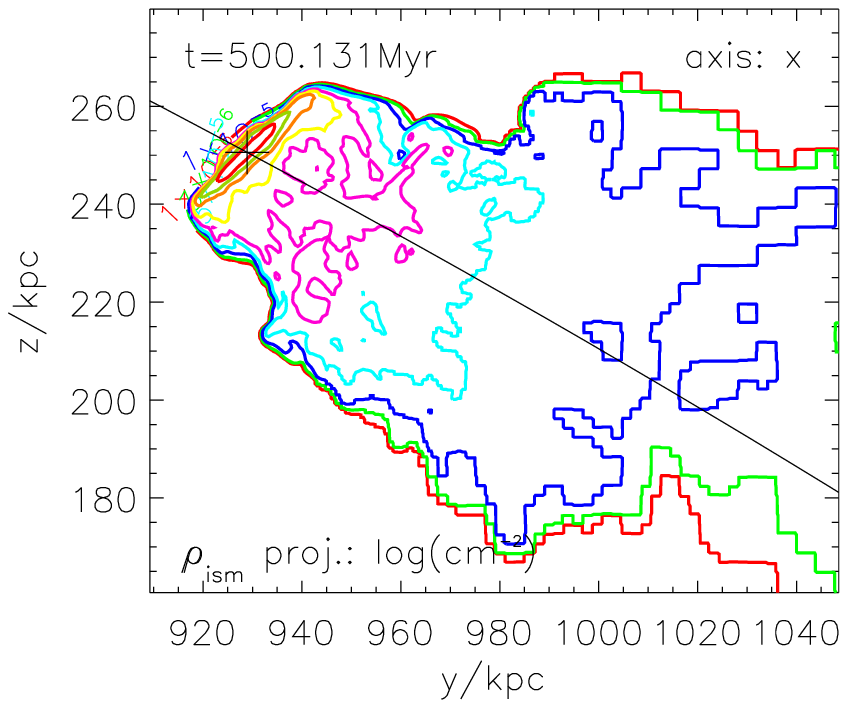}}
\centering\resizebox{0.49\hsize}{!}{\includegraphics[angle=0,width=0.32\textwidth]{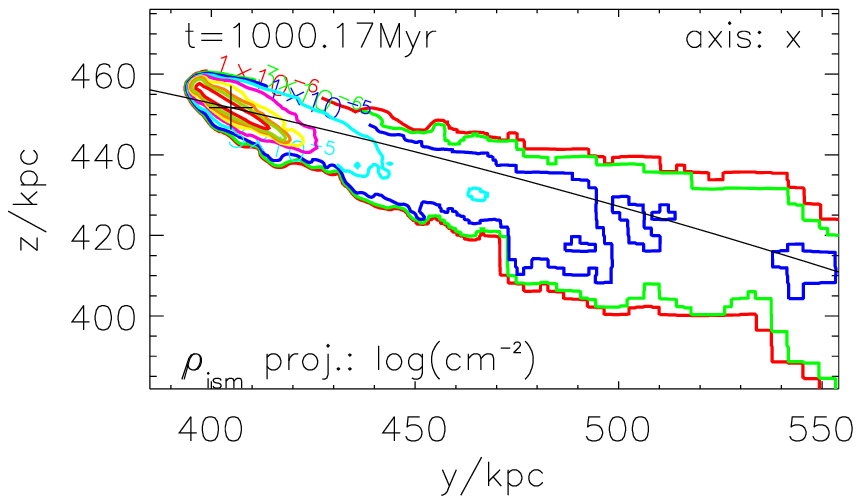}}
\centering\resizebox{0.49\hsize}{!}{\includegraphics[angle=0,width=0.32\textwidth]{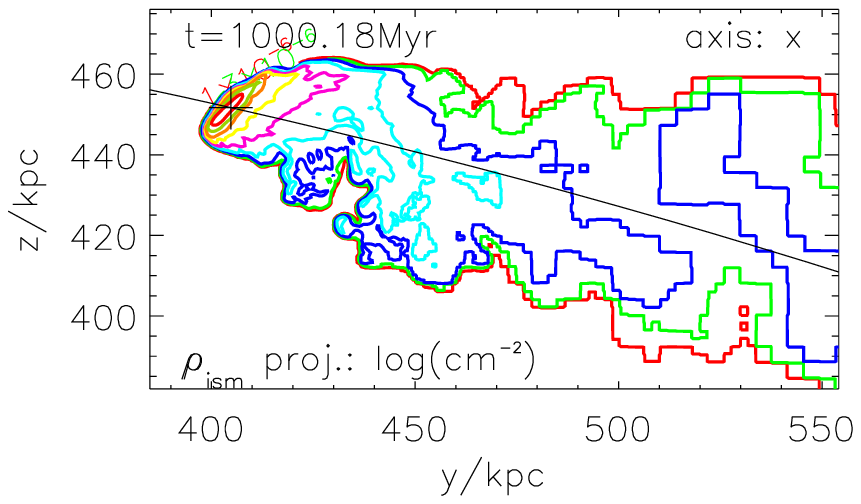}}
\centering\resizebox{0.49\hsize}{!}{\includegraphics[angle=0,width=0.32\textwidth]{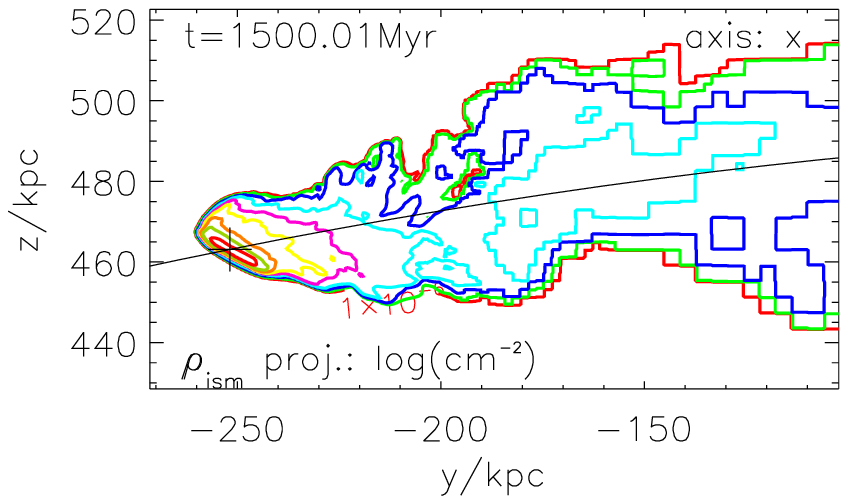}}
\centering\resizebox{0.49\hsize}{!}{\includegraphics[angle=0,width=0.32\textwidth]{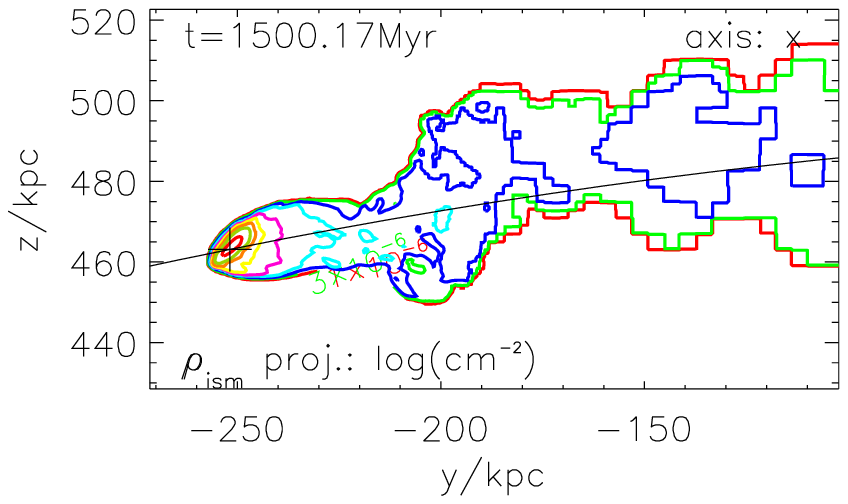}}
\centering\resizebox{\hsize}{!}{\includegraphics[angle=-90,width=\textwidth]{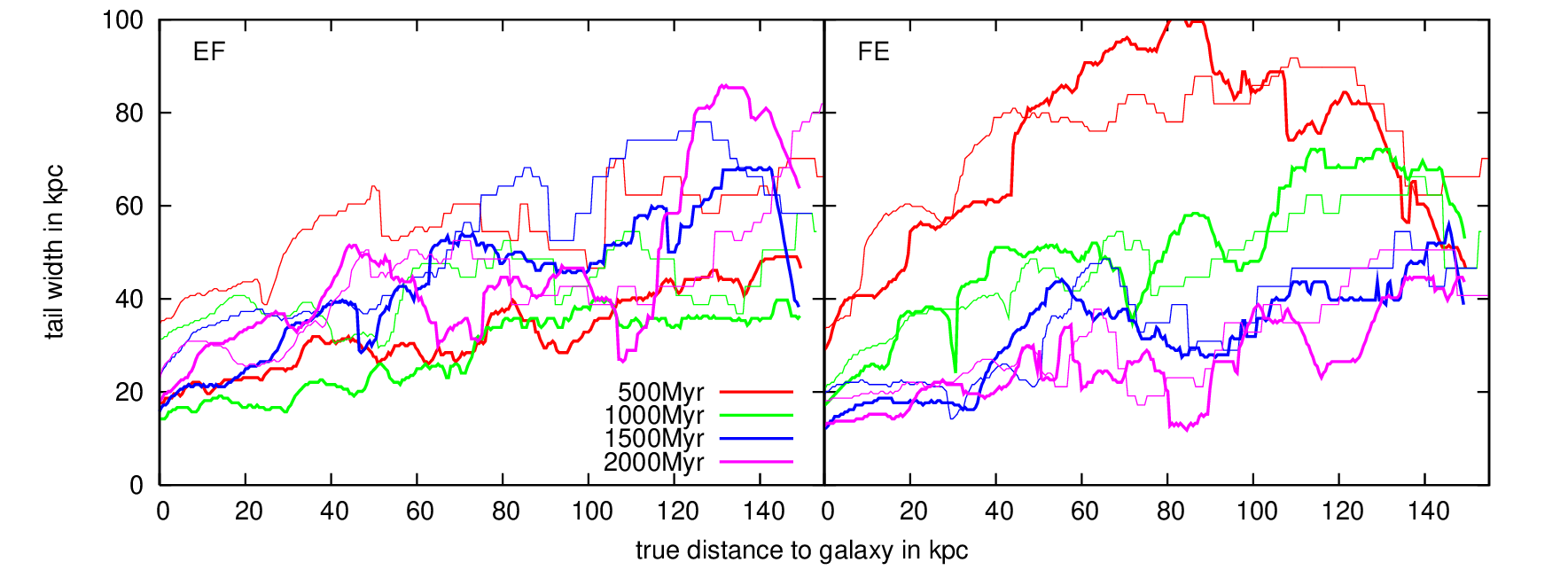}}
\caption{Projected galactic gas density for runs C1-LG-FST-EF (left column)
  and C1-LG-FST-FE (right column; same orbit, different inclination). Lowest
  contour is $10^{-6}\mathrm{g}\,\CM^{-2} \hat{=} 6.0\cdot
  10^{17}\,\CM^{-2}$. The contour spacing is half an order of magnitude. The
  time of each panel is denoted in its upper left corner. The label ``axis:
  x'' means the projection is done along $x$-axis, which means perpendicular
  to the galaxy's orbital plane in our simulations (see
  Fig.~\ref{fig:sketch}). The galactic orbit is marked by the black line. The
  plus-sign marks the galactic centre. The coordinates at the axes are in the
  cluster-centric system. The bottom row displays the tail width as a function
  of distance to the galaxy at different time-steps as measured in the
  snapshots. Same colours are for the same time (see legend), thin lines are
  for projection along grid's $z$-axis, thick lines for projection along
  $x$-axis (like this Figure).}
\label{fig:wakes_LG_FST}
\end{figure}

\begin{figure}
\centering\resizebox{0.49\hsize}{!}{\includegraphics[trim=0 25 0 25,clip,angle=0,width=0.32\textwidth]{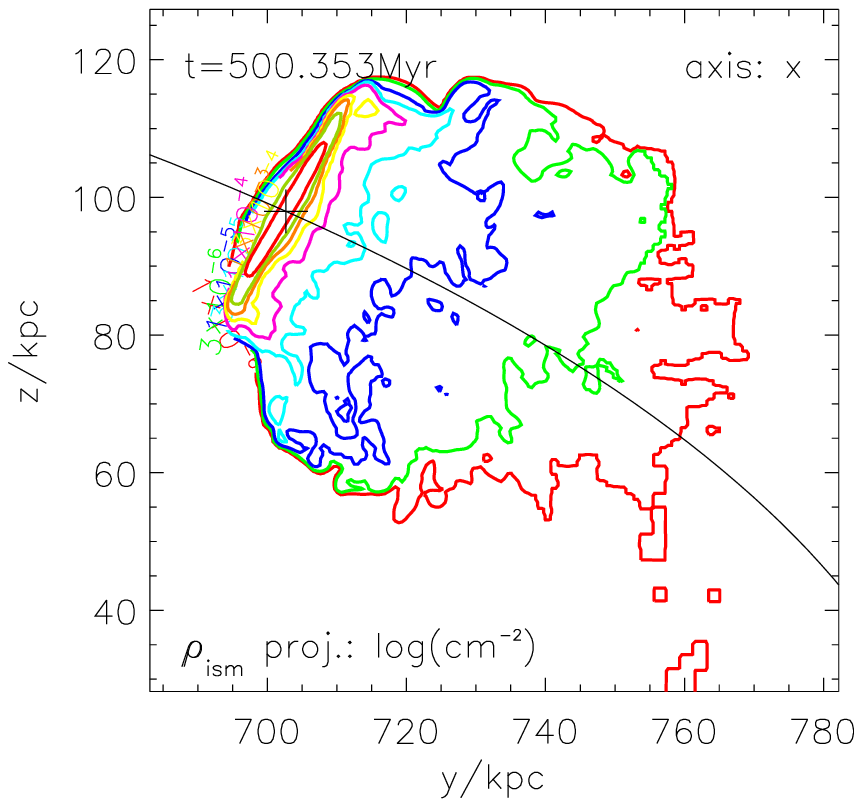}}
\centering\resizebox{0.49\hsize}{!}{\includegraphics[trim=0 25 0 25,clip,angle=0,width=0.32\textwidth]{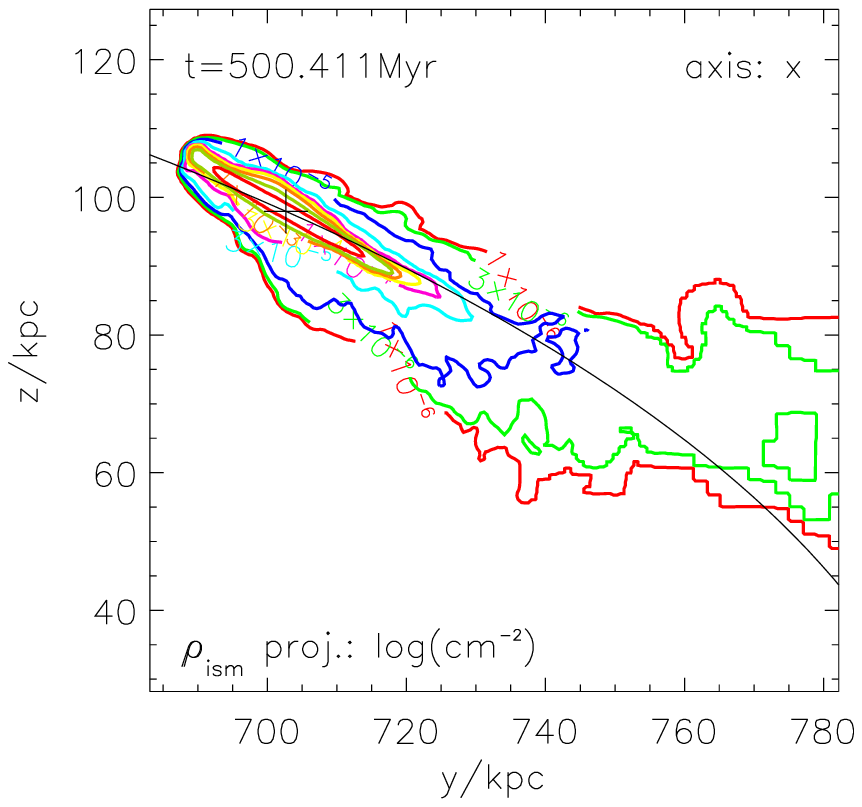}}
\centering\resizebox{0.49\hsize}{!}{\includegraphics[angle=0,width=0.32\textwidth]{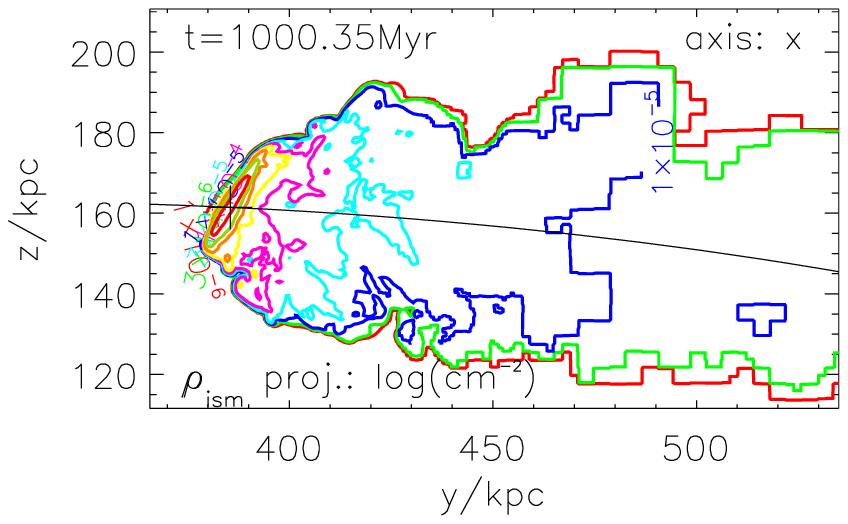}}
\centering\resizebox{0.49\hsize}{!}{\includegraphics[angle=0,width=0.32\textwidth]{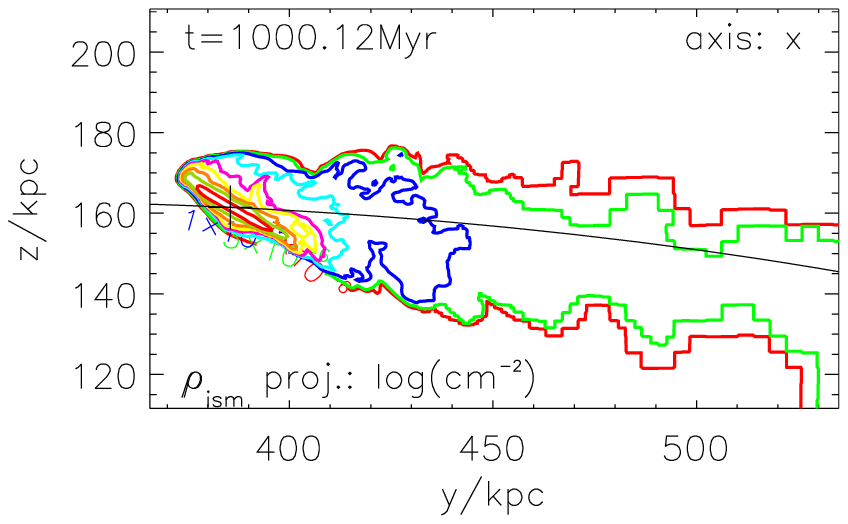}}
\centering\resizebox{0.49\hsize}{!}{\includegraphics[trim=0 25 0 25,clip,angle=0,width=0.32\textwidth]{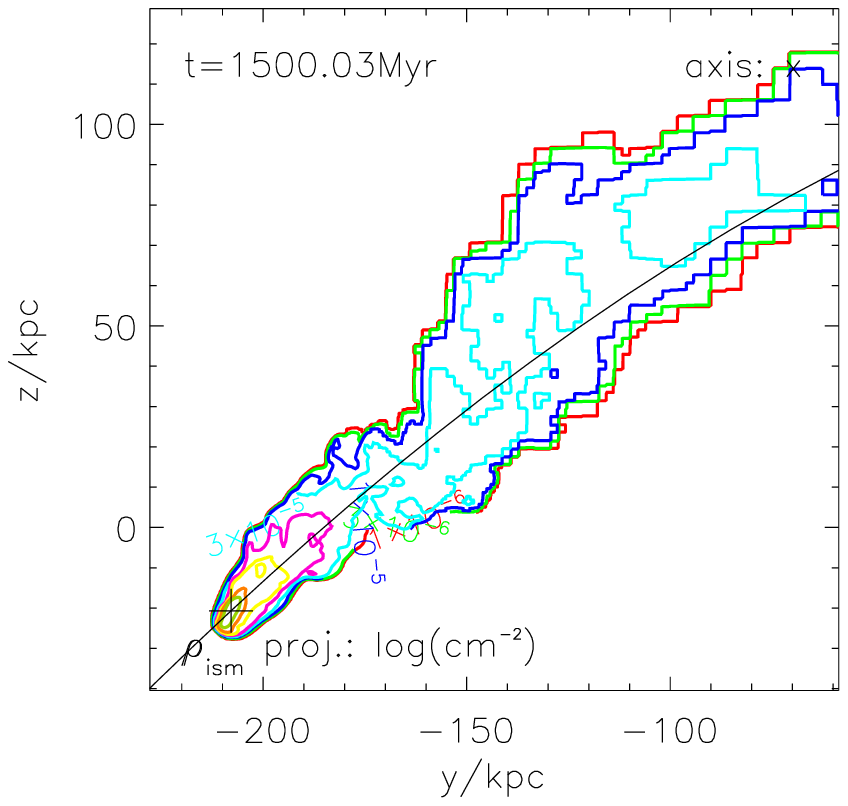}}
\centering\resizebox{0.49\hsize}{!}{\includegraphics[trim=0 25 0 25,clip,angle=0,width=0.32\textwidth]{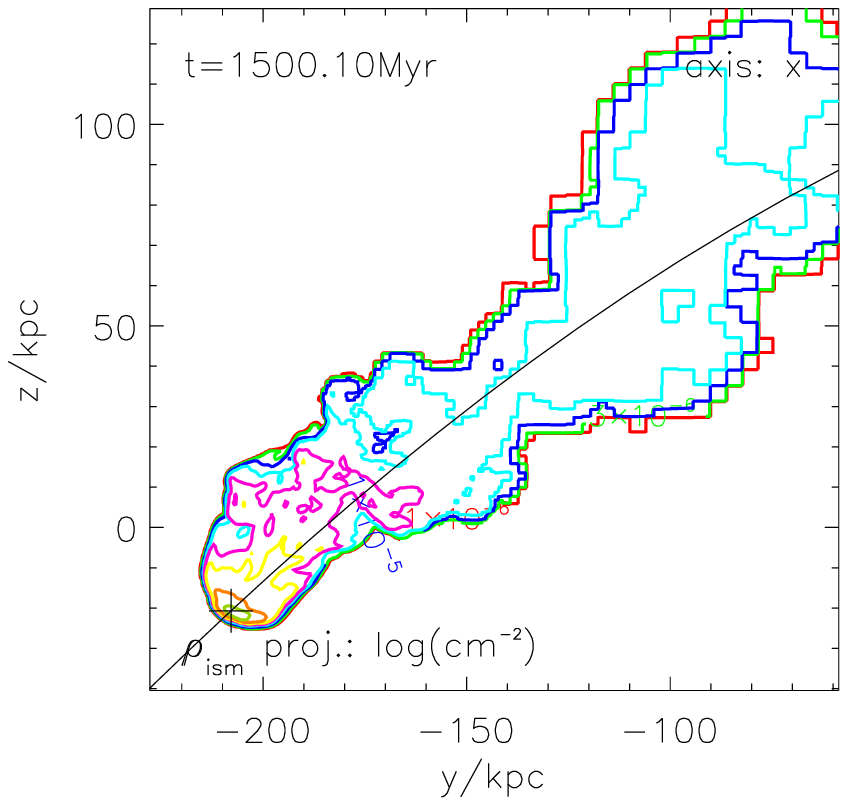}}
\centering\resizebox{\hsize}{!}{\includegraphics[angle=-90,width=\textwidth]{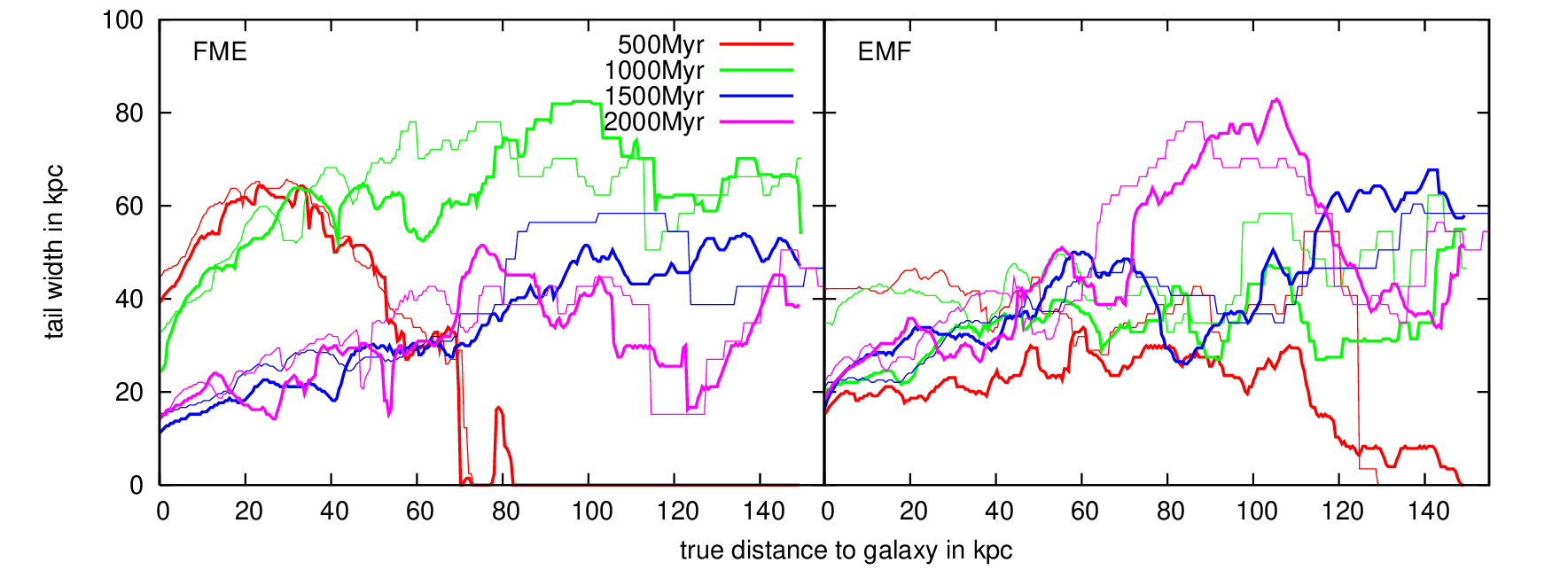}}
\caption{Same as Fig.~\ref{fig:wakes_LG_FST} but for runs
  C2-SM-SLW-FME (left column) and C2-SM-SLW-EMF (right column).}
\label{fig:wakes_SM_SLW}
\end{figure}

\begin{figure}
\centering\resizebox{0.49\hsize}{!}{\includegraphics[origin=c,angle=0]{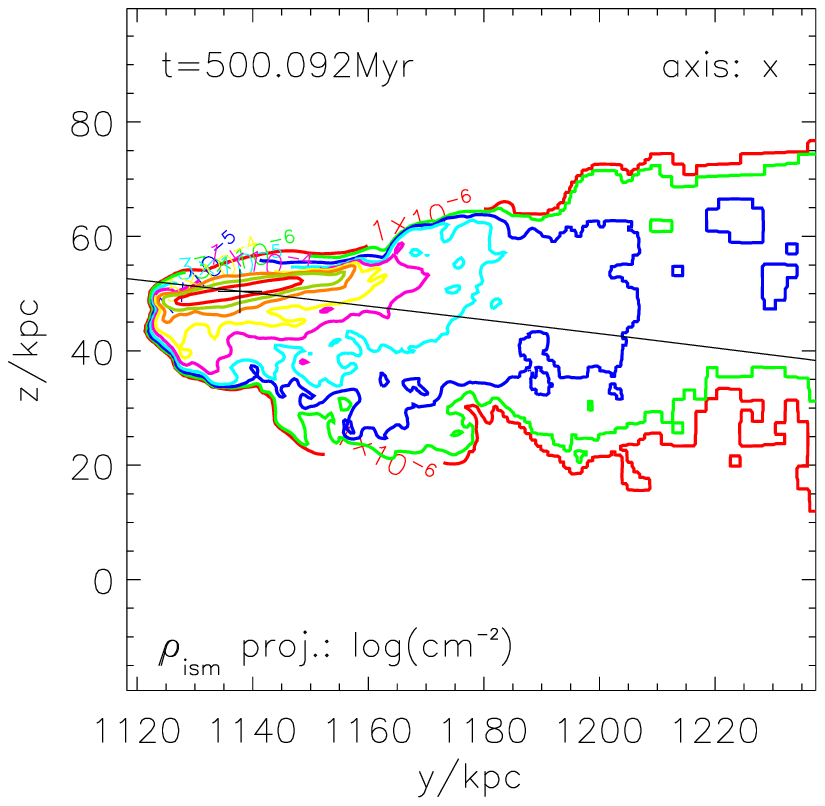}}
\centering\resizebox{0.49\hsize}{!}{\includegraphics[origin=c,angle=-90]{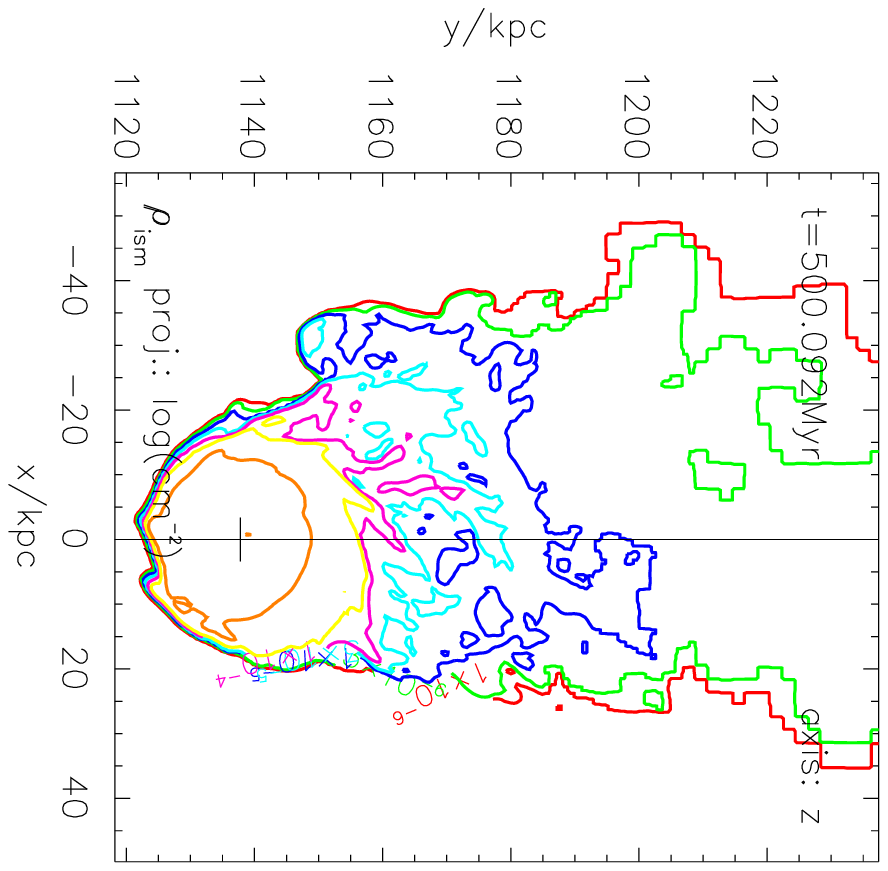}}
\centering\resizebox{0.49\hsize}{!}{\includegraphics[origin=c,angle=0]{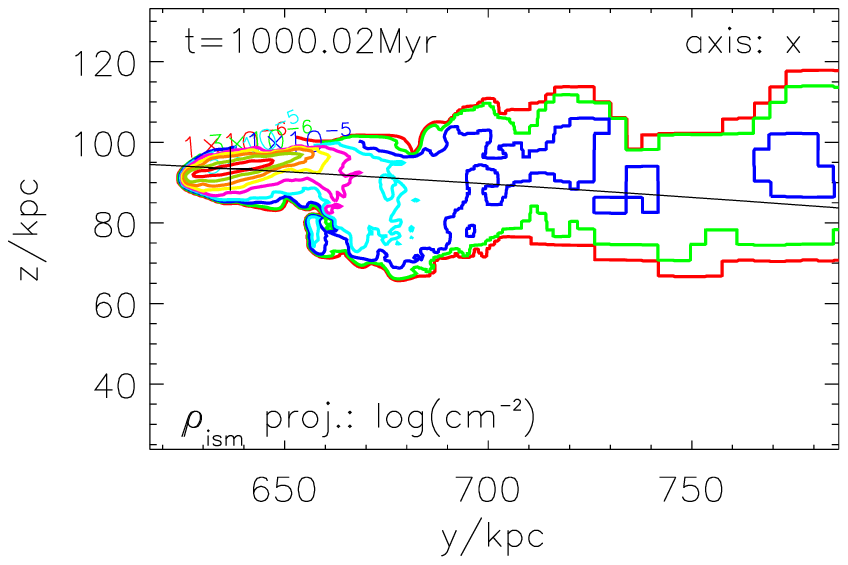}}
\centering\resizebox{0.49\hsize}{!}{\includegraphics[origin=c,angle=-90]{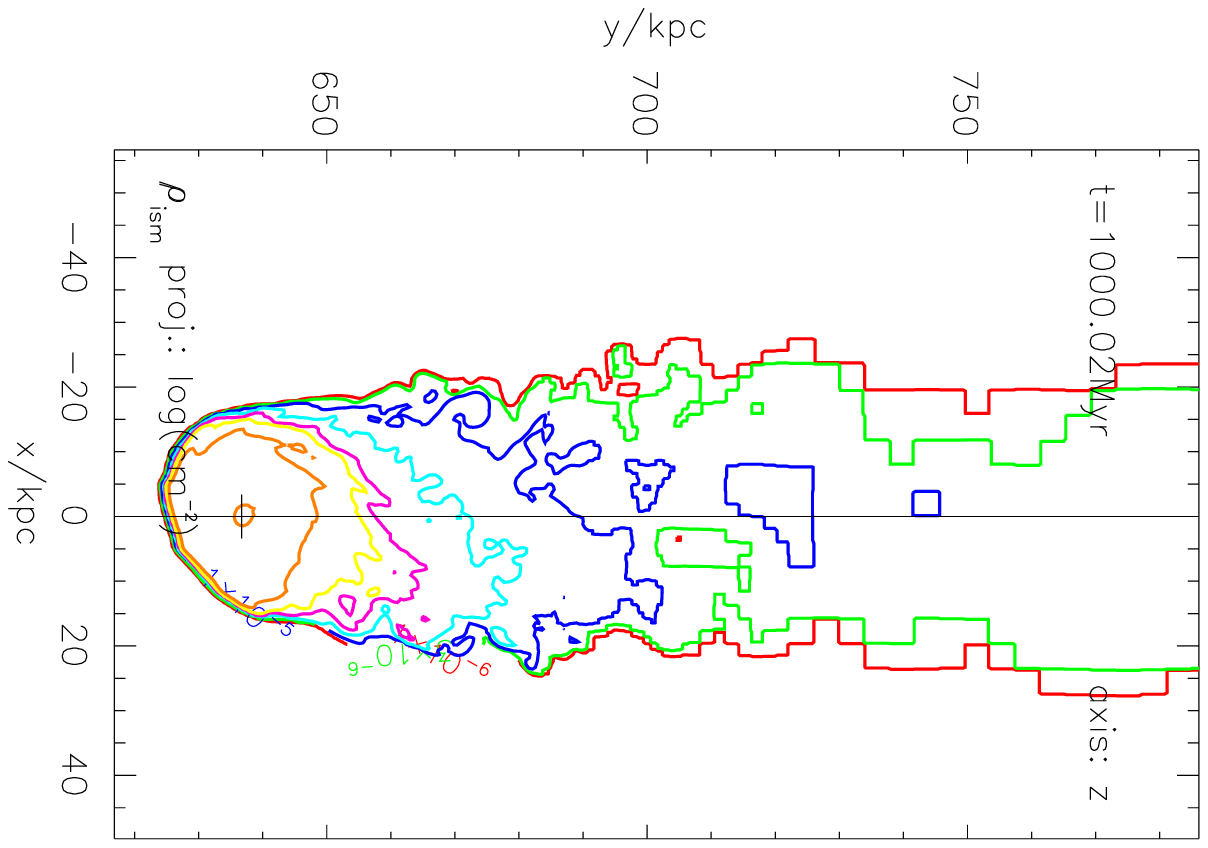}}
\caption{Like Fig.~\ref{fig:wakes_LG_FST}, but for run C1-SM-FST-E. Here, the
 lhs column shows the projection along the grid's $x$-axis, and the rhs column
 shows projections along the grid's $z$-axis.
}
\label{fig:wakes_SM_FST_xz}
\end{figure}

Figures~\ref{fig:wakes_LG_FST}, \ref{fig:wakes_SM_SLW} and \ref{fig:wakes_SM_FST_xz} show snapshots
of the projected ISM density for several runs. 
%
%
%
%
Each figure is for a different orbit. In Figs.~\ref{fig:wakes_LG_FST} and
\ref{fig:wakes_SM_SLW}, the columns compare cases with different inclination
but identical orbit. Figure~\ref{fig:wakes_SM_FST_xz} compares two different
lines-of-sight (LOS) for the same run. The bottom rows of
Figs.~\ref{fig:wakes_LG_FST} and \ref{fig:wakes_SM_SLW} summarise the temporal
evolution of the tail width as a function of distance to galaxy (see
discussion below).

The tails of stripped gas stretch along the galaxy's orbit. Three features are
prominent in nearly all snapshots: the tails show a flaring width, they
oscillate along the orbit, and are densest close to the galaxy. Details of the
structure depend on the galaxy's current inclination and the stripping
stage. In the following subsections, we describe different aspects of the tails.

\subsubsection{Tail length} \label{sec:taillength}
Typical column densities are a few $10^{19}\,\CM^{-2}$ near the galaxy and
$\sim 2\cdot 10^{18}\,\CM^{-2}$ at large distances ($>100\Kpc$).  A typical
column density sensitivity limit for current HI observations is
$10^{19}\,\CM^{-2}$. Thus, we will define the length of the tail as the extent
of the light blue contour ($\sim 2\cdot 10^{19}\,\CM^{-2}$) behind the galaxy.
In this sense, a typical tail length is $40\Kpc$.  Tails of this length can be
found at early times ($\lesssim 1\Gyr$) of simulations C1-LG-FST-\ldots (two
top rows of Fig.~\ref{fig:wakes_LG_FST}), in runs C1-SM-FST-\ldots
(Fig.~\ref{fig:wakes_SM_FST_xz}) and at early times ($\lesssim 1\Gyr$) of run
C3-SM-FST-MF. We wish to draw attention to the fact that at these moments the
galaxy is still far, namely 400 kpc to 1 Mpc, from the cluster centre. In the
Coma-like cluster C3 (run C3-SM-FST-MF), at $t=500\Myr$, the galaxy is even
$1800\Kpc$ from the cluster centre, and still shows a $50\Kpc$ long
tail. Shorter tails are found during the first Gyr of runs C2-SM-SLW-\ldots
(Fig.~\ref{fig:wakes_SM_SLW}). The longest tails are found during peri-centre
passage of the same run, they extend to beyond $150\Kpc$. Also the snapshots
at $1.5\Gyr$ of runs C1-LG-FST-\ldots reveal tails of $\sim 100\Kpc$ length,
although -- due to the large impact parameter of this orbit -- also in this
case the galaxy is still more than $400\Kpc$ from the cluster centre.

The length of the tail is influenced mainly by two parameters: the ram
pressure and the galaxy's velocity. The ram pressure determines how much gas
the galaxy loses, whereas the orbital velocity determines over which volume
the lost gas is spread. Clearly, high ram pressures lead to a high mass loss
rate. However, as high ram pressures are usually associated with high
velocities, the stripped gas is also distributed over a larger orbital length
or volume. These two parameters can be combined into the mass loss per orbital
length, which is the crucial quantity that characterises many tail properties.
%
\begin{figure*}
\includegraphics[angle=-90,width=\textwidth]{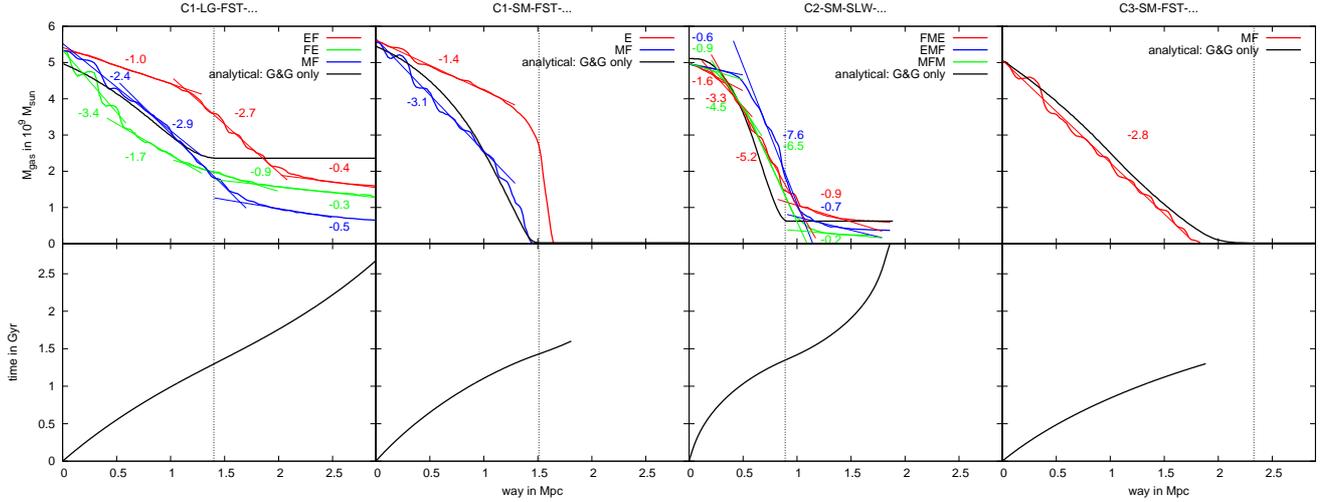}
\caption{Evolution of gas disc mass as a function of distance covered by the
  galaxy (top row). Each column is for one orbit (see title of column). The
  colours code runs with different inclinations, see legend. The thin vertical
  lines mark the peri-centre passage. The bottom row displays the relation of
  time and covered distance. We applied piecewise linear fits to the gas disc
  mass as a function of covered distance. Each piece is labelled with its
  slope in $10^6 M\Sun /\Kpc$.}
\label{fig:mass_way}
\end{figure*}
%
Figure~\ref{fig:mass_way} displays the evolution of the remaining gas disc
mass for all runs as a function of distance covered by the galaxy. The slope
of these functions is the mass loss per orbital length.  We fitted them
piecewise with linear functions.  Each piece is labelled with its slope in
$10^6 M\Sun /\Kpc$. The mass loss per orbital length ranges between a little
below $10^6M\Sun/\Kpc$ to about $7\cdot 10^6M\Sun/\Kpc$. Interestingly, this
quantity is largest  in runs C2-SM-SLW-\ldots during the peri-centre passage,
although the ram pressure along most parts of this orbit is smaller than for
all other orbits. Even during peri-centre passage, the ram pressure for this
orbit is the second lowest. Here, the low orbital velocity causes a high mass
loss per orbital length. The case of the Coma-like cluster C3 illustrates the
opposite extreme of the same issue: Here the ram pressure is always higher
than in all other runs, but the high orbital velocity of the galaxy leads to
only a medium mass loss per orbital length. 
%
\begin{figure}
\centering\resizebox{0.7\hsize}{!}{\includegraphics[angle=-90]{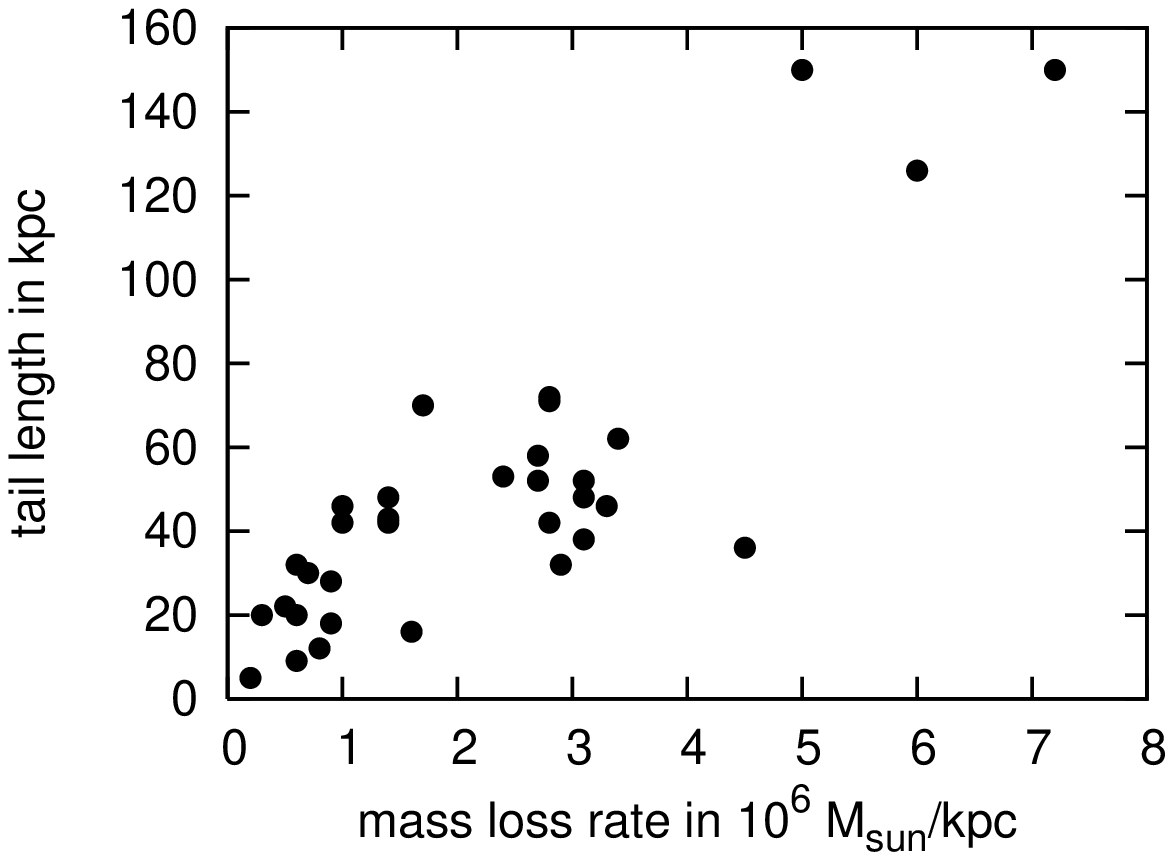}}
\caption{Tail length versus mass loss per orbital length. We define
    the tail length as the extent of the light blue contour ($\sim 2\cdot
    10^{19}\,\CM^{-2}$) behind the galaxy. See also text (Sect.~\ref{sec:taillength}).}
\label{fig:length_massloss}
\end{figure}
%
As expected, we find that the mass loss per orbital length correlates with the
tail length (see Fig.~\ref{fig:length_massloss}). However, additional effects
like the tail width play a role and lead to some scatter in
  this relation. Also a projection non-perpendicular to the
galaxy's direction of motion can enhance the projected density of the tail
while making it appear shorter.
%

\subsubsection{Tail width} \label{sec:tailwidth}
According to simple analytical estimates (\citealt{landau}), the width, $w$,
of the wake behind a body moving through a fluid scales with distance to the
body, $d$, as $\sqrt{d}$ in the case of a laminar flow and as $d^{1/3}$ in the
case of a turbulent flow. In our simulations, the flow is clearly
turbulent. More precisely, at large distances behind the body, the width of
the wake scales as
\begin{equation}
w\sim 2\left( \frac{F d}{\rho U^2}  \right)^{1/3},
\end{equation}
where $F$ is the drag force the body experiences, $\rho$ is the density
of the fluid and $U$ is the flow velocity. The drag force can be written
as $F=0.5 \CW A \rho U^2$, where $\CW$ is the drag coefficient and $A$ the
cross-section of the body. This leads to 
\begin{eqnarray}
w &\sim& 2\left(0.5 \CW A d  \right)^{1/3} \nonumber \\
&= & 50\Kpc \left(\frac{\CW}{1} \right)^{1/3} \left(
\frac{r}{10\Kpc} \right)^{2/3} \left( \frac{d}{100\Kpc} \right)^{1/3}, 
\label{eq:wakewidth}
\end{eqnarray}
where we have inserted values typical for our case. Here, $r$ is the radius of
the remaining gas disc.

In our simulations, tail widths range between 20 to 50 kpc at a distance of
$\sim 25\Kpc$ to the galaxy centre and 30 to 80 kpc at a distance of $\sim
100\Kpc$ to the galaxy centre. Near the galaxy, the tail width is similar to
the galaxy's cross-section with respect to the ICM wind direction. In an
order-of-magnitude comparison, the simulation results match the analytical
expectation. However, a direct application of the analytical scaling relation
to our simulations is not possible, as in our simulations many assumptions
that went into the analytical relation are not fulfilled: The flow past the
galaxy is not static but, both, flow density and velocity change, the body
(the galactic gas disc) varies in size. Furthermore, here we are interested in
the wake close to the body and not far away from it. Thus it is not reasonable
to fit the $d^{1/3}$ law to the galaxy's tail.

Nonetheless, in order to quantify the flaring ratio of the tails, we
applied linear fits to the tail widths as a function of distance to the
galaxy, where examples are given in the bottom rows of
Figs.~\ref{fig:wakes_LG_FST} and \ref{fig:wakes_SM_SLW}. We fixed the $y$-axis
cut to the galaxy's width and fitted the slope to the data. This slope
is the flaring ratio.

For each run, we have derived the flaring ratio for two lines-of-sight: once along the grid's $x$-axis and once along
the grid's $z$-axis. The $x$-axis is always perpendicular to the galaxy's
orbit, whereas the $z$-axis is in the galaxy's orbital plane.
%
\begin{figure}
\centering\resizebox{0.7\hsize}{!}{\includegraphics[angle=-90]{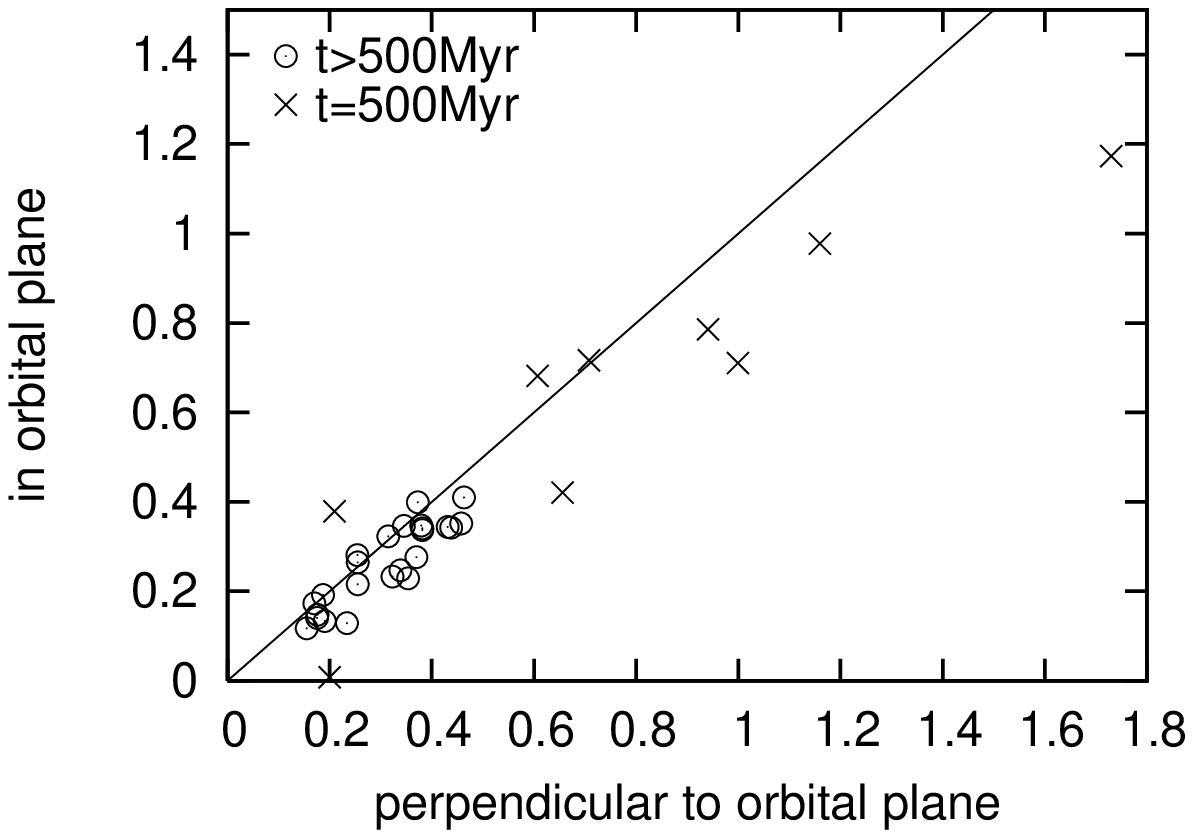}}
\caption{Comparison of measured tail flaring ratios along the two
  different line-of-sights: perpendicular to the orbital plane and in the
  orbital plane. For point symbols see legend. For the definititon of the
  flaring ratio see text, Sect.~\ref{sec:tailwidth}.}
\label{fig:slopes_xz}
\end{figure}
%
Figure~\ref{fig:slopes_xz} demonstrates that the flaring ratios of the tails
are generally independent of the direction of the line-of-sight. The only
exceptions are some of the early ($t=500\Myr$) snapshots, where the galaxy's
remaining gas disc is still large.
%
\begin{figure}
\centering\resizebox{0.7\hsize}{!}{\includegraphics[angle=-90]{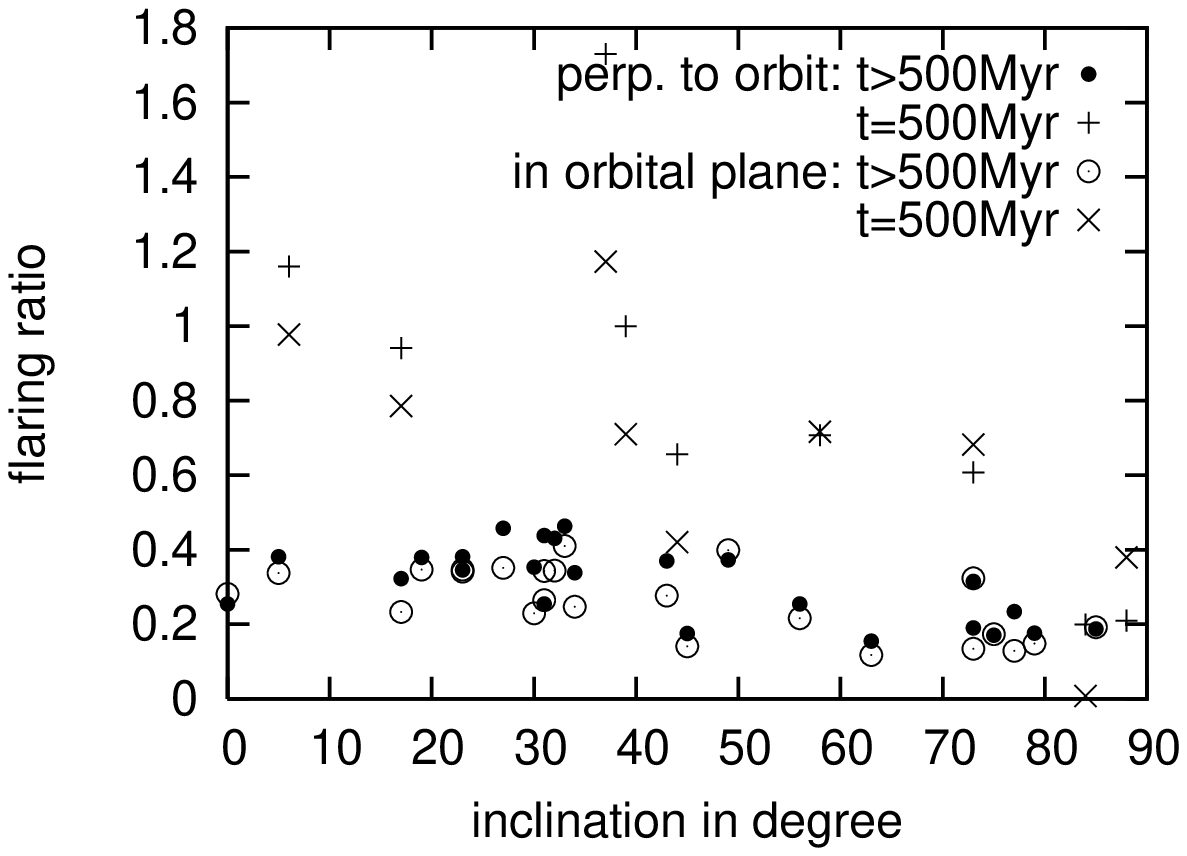}}
\centering\resizebox{0.7\hsize}{!}{\includegraphics[angle=-90]{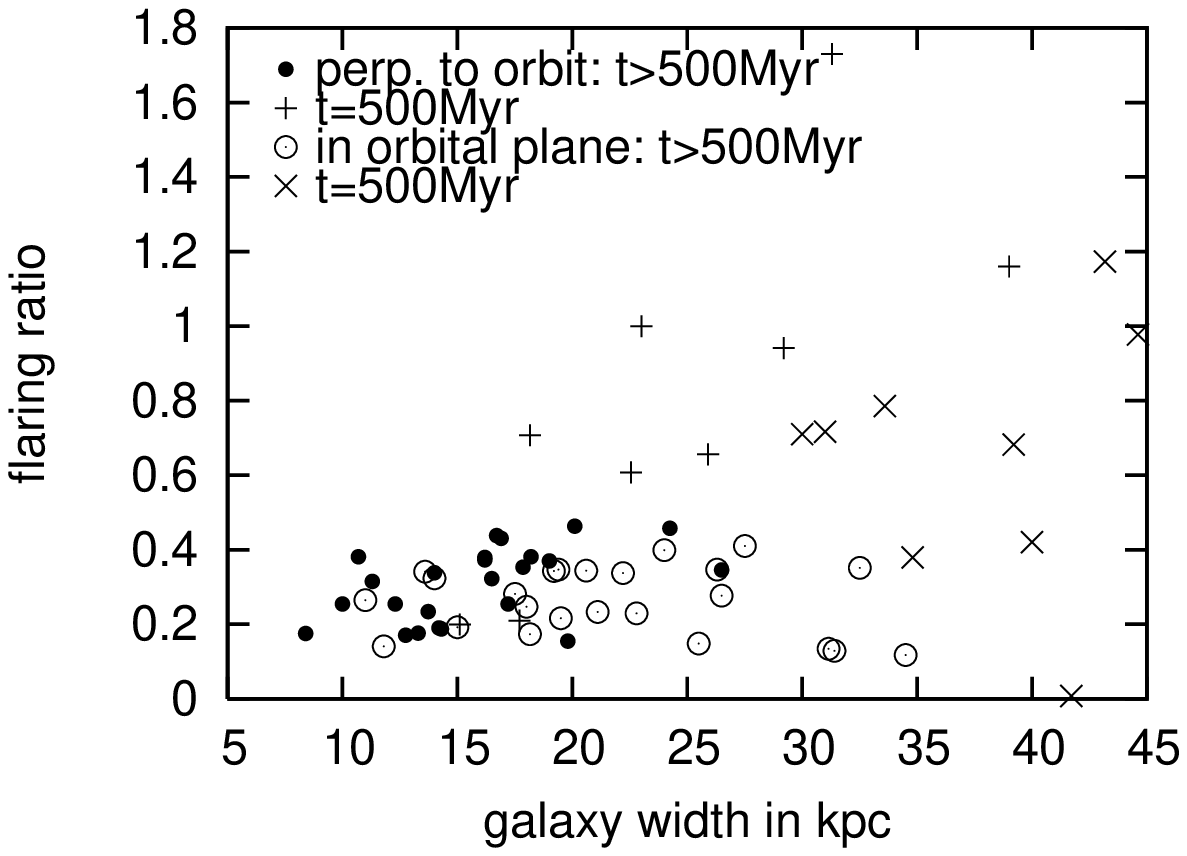}}
\centering\resizebox{0.7\hsize}{!}{\includegraphics[angle=-90]{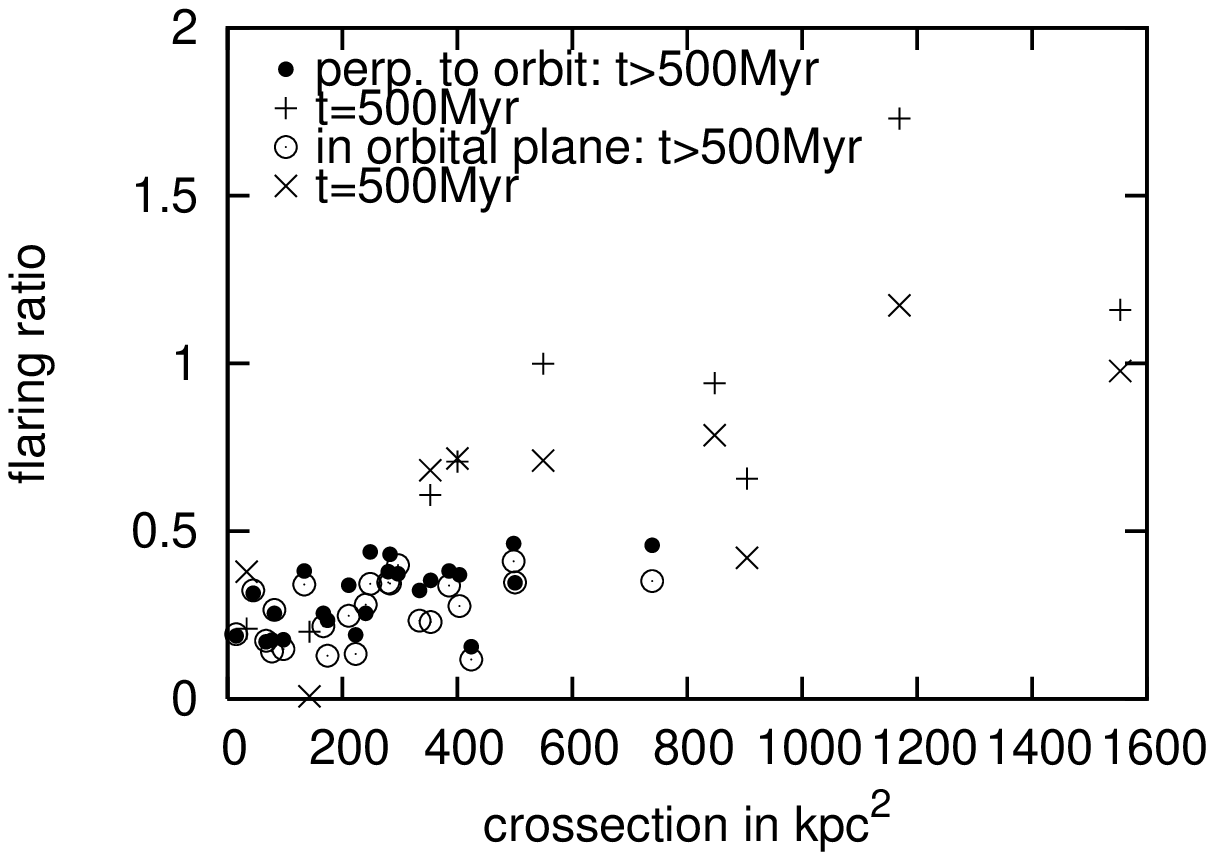}}
\centering\resizebox{0.7\hsize}{!}{\includegraphics[angle=-90]{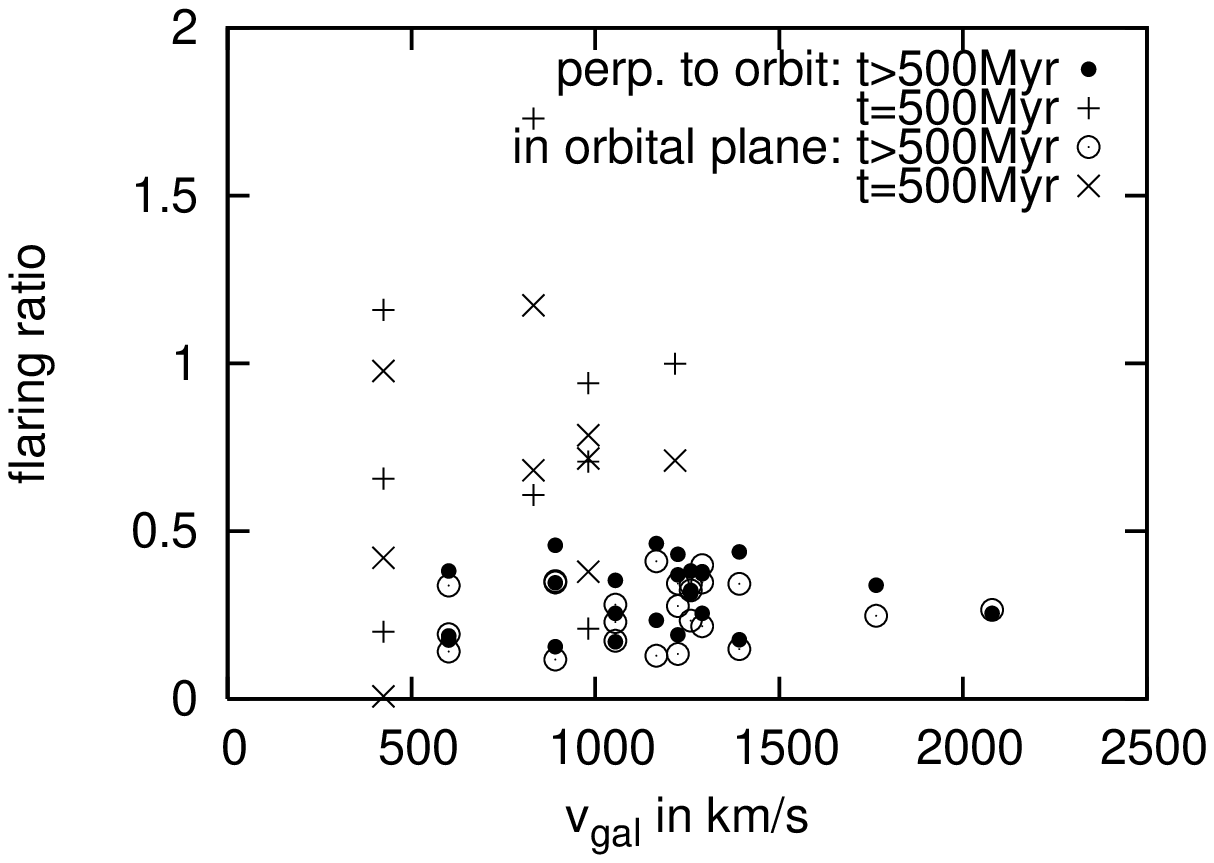}}
\centering\resizebox{0.7\hsize}{!}{\includegraphics[angle=-90]{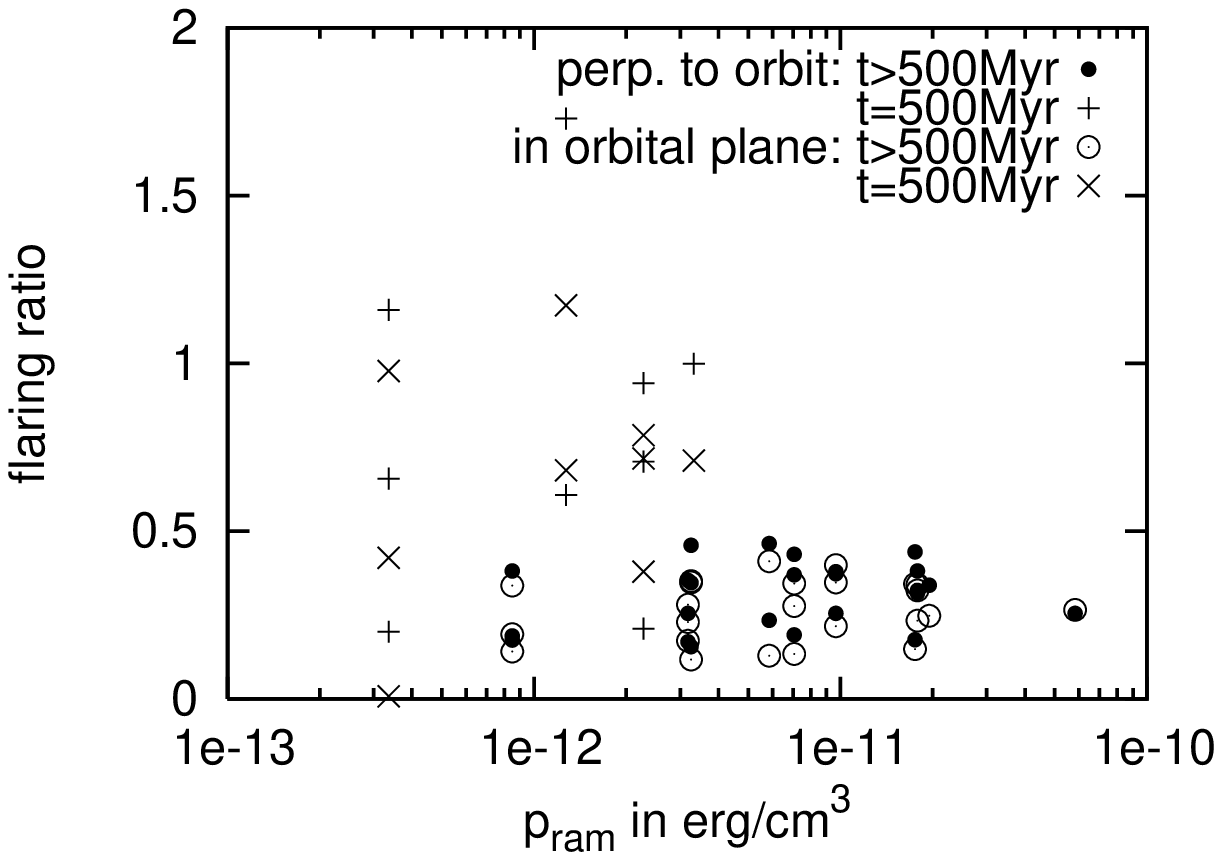}}
\caption{Tail flaring ratios as a function of inclination angle (top panel),
  as a function of galaxy diameter (second panel), as a function of galactic
  cross-section $A=\pi r\Disc^2 \cos(i)$ (third panel), as a function
  of the galaxy's velocity (fourth panel), and as a function of ram pressure
  (bottom panel). For point symbols see legend. For the definititon of the
  flaring ratio see text, Sect.~\ref{sec:tailwidth}.}
\label{fig:slopes}
\end{figure}
%
Figure~\ref{fig:slopes} displays the dependence of the flaring ratio on
several quantities. For high ram pressures and high galaxy velocities,
the flaring ratios are rather small, which reflects that for these cases the
gas disc is already stripped heavily and thus is small. For smaller velocities
and ram pressures, the flaring ratios show a large scatter. In addition to
this,
figure~\ref{fig:slopes} displays the dependence of the flaring ratio on
the galaxy's inclination with respect to the ICM wind direction (top panel)
and to the galaxy's current diameter (second panel). Both plots reveal
correlations in the sense that galaxies moving near face-on (small inclination
angle) and galaxies with large diameters produce tails with stronger
flaring. The only ``exceptions'' are again the early cases
($t=500\Myr$). Combining these two correlations, we plotted the flaring ratios
over the galaxy's cross-section, $A=\pi r\Disc^2 \cos(i)$, with respect to the
ICM wind direction, where $r\Disc$ is the radius of the remaining gas
disc. This plot reveals a clear correlation that small cross-sections lead to
little flaring and vice versa. 
A galaxy can have a small cross-section either
if it is already heavily stripped or if it is moving near edge-on. The latter
cases are especially interesting. If such a galaxy is seen perpendicular to
the disc, the tail is broader than if it is seen edge-on. However, the flaring
for both line-of-sights is small.
This reflects the known dynamics of turbulent wakes: according to
  Eq.~\ref{eq:wakewidth}, at a given distance, $d$, behind an object, the
  flaring of the wake scales as $\frac{\partial w}{\partial d}\propto
  A^{1/3}$, i.e. the wakes of larger objects show stronger flaring. As
  emphasized above, however, several assumptions that went into
  Eq.~\ref{eq:wakewidth} are not fulfilled in our simulations, thus there is
  little hope of recovering the $A^{1/3}$ relation. We only find that a larger
  galactic crossection leads to stronger flaring. Nonetheless, 
these characteristics suggest that the flaring is determined by the turbulence
in the ICM flow past the galaxy.

Effects commonly seen in simulations of harassment and tidal
  interactions are tilts and warps of the galactic discs (e.g. \citealt{moore98,mastropietro05a}). These may change the
  orientation of the galactic disc, i.e. stars and gas, continuously. This
  differs from what we have assumed here. However, the effect of the cluster's
  tidal field is strong only near the cluster centre, where the gas disc is
  already heavily ram pressure stripped. Of course interactions with other
  cluster galaxies may modify the gravitational potential of both galaxies and
  thus also influence the ram pressure stripping efficiency and
  characteristics of the wake. In this work here, we have isolated the effect
  of ram pressure stripping.

The description of the tail widths above considered the full tails.  If we
again adopt a column density limit of $\sim 10^{19}\,\CM^{-2}$ (light blue
contour), the tail flaring is much less pronounced, often not
even detectable. Generally, then the tail widths are similar to the
cross-section of the remaining as disc.
%

\subsubsection{Mass distribution along the tail}
%
%
\begin{figure*}
\includegraphics[angle=-90,width=\textwidth]{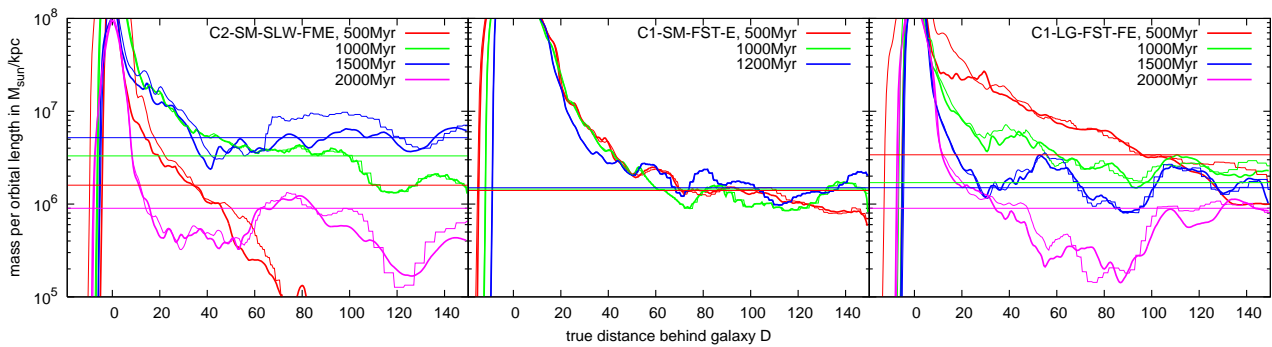}
\caption{Galactic gas mass per orbital length as a function of distance to the
  galaxy for three exemplary runs. Each panel is for one run (see legends),
  the line colours code different time-steps, see also legends. The thin
  horizontal lines mark the predicted mass per orbital length as derived in
  Fig.~\ref{fig:mass_way}.}
\label{fig:massperlength}
\end{figure*}
%
Fig.~\ref{fig:massperlength} shows the galactic gas mass per orbital length as
a function of distance to the galaxy, $D$, for three exemplary cases: we have
integrated the projected gas densities perpendicular to the galaxy's current
direction of motion. As the orbits are only slightly curved, for the $150\Kpc$
behind the galaxy, this corresponds to very good approximation to an
integration perpendicular to the orbit. The peak at $D=0$ is the gas still
trapped in the galaxy. With increasing $D$, the mass per orbital length
decreases and often saturates at $D\sim 70\Kpc$. However, this general
behaviour is superimposed with substantial substructure. Especially in the
tails at early simulation times ($t=500\Myr$), the mass per orbital length
does not saturate but continues to decrease.

The gas mass per orbital length should record the gas loss history of the
galaxy. In Fig.~\ref{fig:mass_way} we have applied piecewise linear fits to
the evolution of the galaxies' gas disc masses as functions of covered
distance. The slopes give the mass loss per orbital length and thus should be
the same as the gas mass per orbital length in the tails. To
compare both quantities, we have plotted the mass losses per orbital length as
derived from Fig.~\ref{fig:mass_way} as thin horizontal lines in
Fig.~\ref{fig:massperlength}, where the colours code the time-steps according
to the legend. Indeed, when the mass per orbital length saturates, it
saturates at approximately the level marked by the thin horizontal
lines. There are several reasons for the differences between the derived mass
loss per orbital length and observed gas mass per orbital length: First of
all, the mass loss per orbital length is no constant but also varies on short
time- and length-scales. Secondly, the stripped gas has to be accelerated away
from the galaxy, which also influences the gas distribution along the tail.
Moreover, gas can be temporarily trapped in turbulent eddies, which places it
in a position that cannot be predicted by the mass loss and acceleration.

\subsection{Velocities in the wakes}
Some slices in the orbital plane showing the gas density and the velocity
field can be found in Fig.~\ref{fig:res_flow}. The ICM is flowing around the
gas disc. In cases where the galaxy moves supersonically, a bow shock is
prominent. In the wake, the velocities are generally smaller than the ICM wind
velocity. Additionally, the velocity field shows a turbulent structure. 

\begin{figure*}
\includegraphics[angle=-90,width=0.8\textwidth]{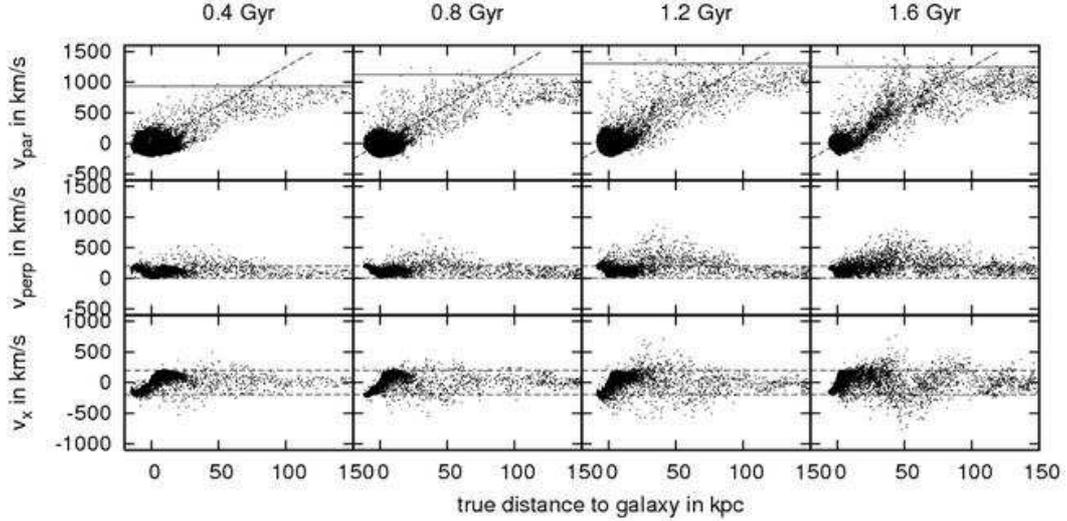}
\caption{Three different velocity components in the galaxy's tail, as a
 function of distance to the galaxy. For run C1-LG-FST-EF. The top row shows
 the velocity component along the tail. The middle row shows the component
 perpendicular to the galaxy's direction of motion, and the bottom row the
 component along the grid's $x$-axis, i.e. perpendicular to the galaxy's
 orbital plane. Each column is for one time-step, as indicated. The velocities
 are given in the galactic rest frame. An ISM density-weighted random subset
 of grid cells is shown.  \newline The solid horizontal lines in the top row
 panels mark the current ICM wind velocity, i.e. the negative of the galaxy's
 current velocity. An identical diagonal dashed line is plotted in the top row
 panels in order to aid the eye to compare the slope of the velocity
 gradient. The horizontal lines in the middle row panels mark 0 and $200\Kms$,
 the horizontal lines in the bottom row panels mark $\pm 200\Kms$.}
\label{fig:vel_L_EF}
\end{figure*}

\begin{figure*}
\centering\resizebox{0.8\hsize}{!}{\includegraphics[angle=-90,width=\textwidth]{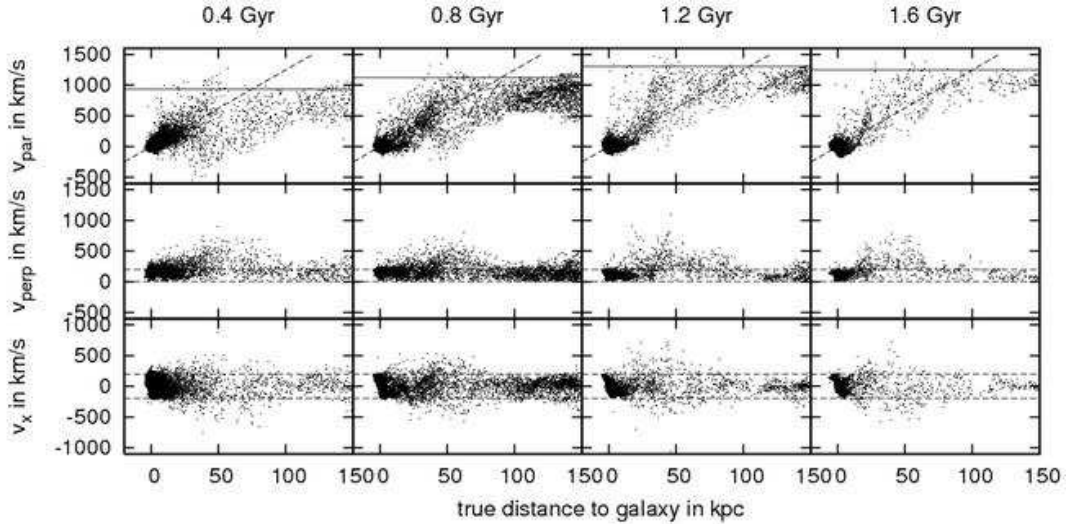}}
\caption{Same as Fig.~\ref{fig:vel_L_EF}, but for run C1-LG-FST-FE.}
\label{fig:vel_L_FE}
\end{figure*}

\begin{figure*}
\centering\resizebox{0.7\hsize}{!}{\includegraphics[trim=0 0 50 0,clip,angle=-90,width=0.5\textwidth]{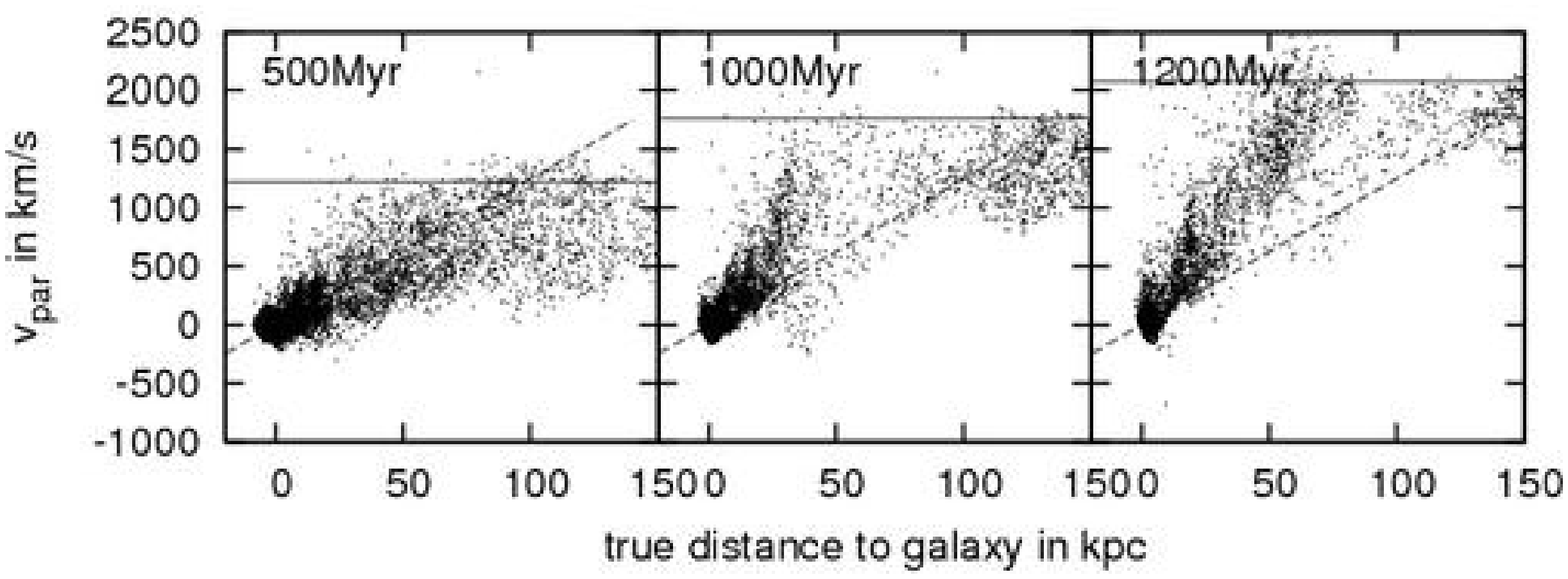}}
\centering\resizebox{0.7\hsize}{!}{\includegraphics[trim=25 0 50 0,clip,angle=-90,width=0.5\textwidth]{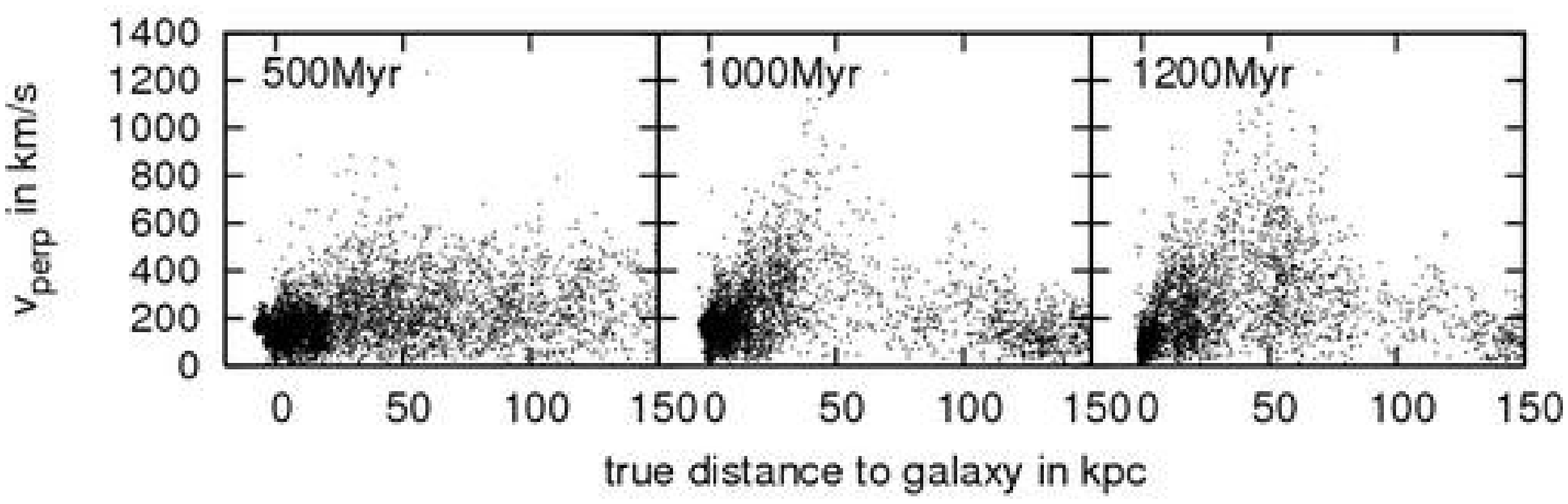}}
\centering\resizebox{0.7\hsize}{!}{\includegraphics[trim=25 0 0 0,clip,angle=-90,width=0.5\textwidth]{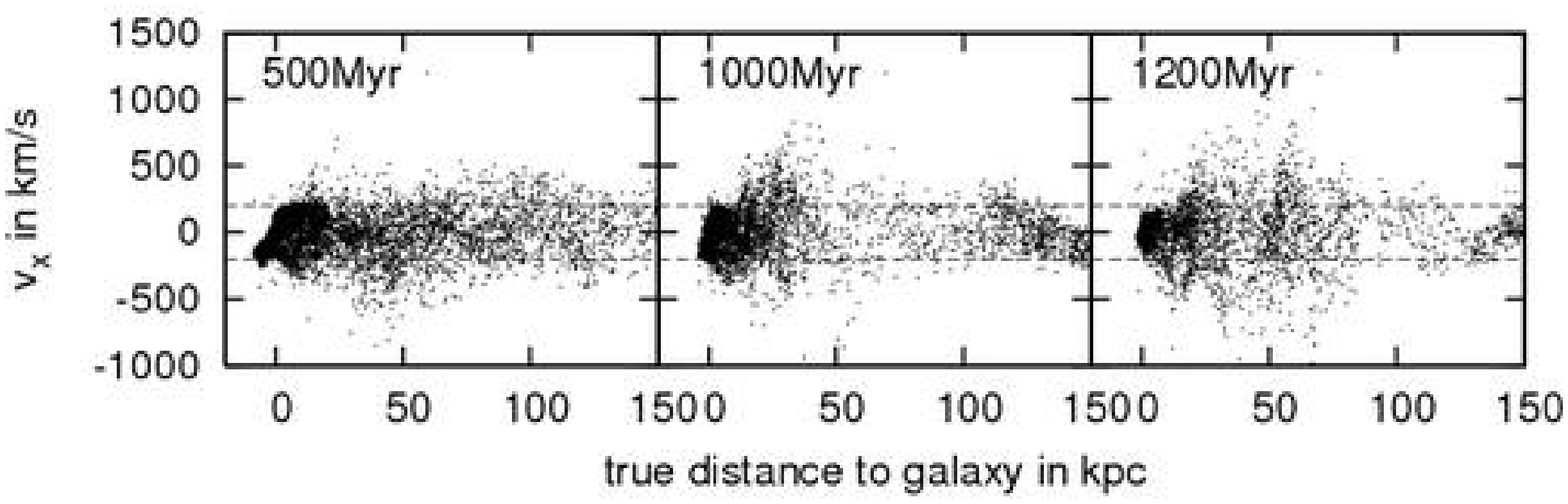}}
\caption{Same as Figs.~\ref{fig:vel_L_EF}, but for run C3-SM-FST-MF.}
\label{fig:vel_C3-SM_FST_MF}
\end{figure*}

Figures~\ref{fig:vel_L_EF} to \ref{fig:vel_C3-SM_FST_MF} summarise the
information about the velocity in the wakes for some exemplary cases. For an
ISM-density-weighted random subset of grid cells we calculate $v\Par$, the
velocity component anti-parallel to the galaxy's direction of motion,
$(v_x{}\Gal,v_y{}\Gal,v_z{}\Gal)$. We also calculate $v\Perp$, the component
perpendicular to $v\Par$.  By definition, $v\Perp$ is always positive. Both
velocity components are plotted as a function of distance to the galaxy, $D$,
i.e. the projection of each grid cell's position vector in the galactocentric
frame onto $(-v_x{}\Gal,-v_y{}\Gal,-v_z{}\Gal)$.  Additionally, we show the
velocity component along the grid's $x$-axis, $v_x$, which in our simulations
is always perpendicular to the galaxy's orbital plane. All velocity components
are given in the galaxy's rest frame, where the galaxy's motion translates
into an ICM wind flowing past the galaxy.
%
%
%
%

\subsubsection{Velocity component anti-parallel to $\vec{v}\Gal$}
The overall structure of the $v\Par$-plots is the same for all runs and all
time-steps. The gas in the galactic disc appears as a blob between $D=\pm
20\Kpc$ and $v\Par=\pm 200\Kms$, which is due to the disc's rotation. With
increasing distance to the galaxy, $v\Par$ increases towards the ICM wind
velocity, which reflects the acceleration of the stripped gas by the ICM
wind. At a distance of about 80 to $100\Kpc$ behind the galaxy, the
acceleration of the stripped gas is completed, $v\Par$ does not increase any
further. For a fixed distance to the galaxy, the velocity range in the tail is
large. Inside the galactic disc, a width of $400\Kms$ is expected due to the
rotation. This behaviour is also reflected in the $v\Par$-plots. In the region
where the stripped gas is accelerated, the velocity width is larger than this,
it can reach $1000\Kms$. Beyond distances of $\sim 40\Kpc$ behind the galaxy,
the velocity width decreases. At distances beyond $100\Kpc$, the velocity
width is again 400 to $500\Kms$ and remains constant.

We have marked the current ICM wind velocity (or orbital velocity of the
galaxy) in each plot by a horizontal line. Only very few grid cells reach
$v\Par$ above this line, which is not surprising, as the ICM wind cannot
accelerate gas beyond its own velocity. The acceleration of the stripped gas
always proceeds in a rather similar fashion.  We have plotted the same dashed
diagonal line in all plots in order to aid the eye to compare the slope of
$v\Par$ between different snapshots. The slope of the dashed line is
representative for snapshots with ram pressures around $10^{-12}\Erg\ccm$. For
stronger ram pressures, which are also associated with higher orbital
velocities, the slope is somewhat larger, for weaker ram pressures somewhat
smaller. Especially for the later time-steps of run C3-SM-FST-MF
(Fig.~\ref{fig:vel_C3-SM_FST_MF}), where the orbital velocity is much larger
than in the other cases, the velocity slope along the tail is steeper. The
large velocity width in the tail near the galaxy is caused by a superposition
of several processes. Firstly, also here the velocities originating from the
rotation of the gas disc add to the velocity width. Secondly, denser ISM
clouds need a longer time to be accelerated, while less dense structures are
accelerated more easily. A third effect is the turbulence in the wake.  As in
our simulations the spatial resolution beyond $150\Kpc$ from the galactic
centre is limited to $15\Kpc$, these simulations cannot provide reliable
information beyond this distance.

We observe some occasional backfall of stripped gas well before the ram
pressure peak. The backfall is temporal and becomes evident as negative
$v\Par$ closely behind galaxy (e.g.~early panels of
Fig.~\ref{fig:vel_L_FE}), and as local maxima/plateaus in temporal
evolution of gas mass in disc region (see paper I, or
Fig.~\ref{fig:mass_way}). 
A corresponding feature is evident in earlier simulations with
  constant ICM wind. There, the gas disc mass as a function of time shows a
  local maximum or plateau immediately after the instantaneous stripping
  phase (e.g.~grid codes: \citealt{marcolini03,roediger05,roediger06}, but
  also  in SPH: \citealt{schulz01}). Of couse, this feature is only observable
if the evolution of the model galaxy was followed for a sufficiently long
time.
 However, this backfall is not observed in all runs,
but mainly while the galaxy moves near face-on. While the galaxy moves near
edge-on, hardly any backfall is observed. Thus we conclude that the temporal
backfall in our simulations is caused by the turbulent flow around the
remaining gas disc.
Also the shape of the galaxy's total potential supports such this
  backfall behaviour, as, in the face-on case, the ram pressure meets the
  strongest gravitational restoring force a few kpc behind the disc plane (see
  e.g.~\citealt{schulz01,roediger05,jachym07}, but it is also the case for our
  model galaxy).

\subsubsection{Velocity component perpendicular to $\vec{v}\Gal$}
Velocities perpendicular to galaxy's direction of motion, $v\Perp$, can be due
to the disc's rotation as well as the ICM flow around the galaxy. Moreover,
turbulence in the wake adds to $v\Perp$. The component $v\Perp$ causes the
tail flaring.

If all three sources were responsible for the tail flaring, we should observe
that in cases where the remaining gas disc is still large but the galaxy is
moving edge-on, the flaring differs between LOS perpendicular and parallel to
the galactic disc. If such a galaxy is seen face-on, the rotation and the ICM
flow around the galaxy contribute to the velocity component perpendicular to
the LOS and tail direction. In contrast, if such a galaxy is seen edge-on,
neither ICM flow nor rotation contribute to the velocity component
perpendicular to the LOS and tail direction. Thus, one could expect that the
tail flaring should appear stronger in the case where the galaxy is seen
face-on than in the case where the galaxy is seen edge-on. However, we observe
similar flaring ratios for both LOS. Thus we can conclude that mainly
turbulence causes the flaring.

The velocities perpendicular to the galaxy's direction of motion range between
0 and up to $700\Kms$. In near edge-on cases, $v\Perp$ ranges only between 0
and $500\Kms$. The range of $v\Perp$ is also smaller for small ram pressures
and galaxy velocities. This behaviour corresponds to the rate of tail
flaring. Again we observe that in cases where the galaxy's cross-section
is large, the wake is more turbulent.  

The velocity component along the grid's $x$-axis, $v_x$, is always
perpendicular to the orbital plane and thus perpendicular to the galaxy's
direction of motion. However, unlike $v\Perp$, it does not only contain
information about the amplitude of the perpendicular velocity components, but
also directional information. It often shows an oscillating behaviour, which
causes the oscillation of the galactic tails along the orbits as described in
Sect.~\ref{sec:projections}. These oscillations resemble von-Karman vortex
streets.

The amplitude of the turbulence in the wakes seen in our simulations is
comparable to the ICM turbulence generated by other processes:
e.g.~\citet{norman99}, \citet{dolag05} and \citet{vazza06} find turbulent
velocities due to structure formation and subclump infall of some
$100\Kms$. \citet{kim07} find that gravitational wakes of galaxies produce
turbulence at a level of $\lesssim 220\Kms$.

\subsection{Distribution of stripped material throughout the cluster}
\begin{figure*}
\includegraphics[angle=0,width=0.49\textwidth]{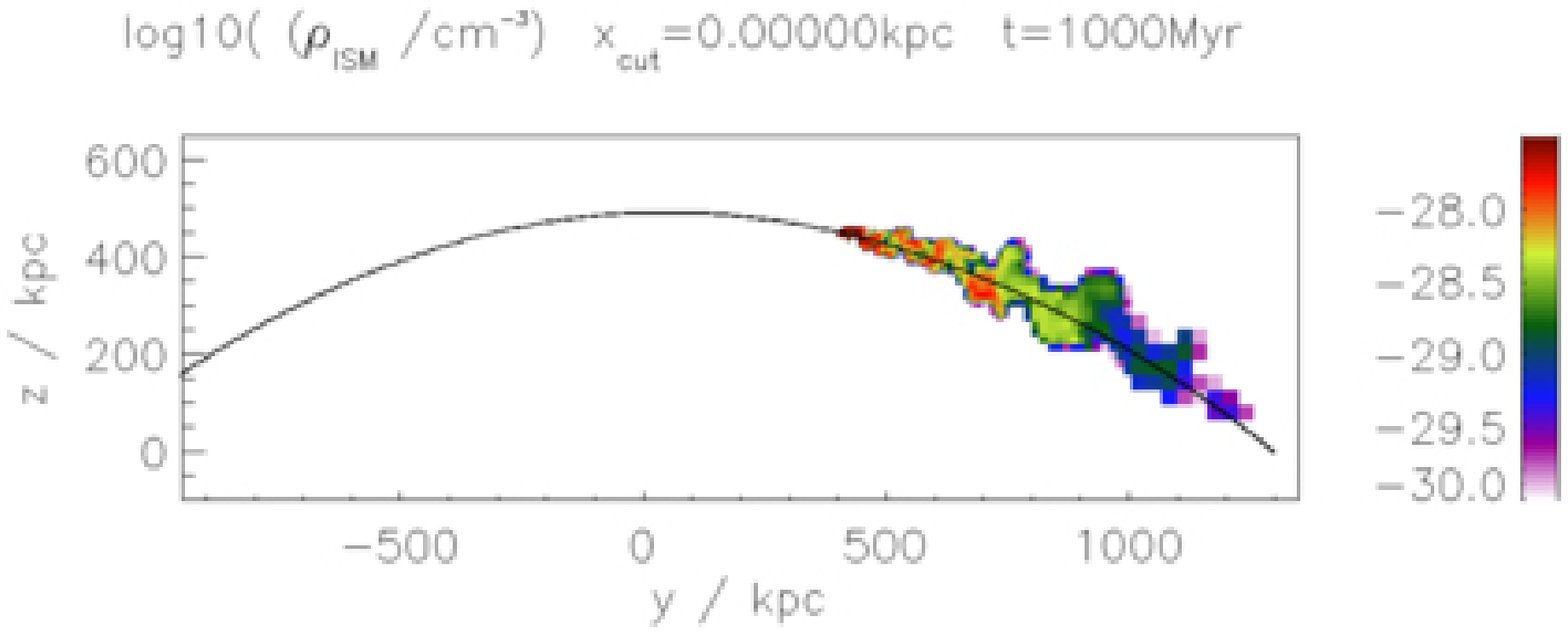}
\includegraphics[angle=0,width=0.49\textwidth]{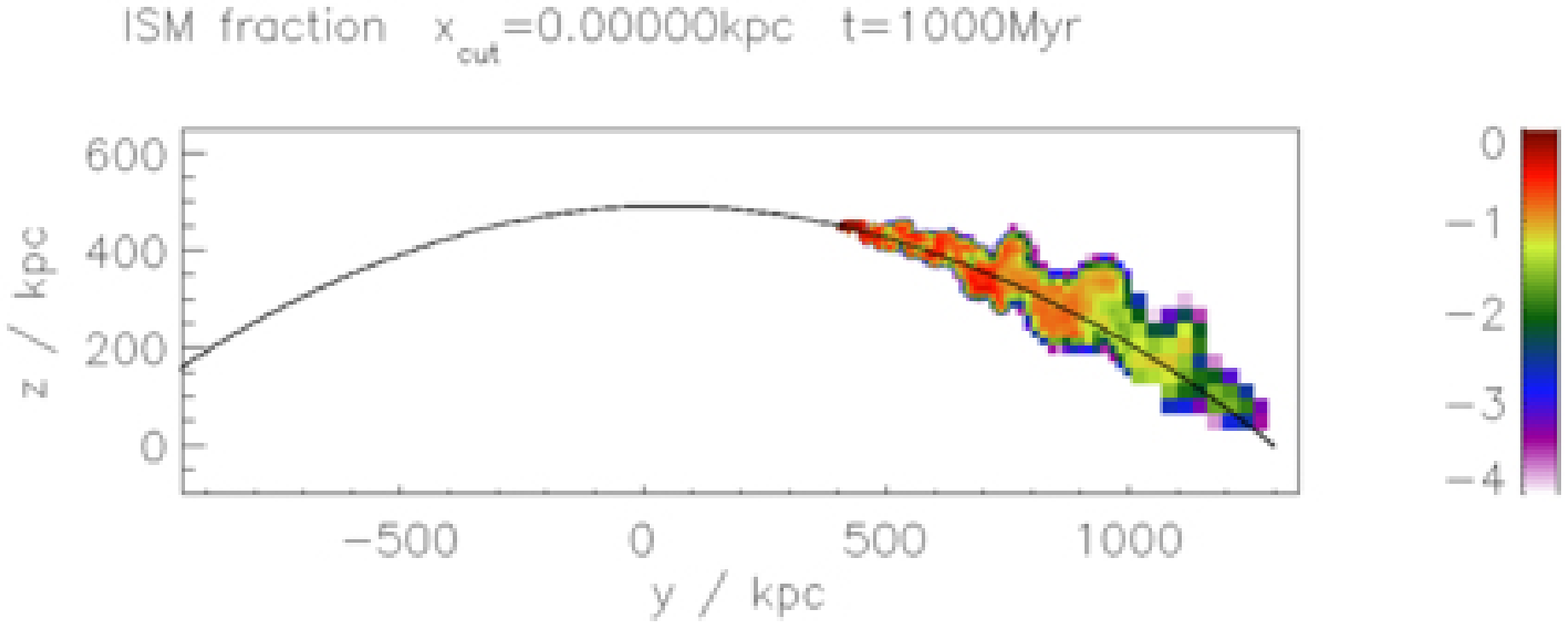}
\includegraphics[angle=0,width=0.49\textwidth]{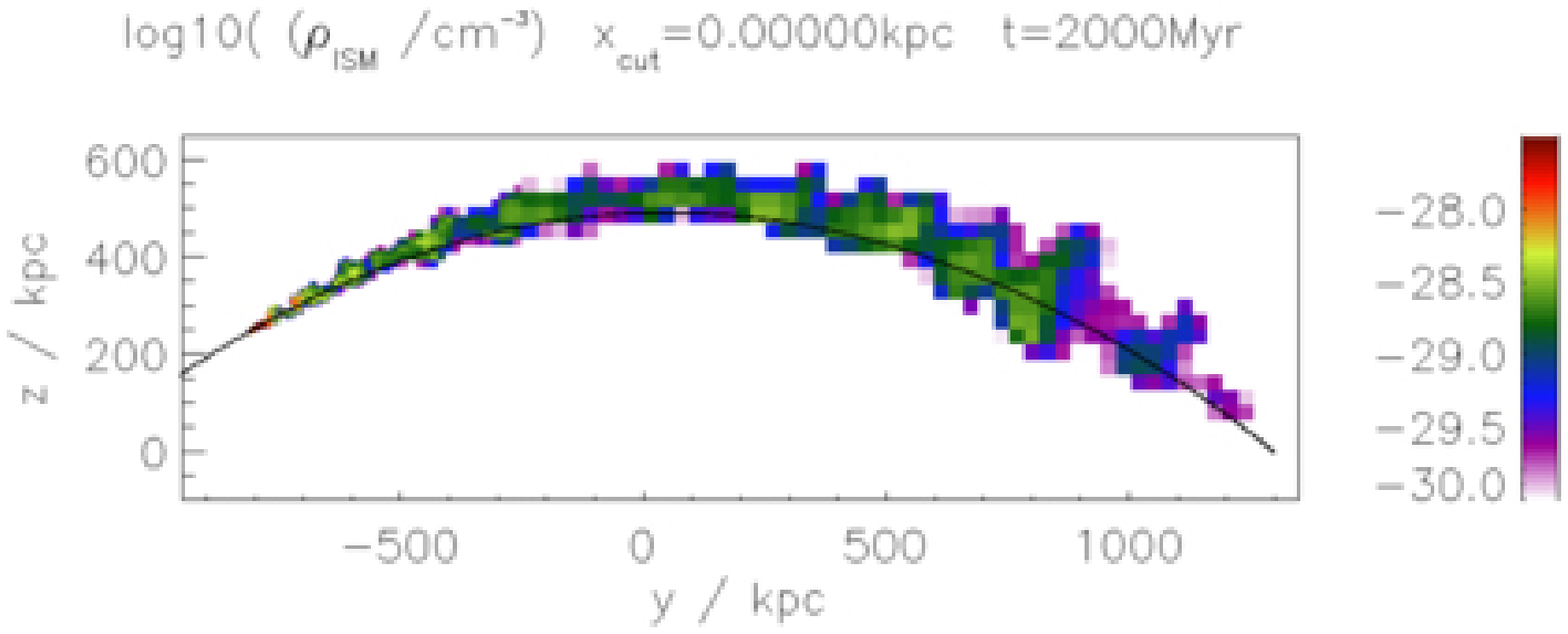}
\includegraphics[angle=0,width=0.49\textwidth]{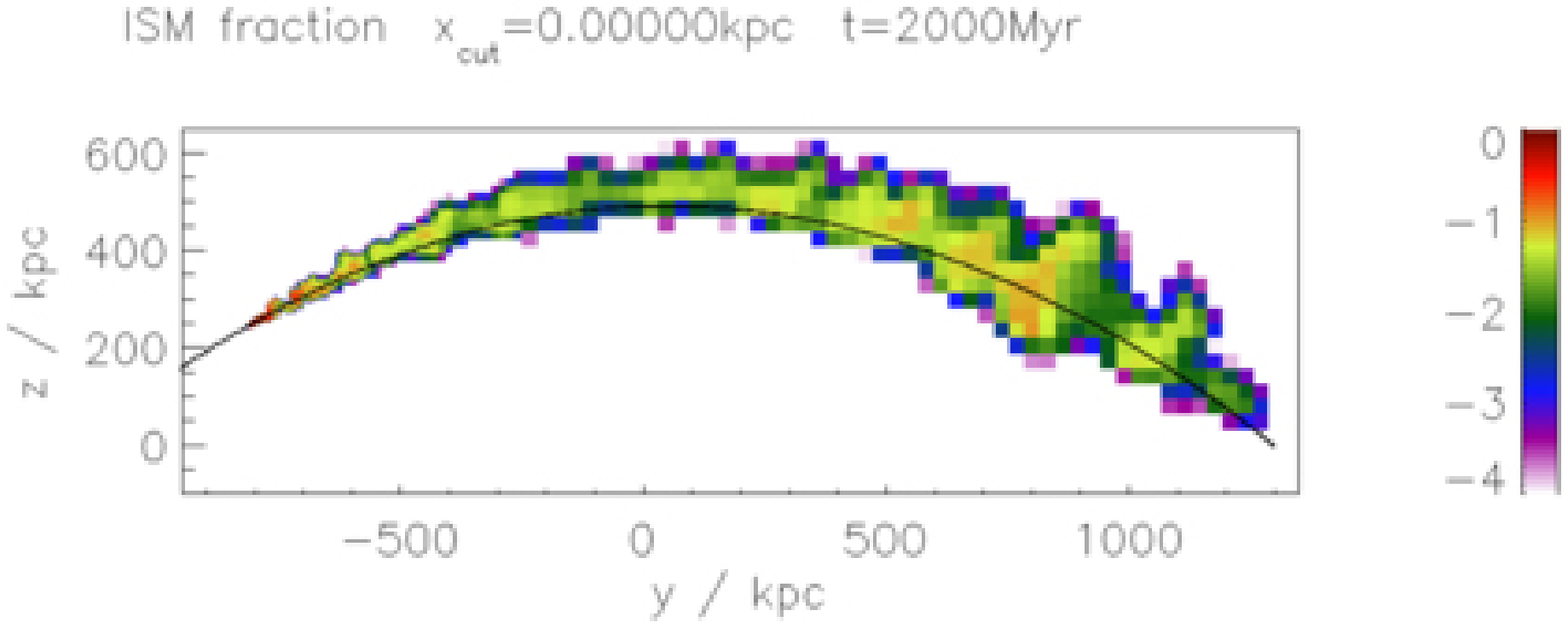}
\vspace*{0.5cm}\rule{\textwidth}{0.2mm}\vspace*{0.5cm}
\includegraphics[angle=0,width=0.49\textwidth]{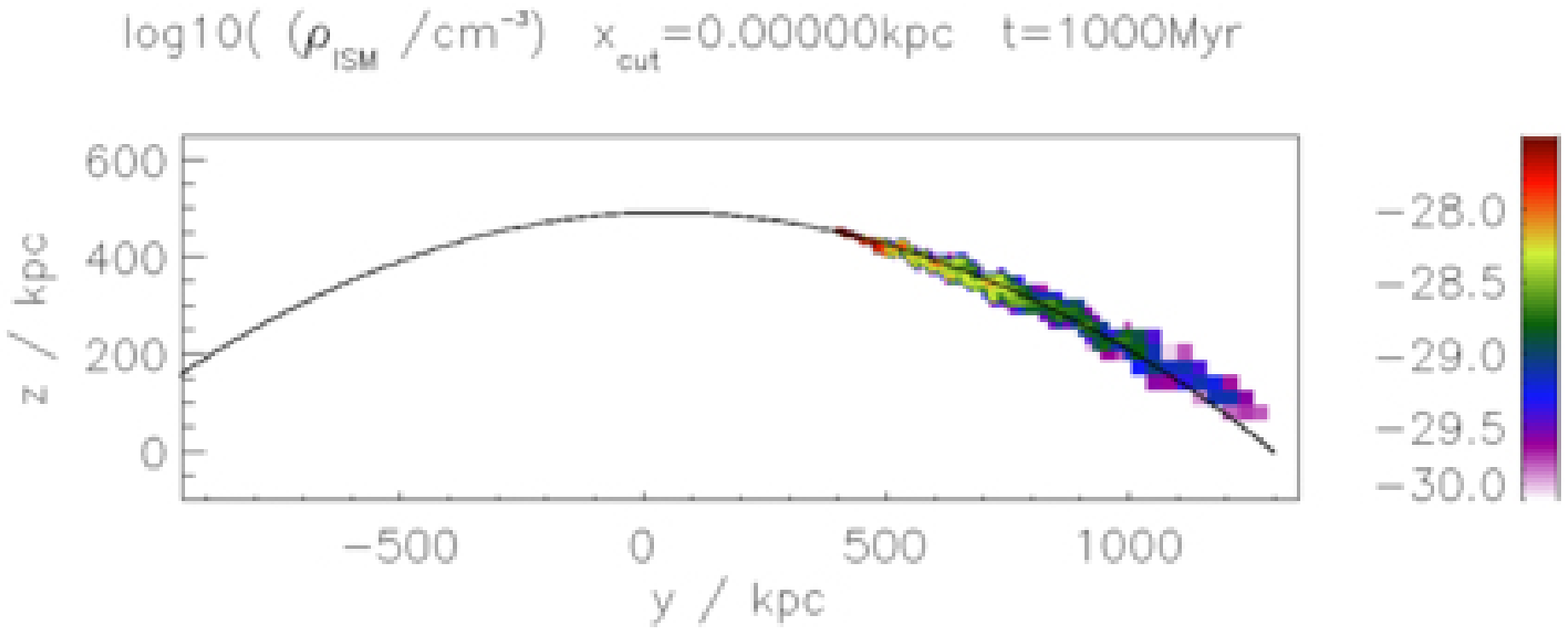}
\includegraphics[angle=0,width=0.49\textwidth]{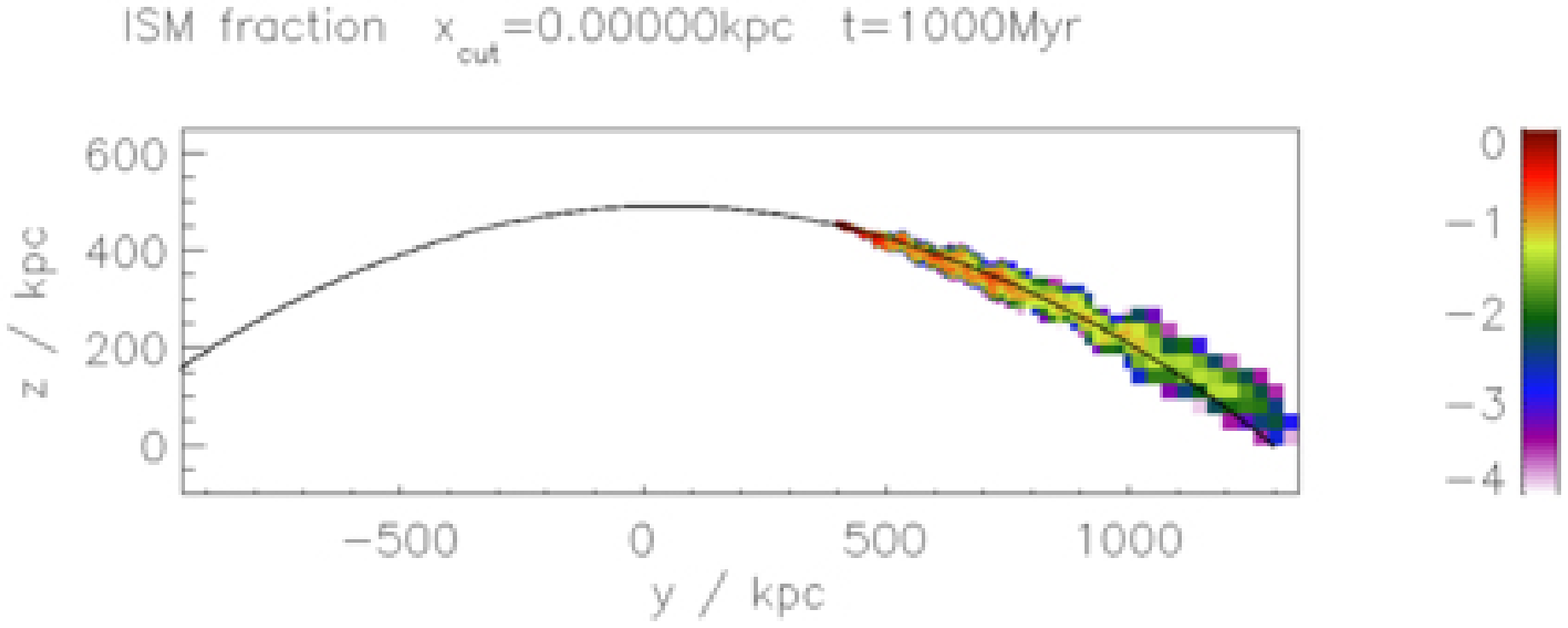}
\includegraphics[angle=0,width=0.49\textwidth]{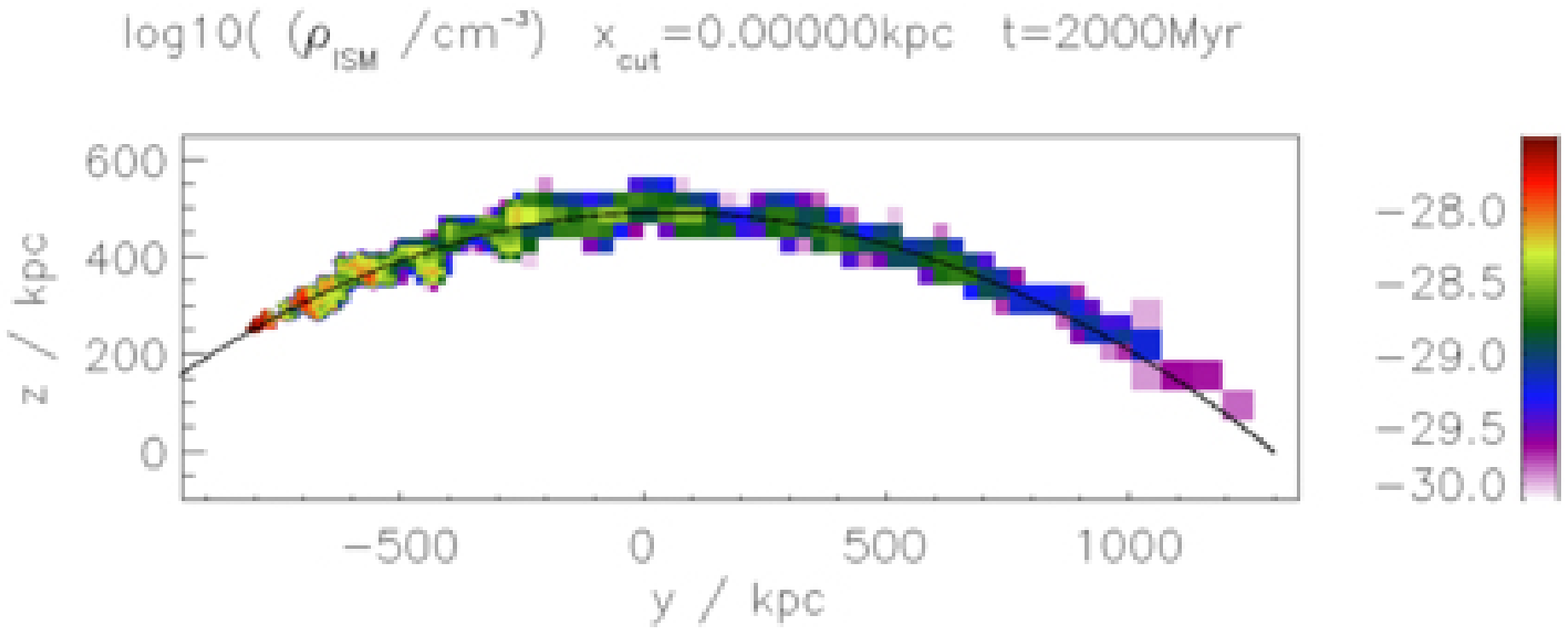}
\includegraphics[angle=0,width=0.49\textwidth]{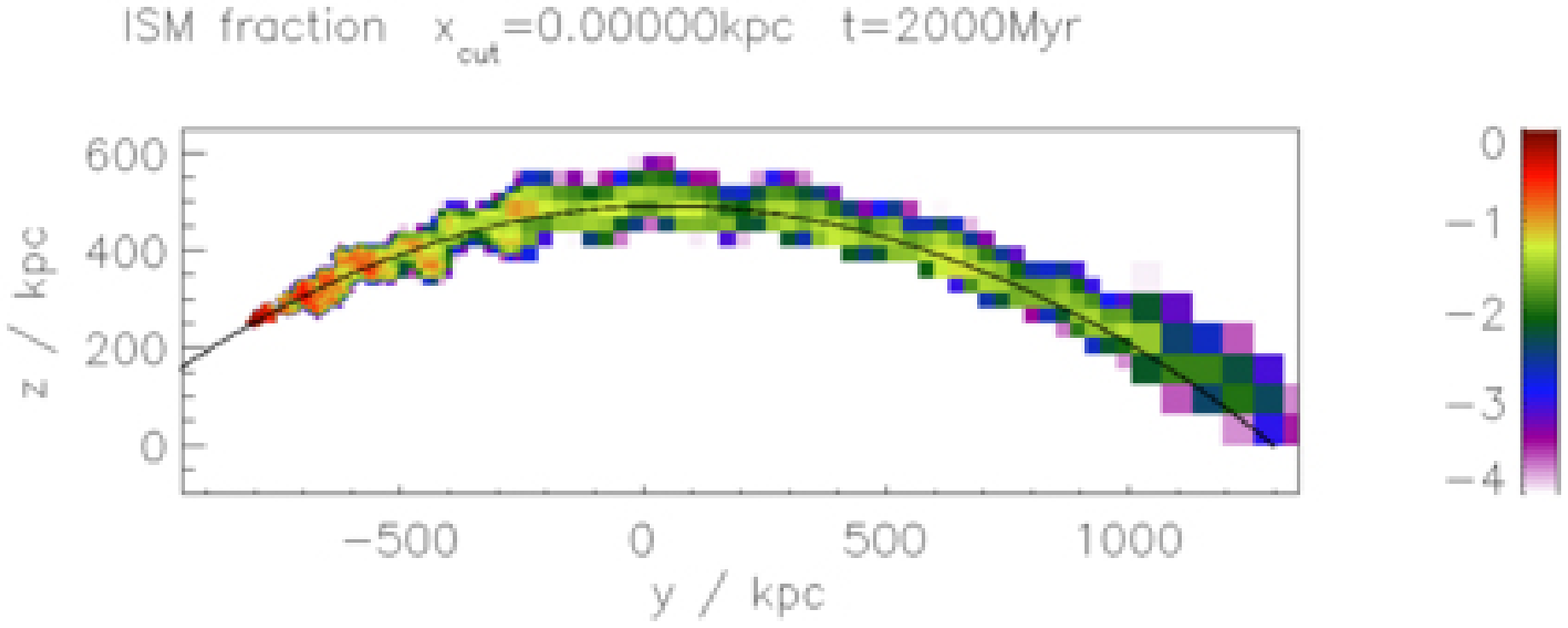}
\caption{Slices in the orbital plane showing the ISM distribution throughout
  the cluster colour-coded for different time-steps. The two top rows are for
  run C1-LG-FST-FE, the two bottom rows for run C1-LG-FST-EF. The
  lhs column displays the density of (stripped) ISM, $\rho\ISM$, the rhs
  column displays the local ISM fraction, $\rho\ISM/\rho\ICM$. The black line
  marks the galaxy's orbit.}
\label{fig:slice_ism_L_I45}
\end{figure*}

\begin{figure}
\centering\resizebox{0.7\hsize}{!}{\includegraphics[angle=-90]{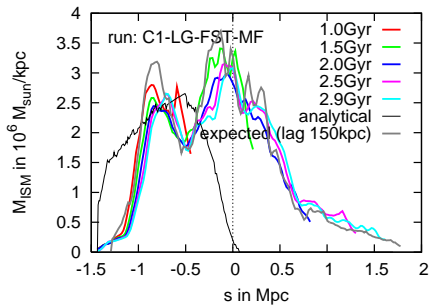}}
\caption{Distribution of stripped gas along orbit at different time-steps for
  run C1-LG-FST-MF, averaged over
  $300\Kpc$. The thin black line is the prediction based on the
  analytical estimate of the stripping mass (see paper I). The gray line is
  the prediction based on the numerical result of the bound gas mass as a
  function of covered distance --  however -- shifted by $150\Kpc$ along the
  orbit. The zero-point of the $x$-axis is shifted to peri-centre passage.}
\label{fig:ism_along_orbit_evol}
\end{figure}

\begin{figure}
\centering\resizebox{\hsize}{!}{\includegraphics[angle=-90]{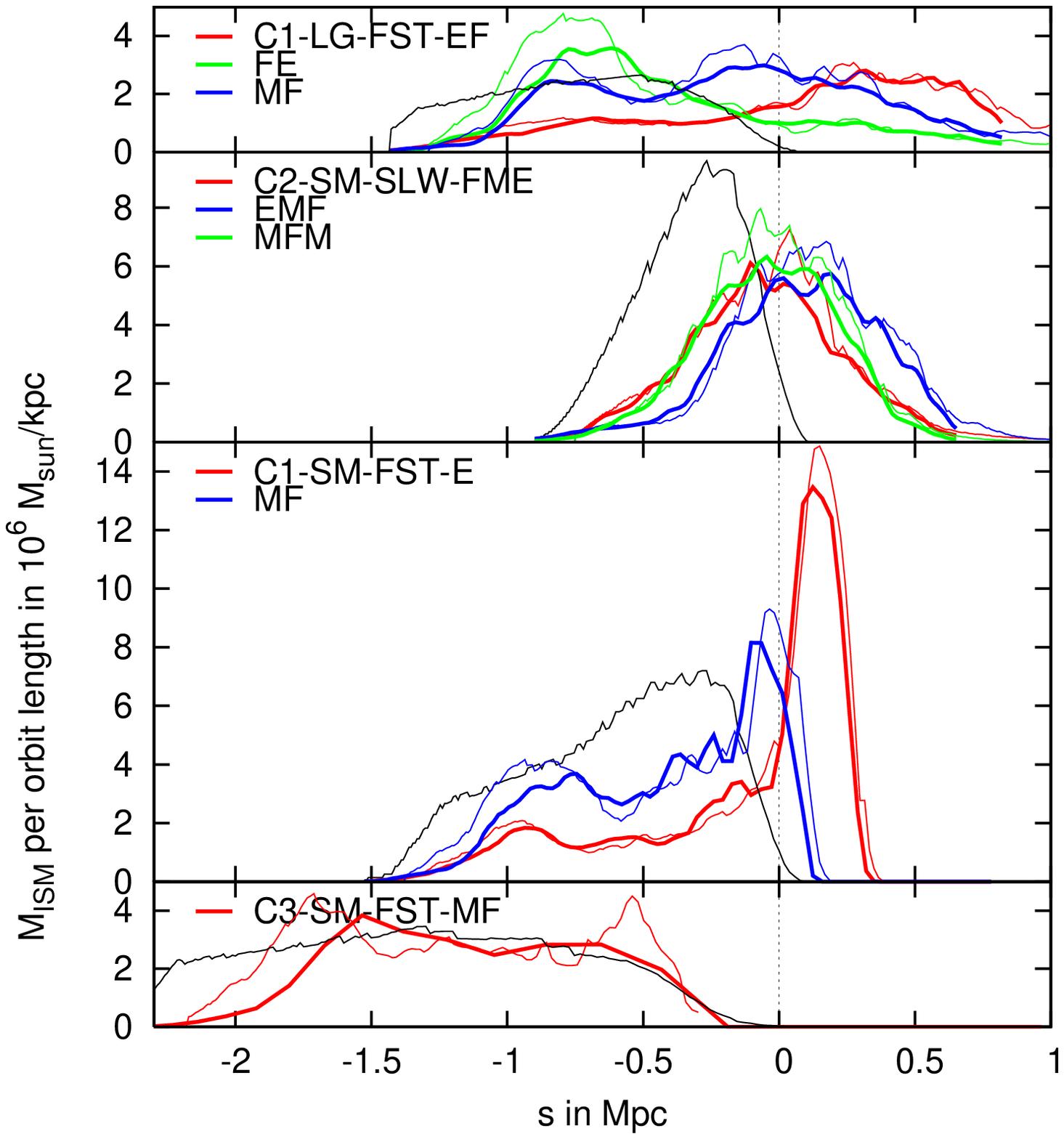}}
\caption{Distribution of stripped gas along orbit at $t=2\Gyr$, averaged over
  $300\Kpc$. Each panel is for one orbit, different inclinations are
  colour-coded (see legend). For each run, the thick line is the distribution
  of stripped gas as seen in the simulation. The thin line of the same colour
  is the prediction based on the numerical result of the bound gas mass as a
  function of covered distance -- however -- shifted by $150\Kpc$ along the
  orbit.  The black line is the prediction from the analytical estimate. The
  zero-point of the $x$-axis is shifted to peri-centre passage.}
\label{fig:ism_along_orbit}
\end{figure}

The galaxy's gas loss history is reflected in the distribution of the stripped
gas throughout the galaxy cluster. In Fig.~\ref{fig:slice_ism_L_I45} we show
this distribution for the orbit C1-LG-FST-\ldots for two different galaxy
inclinations. 
%
%
Clearly, different inclinations lead to different gas loss histories and
thus differing distributions of the stripping gas throughout the cluster. In
order to describe the distribution of stripped gas along the galaxy's orbit,
we have calculated the ISM mass inside a sphere of radius $150\Kpc$ around
each orbit point. Dividing this mass by the orbit length inside the sphere
(which is $300\Kpc$) yields the local ISM mass per orbital length, averaged
over $300\Kpc$. This quantity is plotted in
Fig.~\ref{fig:ism_along_orbit_evol} for different time-steps of simulation run C1-LG-FST-MF. 
%
Figure~\ref{fig:ism_along_orbit} displays the same quantity for all runs at a
fixed time-step $t=2\Gyr$.
%
%
We had to use a rather large averaging length due to the large wake width at
large distances from the galaxy. Using a smaller smoothing length would have
meant to miss some gas. 

Figure~\ref{fig:ism_along_orbit} again reflects the fact that the amount of
gas that the galaxy deposits locally depends on its mass loss per orbital
length. E.g.~in the simulation run corresponding to the bottom panel, the ram
pressure is always larger than in the other simulations. However, here the
galaxy is also moving faster than in all other simulations, and thus spreads
its gas over a larger volume and larger orbital length. Consequently, the local
amount of stripped gas is not higher than in the other simulations.

The amount of stripped gas found per orbital length has to be closely linked
to the galaxy's mass loss rate per orbital length discussed in connection with
Fig.~\ref{fig:mass_way}. In our simulations, we know the amount of ISM bound
to the galaxy potential.  If we assume that the stripped gas is deposited into
the ICM exactly where it is lost, we can predict the expected ISM mass per
orbital length. The gray line in Fig.~\ref{fig:ism_along_orbit_evol} shows
this function, \emph{however}, shifted by $150\Kpc$ along the orbit. The thin
lines in Fig.~\ref{fig:ism_along_orbit} show the same for the other runs. This
means that the stripped gas is deposited into the ICM about $150\Kpc$
from the position where it was lost from the galaxy's potential. In other
words, the stripped gas follows the galaxy for about $150\Kpc$ before finding
its final position in the cluster centre. Thus, in our simulations, the
stripped gas is deposited into the ICM rather locally. We have also plotted
the prediction for the distribution of the stripped gas based on the
analytical estimate of the galaxy's stripping radius (Gunn\& Gott criterion,
see paper I). If the galaxy lost its gas according to this criterion and
if the stripped gas remained where it was lost, its distribution along the
orbit would be as shown by the thin black lines in
Figs.~\ref{fig:ism_along_orbit_evol} and \ref{fig:ism_along_orbit}. In most
cases, the simulations and the analytical prediction differ, according to the
simulations the stripped gas is deposited later along the orbit.

We need to point out that results regarding the distribution of the stripped
gas throughout the cluster are approximations, as in our simulations we have
limited the spatial resolution outside $150\Kpc$ from the galaxy. However,
according to the velocity plots, the deceleration of the stripped gas is
finished within $150\Kpc$ from the galaxy. Moreover, if the spatial resolution
is improved by a factor of 2 throughout the simulation box, the results are
very similar (see Figs.~\ref{fig:slice_ism_L_I30_HR} and
\ref{fig:ism_along_orbit_res}). Thus, we conclude that our results do not
suffer from insufficient resolution.

Typical ISM fractions in the wake $\sim 100\Kpc$ behind the galaxy are around
20\%. Assuming an ISM metallicity of about solar and an ICM metallicity of about
0.4 solar, this leads to a metallicity of 0.52 in the wake. For an ISM
metallicity of twice solar and ICM metallicity of 0.2 solar, the resulting
metallicity in the wake is 0.56 solar.  In how far the wake
is detectable in X-ray metal maps depends also on the temperature in the
wake. As we do not know the temperatures in the wake, we cannot make
predictions at this point.

Obviously, RPS is a source of metals for the ICM. The cumulative effect of the
whole galaxy population on the ICM is studied by \citet{domainko06} and
compared to other enrichment processes by
\citet{schindler05,kapferer07,moll07}. Here, the enrichment by RPS is modelled
by applying the classical Gunn \& Gott criterion to the galaxy particles and
thus calculating a mass loss rate for each galaxy. The gas predicted to be
stripped is added locally to the ICM. In this treatment, galaxies lose gas
only as long as the ram pressure is increasing, only on their way towards the
cluster centre. Our simulations show that galaxies lose their gas somewhat
more slowly than predicted, stripped gas is also found along the orbit after
peri-centre passage. These characteristics could lead to a slightly broader
distribution of ram pressure induced metals in the ICM.

\subsection{Energy input into the ICM}
The galaxy carries a large amount of kinetic energy. The ram pressure working
on the gas disc is associated with a drag force that decelerates the
galaxy. This mechanism provides a heating source for the ICM. The energy lost
by the galaxy can be thermalised in the ICM either via decaying turbulence or
via viscosity. Here we want to investigate the relevance of this heating
mechanism. 

A body of mass, $M$, and cross-section, $A$, that is subject to the ram
pressure, $p\Ram$, experiences the drag force, $F\Drag=-p\Ram A$ and the
associated deceleration
\begin{equation}
a\Drag = -\frac{p\Ram A}{M}.
\end{equation}
The energy loss rate for this body is 
\begin{equation}
\frac{\De}{\De t} E\Kin =-M v a\Drag = -v p\Ram A.
\end{equation}
The more
relevant quantity in our context is the energy loss per orbital length,
\begin{equation}
\frac{\De}{\De s} E\Kin = F\Drag = -p\Ram A
\end{equation}
We can compare this quantity to the local turbulent energy per orbital length
in the galaxy's wake, which is 
\begin{equation}
\frac{\De}{\De s} E\Turb = \frac{1}{2} \rho\ICM v\Turb^2 A\Wake,
\end{equation}
where $v\Turb$ is the typical turbulent velocity in the wake and $A\Wake$ the
wake's cross-section. Assuming that the loss of kinetic energy is converted
locally into turbulence, energy conservation requires that these two quantities
are the same. Making use of the relation $p\Ram=\frac{1}{2}\CW \rho\ICM
v\Gal^2$, this leads to the following relation between the body's (the
galaxy's) orbital velocity, $v\Gal$, and the ratio of the cross-sections of the
body and the wake:
\begin{equation}
\frac{v\Turb}{v\Gal} = \sqrt{\CW \frac{A}{A\Wake} }.
\end{equation}
As usual, $\CW$ is the drag coefficient that parametrises the response of the
body in question to a flow, it depends on the body's shape and surface
properties. For a disc-like body we have $\CW\sim 1$ if the disc is moving
face-on. In case of our galaxy, the ratio between the gas disc's cross-section
and the wake's cross-section is $\sim 4$ to 9. For a typical orbital velocity
of $1000 \Kms$ this leads to turbulent velocities of the order of $400\Kms$,
which is what we observe in our simulations.

\begin{figure}
\centering\resizebox{\hsize}{!}{\includegraphics[angle=-90,width=0.45\textwidth]{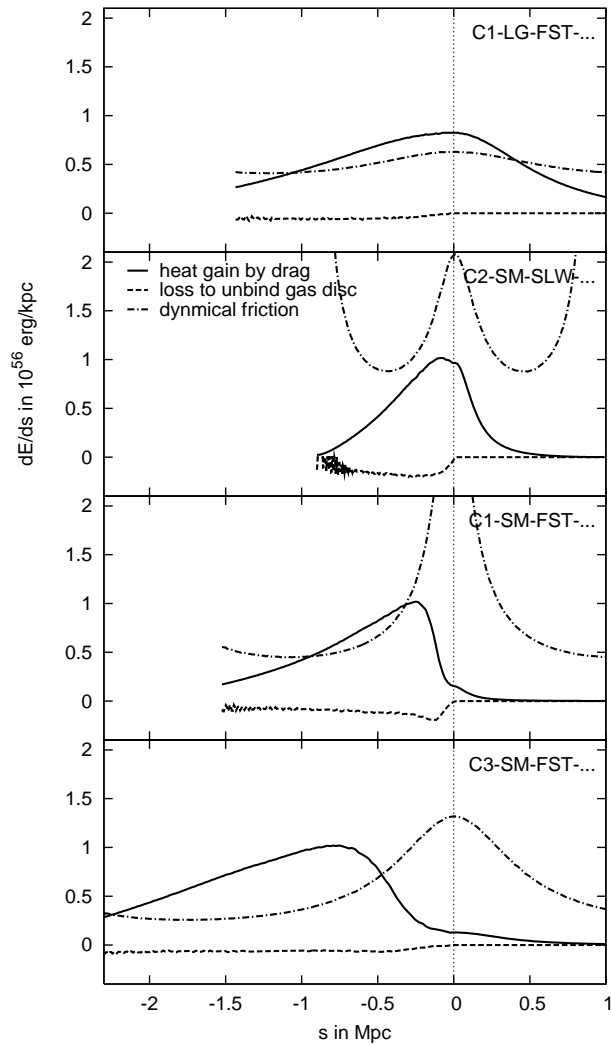}}
\caption{Energy gain for the ICM along the orbit. For four different
  orbits (see label in upper right corners). The zero-point of the $x$-axis is shifted to peri-centre passage.}
\label{fig:energies}
\end{figure}

Now we want to address the question of the relevance of this process as a
heating mechanism for the ICM. Given that the analytical estimate for the
stripping radius gives a reasonable result for a large range of situations (see
paper I), we can use $A=\pi r\Disc^2$ as an estimate for the galaxy's
cross-section, where $r\Disc$ is the stripping radius derived from the analytical
estimate. Furthermore, we assume that the kinetic energy lost by the galaxy is
available for ICM heating locally. Thus, we
can calculate the local heat gain per orbital length to be 
\begin{equation}
\frac{\De}{\De s} E\Heat =-\frac{\De}{\De s} E\Kin = p\Ram \pi r\Disc^2.
\end{equation}
However, not all of this energy can really be used to heat the ICM. Some part
is consumed in removing the galaxy's gas disc from the galactic potential. The
total binding energy of the gas disc is $\sim 1.5\cdot 10^{58}\Erg$, and it
scales approximately linear with disc radius with a slope of $\frac{\De
E\Pot}{\De r\Disc} \sim 8\cdot 10^{56}\Erg / \Kpc$. Thus, the amount of energy
consumed in this process per orbital length is 
\begin{equation}
\frac{\De}{\De s}E\Pot = \frac{\De E\Pot}{\De r\Disc} \frac{\De r\Disc}{\De s}
\approx 8\cdot 10^{56} \frac{\Erg}{\Kpc} \frac{\De r\Disc}{\De s},
\end{equation}
where the change of disc radius per orbital length, $\frac{\De r\Disc}{\De
s}$, is also given by the analytical estimate. 
Figure~\ref{fig:energies} compares the energy input into the ICM due to the
drag deceleration of the galaxy and the energy lost by the removal of the
stripped gas from the galaxy's potential for
the four different orbits.
%
%
The energy budget is given locally along the galaxy's orbit, as energy per
orbital length. Clearly, the input of energy into the ICM due to the galaxy's
drag deceleration dominates. The radiative loss from the ICM
by thermal bremsstrahlung was not considered here as due to its strong
dependence on ICM density it can be relevant only close to cluster centres. 

In order to judge the importance this energy gain for the ICM, we can compare
the original amount of thermal energy along the wake, $e A\Wake$, with this
heat gain. The thermal energy density, $e$, can be expressed in terms of the
sound speed, $\CS$:

\begin{equation}
e\approx \rho\ICM \CS^2,
\end{equation}
where we have made use of the fact that in our case $(\gamma -1)\gamma\approx
1$. Thus, the ratio of heat gain and  thermal energy is
%
\begin{equation}
\frac{\De E\Heat/ \De s}{e A\Wake}=0.5 \CW \frac{v\Gal^2 A\Disc}{\CS^2 A\Wake}
\end{equation}
Usually, the wake diameter is at least a factor of 2 larger than the diameter
of the gas disc. In our simulations, the orbital velocity reaches Mach numbers
of 2 only in the cluster centre. Thus, the energy gain by ram pressure
deceleration would correspond to a local energy (and temperature) increase by
a factor of 2 at maximum. In principle, the heating could be strongest near
the cluster centre as there the galaxy moves fastest, but due to ram pressure
stripping also here the gas disc is rather small or even gone
completely. Thus, the process of drag deceleration is unlikely to be able to
stop cooling flows. In summary, this means that ICM heating due to ram
pressure deceleration of galaxies is not expected to play a crucial role. This
conclusion is not surprising if we remember that the ICM temperature as well
as the galaxies' velocity dispersion are determined by the same gravitational
potential and the total mass of ICM is larger or comparable to the total mass
of all galaxies. This means that the total amount of thermal energy in the ICM
is comparable or larger than the total kinetic energy of all cluster
galaxies. Moreover, the galaxies lose only a fraction of their kinetic energy
by drag deceleration. In our calculation we have also assumed that the galaxy
is moving (near) face-on. For galaxies moving near edge-on, the drag force as
well as the heating rate will be somewhat smaller.

Another process to convert the kinetic energy of the galaxies into thermal
energy of the ICM is dynamical friction (e.g.~\citealt{kim05} and references
therein). The decelerating force working on a galaxy of mass, $M\Gal$ as it
moves through a homogeneous medium of density, $\rho$, with velocity, $v\Gal$,
is
\begin{equation}
F\DF = \frac{4\pi \rho G^2 M\Gal^2}{v\Gal^2} I,
\end{equation}
According to \citet{ostriker99}, the efficiency factor $I$ summarises the
dependence on Mach number and impact parameters, it is in the range of 1 to
10. The density, $\rho$, of the surrounding medium is the ICM density plus the
DM density. As an approximation, we will use $I=1$ and $\rho=10\rho\ICM$, and
a total galaxy mass of $5\cdot 10^{11}M\Sun$. Then
we can calculate the dynamical friction force, which is equivalent to the
galaxy's energy loss per orbital length. The result is also shown in
Fig.~\ref{fig:energies}. The energy available from dynamical friction is
comparable to the energy available from hydrodynamical drag. However,
dynamical friction is stronger in cluster centres. At first glance, also the
hydrodynamical drag should be strong in cluster centres as the ram pressure is
highest there, but the high ram pressure causes ram pressure stripping and
thus small galactic gas discs and, consequently, small drag forces.

Given the deceleration by dynamical friction, $a\DF=F\DF/M\Gal$, we can
estimate how much the galaxy gets decelerated during one orbit:
\begin{eqnarray}
\Delta v &=& a\DF\Delta t\nonumber\\
 &=& 18\Kms  \left(\frac{\rho}{10^{-26}\gccm}\right) \left(\frac{M\Gal}{5\cdot
 10^{11}M\Sun}\right) \nonumber\\
&&\left(\frac{v\Gal}{1000\Kms}\right)^{-2}
 \left(\frac{I}{1}\right) \left(\frac{\Delta t}{1\Gyr}\right) ,
\end{eqnarray}
%
which implies that deceleration due to dynamical friction is
negligible as long as only the first orbit is concerned. As the strength of
the hydrodynamical drag force is similar to the dynamical friction, the same
applies to the hydrodynamical drag.


\section{Discussion}
\label{sec:discussion}
%
\subsection{Summary of tail properties}

The overall picture of ram pressure induced galactic tails as derived from our
simulations can be summarised as follows (we use the galactic rest
frame). There are two main processes:
\begin{itemize}
\item The ICM wind accelerates the stripped gas along the orbit away from the
  galaxy. The acceleration is finished $\sim 100\Kpc$ behind the galaxy.
\item Additionally, mainly turbulence in the wake leads to a flaring of the
  tail. The turbulence and hence flaring ratios
  depend on the galaxy's cross-section with respect to the ICM wind direction:
  Small cross-sections lead to little flaring, large cross-sections lead to
  stronger flaring.
\end{itemize}
This leads to the following characteristics:
\begin{itemize}
\item The tails of stripped gas stretch along the galaxy's orbit. They
oscillate slightly along the orbit and show a flaring width. 
\item Local ISM densities in the tail are around $10^{-26}\gccm$  close to
  the galaxy ($\sim 20\Kpc$ distance from galaxy centre) and a few
  $10^{-28}\gccm$ at larger distances ($50\Kpc$).
\item Projected ISM densities are a few $10^{19}\,\CM^{-2}$ near the galaxy
  and $\sim 2\cdot 10^{18}\,\CM^{-2}$ at large distances (50 to $100\Kpc$).
\item The density and length of the tail (for a given column density limit)
  are set by the galaxy's mass loss per orbital length.
\item Widths range between 20 to 50 kpc at a distance of $\sim 25\Kpc$ to the
galaxy centre and 30 to 80 kpc at a distance of $\sim 100\Kpc$. If we adopt a
column density limit of $\sim 2\cdot 10^{19}\,\CM^{-2}$, the tail width is
similar to the galaxy's apparent cross-section with respect to ICM wind
direction and does not show flaring.
\item For the same column density limit, typical tail lengths are
  $40\Kpc$. Even in compact clusters like Virgo, such tails can also be found
  for galaxies that are still 0.5 to 1 Mpc from the cluster centre. In
  extended clusters like Coma, 40 kpc long tails can be found even at
  cluster-centric distances of 1800 kpc. For high mass losses per orbital
  length (e.g. a passage close to the cluster centre with moderate velocity),
  tail lengths can reach $150\Kpc$.
\item At a distance of $\sim 70\Kpc$, the galactic gas mass per orbital length
  along the tail converges roughly to the galaxy's mass loss per orbital
  length. Typical values are around $3\cdot 10^6 M\Sun/\Kpc$. The mass per
  orbital length along the tail can also be derived for observed galaxies.  In
  order to disentangle a galaxy's mass loss history from this quantity, the
  tail needs to be observed to at least this distance behind the galaxy. With
  typical tail widths of $\sim 50\Kpc$, a sensitivity limit of a few
  $10^{18}\,\CM^{-2}$ is required.
\end{itemize}

\subsection{Limits of this work}
In our simulations, we neglected cooling and thermal conduction. Thus, we are
unable to predict the temperature in the wakes and the spectral range in which
they are observable. Recent observations of long galactic tails in the Virgo
cluster (\citealt{oosterloo05}, \citealt{chung07}) suggest that the stripped
gas could remain cool for a few $100\Myr$. In contrast, the long X-ray tail
observed by \citet{sun06} -- if it is ram pressure induced -- suggests that at
least some part of the stripped gas is heated to higher temperatures. 
Cooling may also affect the stripping efficiency
  (\citealt{mastropietro05b}). However, in order to include cooling in a
  propper way, also star formation and supernova feedback would have to be
  modelled, i.e. the multiphase nature of the ISM would have to be taken into
  account. This is beyond the scope of this paper but will be a highly
  relevant subject of future work.

Cooling may also influence the morphology
  of the tails. Dense regions may cool and contract further, thus forming a clumpy
  tail. However, for the densities ($<10^{-26}\gccm$) and temperatures ($\sim
  10^7\K$) in the wakes in our current simulations the cooling time is at
  least several 10 Myr and thus comparable or longer than the dynamical
  timescale.

Moreover, \citet{vietri97} have shown that in the case of dense cool clouds in
moving though the warm ISM, radiative cooling can prevent the disruption of
these clouds by the KH instability. The instability of shear flows at the boundary
of an object that moves through some fluid are the main reason for the
generation of a turbulent wake. Thus, one may wonder if the presence of
cooling could also prevent the generation of turbulence in the wake of our
model galaxies. One major difference to the work of \citet{vietri97} is,
however, the density and temperature range under consideration. In the
simulations of \citet{vietri97}, the cooling times inside the clouds are
always much shorter that the dynamical timescale. This is not the case in the
ICM, here cooling times are at least comparable to the dynamical
timescale. Moreover, a close inspection of the snapshots presented in
\citet{vietri97} reveals that the flow of the hot gas still shows some eddies,
even when cooling is included. In a similar context, \citet{vieser07} have
studied the influence of thermal conduction on the survival of cool, dense
clouds. They find that the interplay of cooling and thermal conduction can
stabilise the clouds against KH instability. Again, the densities and
temperatures studied are different from our case.  Thus, the influence of
cooling and thermal conduction on the dynamics of the wakes is another
important task for future studies.

We also neglected viscosity in our simulations. The amplitude of viscosity in
the ICM is still a matter of debate
(e.g.~\citealt{reynolds05,ruszkowski04}). 
We note that the same applies to the
heat conduction. 
A higher viscosity prevents
turbulent motions in the wakes and thus would suppress tail
flaring. Consequently, in a viscous ICM, ram pressure induced tails would be
narrower than the ones in our simulations. Moreover, the velocity width across
the tail would be smaller. A comparison between viscous and non-viscous RPS
simulations and detailed observations could help to measure the viscosity of
the ICM.

Despite the many possibilites to improve our model, this model provides an
important starting point to compare theory and observations.

\subsection{Comparison with other simulations}

\subsubsection{Constant ICM wind simulations}
\citet{roediger06wakes} have studied galactic wakes in ram pressure
simulations using a constant ICM wind. Also these simulations showed the
acceleration of the stripped gas away from the galaxy and the flaring of the
tails. However, in these simulations, the tails were generally broader than
the ones presented here. In static wind simulations, the increase of $v\Par$
due to acceleration takes somewhat longer.  However, the constant wind
simulations suffered from the difficulty that the flow had to be initialised
at full strength, which is artificial. Concerning the dynamical aspect, the
cluster crossing simulations presented here are by far more realistic.

\subsubsection{Sticky-particle simulations}
Also the sticky-particle simulations of Vollmer et al. (see
Sect.~\ref{sec:intro}) describe RPS of galaxies on cluster orbits.  According to their simulations, regarding the
particle density, the stripped gas often forms a dense arm that originates at
one edge of the galaxy and is much narrower than the galaxy's
cross-section. The low particle density extent of the tail has the same width
as the galaxy and shows some flaring depending on stripping geometry.  In
their work, they usually concentrated on the gas distribution close to the
galaxy, thus there are no predicted HI maps for long tails.

For galaxies moving near edge-on, the sticky particle simulations predict a
characteristic backfall of stripped gas some time after peri-centre
passage. We
do not observe such a backfall in any of our simulations, 
which is most likely due to the fact that our model clusters and thus
  ram pressure peaks are somewhat broader than the cases studied by \citet{vollmer01}. 
However, we observe
a temporal backfall of gas for near face-on cases while the galaxy approaches
the cluster centre. This backfall is caused by the turbulent velocity
structure in the wake and can thus not be present in the sticky-particle
simulations.

\subsubsection{SPH simulations}
\citet{jachym07} presented SPH simulations of RPS of disc galaxies on cluster
orbits. 
Also in their work, internal processes in the ISM (cooling, star
formation, feedback) were neglected; the ISM was treated isothermally, while
the ICM was treated adiabatically.
Additionally,
 they restricted the ram pressure interaction to the inner
$300\Kpc$ (in diameter) of the cluster. Most of their model
clusters were more compact than our model clusters. This fact and the
restriction to the inner cluster part may be the reasons why also this group
observes reaccretion or backfall of stripped gas after peri-centre passage,
while our simulations do not show such a behaviour. The galactic wakes in
these simulations also seem to differ from the wakes in our
simulations. However, a detailed comparison is difficult as they mostly show
only the first $20\Kpc$ behind the galaxy. In the simulations of
\citet{jachym07}, the opening angle of the tail is influenced mainly by the
ram pressure strength: when the ram pressure is still low and the galaxy
moving slowly, the tails appear to
be flaring strongly, and with increasing ram pressure and galactic velocity the opening angle
decreases and remains moderate after peri-centre passage. Apparently, the tail
width at a distance of 100 kpc behind the galaxy is less than 50
kpc,
which is not much more than the original diameter of the galaxy.
 Additionally, the velocity in the wake shows mainly a slow motion in ICM
wind direction and 
only very small velocity components perpendicular to the galaxy's
  direction of motion. Thus, the tail opening angle in these simulations seems
not to be caused by a significant perpendicular velocity component as in our
case. It mainly reflects that first the outer parts of the gas disc are
stripped, and, as the galaxy approaches the cluster centre, gas from smaller
and smaller radii is stripped.
The authors also mention that the
stripped gas tails behind the galaxy with a fraction of the galaxy's velocity
and that the tails remain denser than the local ICM by a factor of a few. Our
wakes behave differently. The stripped gas remains within $\sim 150\Kpc$ from
where it was stripped. Also the density in the tails of our galaxies are much
lower, it exceeds 1.2~times the local ICM density only very close to the
galaxy (see Fig.~\ref{fig:slice_ism_L_I45} and \ref{fig:slice_ism_L_I30_HR}).

The different behaviour in the simulations of \citet{jachym07} has several
reasons: The ram pressure interaction is described differently, their
effective resolution in the tails may be lower than in our simulations,
most of their clusters are more compact and have a lower ICM density, and they
restricted the ICM-ISM interaction to the inner 300 kpc of the cluster.

\subsection{Comparison with observations}
%
\subsubsection{Long HI tails in cluster outskirts}
\citet{chung07} have searched for HI tails in the Virgo cluster. Of the $\sim
50$ targeted spiral galaxies, 7 revealed long one-sided HI
tails. Surprisingly, these galaxies are at projected cluster-centric distances
of $0.6$ to $1\Mpc$. For their sensitivity limit of $\sim 2\cdot
10^{19}\,\CM^{-2}$, the tail lengths are $\sim 30$ to $40\Kpc$, while tail
widths are similar to the cross-section of the remaining gas disc with respect
to tail direction. The column density decreases gradually
along the tail. \citet{chung07} estimated likely ram pressures and
gravitational restoring forces for the galaxies. Comparing the two forces, they
found that 5 of these 7 galaxies could indeed suffer RPS.

In our simulations, we have not modelled the Virgo cluster
directly, but our cluster C1 is also a compact cluster. In the literature,
different parameters for the ICM distribution in the Virgo cluster are given
(see Fig.~\ref{fig:cluster_profiles}). For cluster-centric radii of 0.5 to 1
Mpc, our cluster C1 is a factor of about 3.6 denser than the version used in
\citet{vollmer01a}, but agrees well with the parameters given by
\citet{matsumoto00}. Thus, we consider a comparison of the tails observed by
\citet{chung07} with our simulated tails in cluster C1 reasonable.

Adopting a column density limit comparable to the sensitivity limit of the
observations, our simulations are able to produce tails with similar
characteristics regarding length, width and structure also at large distances
to the cluster centre (e.g.~first row of Fig.~\ref{fig:wakes_LG_FST} and first
two rows of Fig.~\ref{fig:wakes_SM_FST_xz}). However, not all simulated
galaxies at large cluster-centric distances show long tails. In the first row
of Fig.~\ref{fig:wakes_SM_SLW}, the tails are rather short, because here the
ram pressure is very small due to the small orbital velocity of this
galaxy. Thus, we conclude that it is possible that the tails observed by
\citet{chung07} are produced by ram pressure stripping although the galaxies
are still a large distance from the cluster centre. However, these galaxies
need to have a substantial velocity component, i.e. about $1000\Kms$, in the
plane of the sky. Not all galaxies at large cluster-centric distances observed
by \citet{chung07} show long tails.  The reason can simply be that not all
galaxies have a sufficient velocity component in the plane of the sky.  We
expect that in more extended clusters like Coma, long ram pressure induced
tails can be found at even larger distances to the cluster centre.

Our velocity plots (Figs.~\ref{fig:vel_L_EF} to
\ref{fig:vel_C3-SM_FST_MF}) are the analogue to the position-velocity diagrams
(PVD) presented by \citet{chung07}. However, given the sensitivity limit of
observations, only the densest features of our velocity diagrams will be
observable.  In the observations presented by \citet{chung07}, the gas tails
appear as a ``hook'' that is attached to one edge of the galaxy and shows a
slight slope towards the galaxy's systemic velocity. This hook has a small
velocity width, $\lesssim 200\Kms$, in some cases even less. If this was the
whole velocity width of the tail, this would be surprising, as one would
expect a velocity width like in the galactic disc ($\sim 400\Kms$).

In our simulations, we can observe a similar ``hook''-feature in our
$v_x$-plots, if we consider only densest parts, in cases where the galaxy is
not moving close to face-on. Also here, the velocity width in the hook is only
$\lesssim 200\Kms$, which agrees with the observations. In the simulations,
the ``hook'' is not as long as in observations.  This feature is
generated by the fact that in these cases the gas loss happens mainly at one
side of the galaxy, namely the part of the leading edge where the galaxy
rotates along with the wind. Also most galaxies with tails observed by
\citet{chung07} are not moving close to face-on. However, in our simulations,
this hook feature is not as clear as in the observations. Moreover, nearly all
of the tailed galaxies observed by \citet{chung07} show this hook, whereas in
our simulations the hook appears only occasionally and along preferable
line-of-sights. In general, in the PVDs of our simulations, the tail is much
more diffuse than the observed ones. A reason for this difference could be
that real RPS does not proceed as turbulent as in our simulations, but that
viscosity leads to a smoother flow.

We note that in the observed galaxies, in almost all cases the velocity
gradient along the tail/hook is not only towards the cluster mean, but also
towards the galaxy's systemic velocity. Moreover, some of the observed
galaxies have a very small radial velocity with respect to the Virgo cluster
mean, so that at maximum a weak velocity gradient along the tail could be
expected, if the gradient is due to the acceleration of the stripped gas away
from the galaxy. We speculate that the gradient in the tail does not simply
show the acceleration of stripped gas away from the galaxy, but that the
deceleration of the gas disc's rotational component along the tail plays a
crucial role.

\subsubsection{The case of NGC~4388}
The Virgo spiral galaxy NGC~4388 seems to be an excellent example of a ram
pressure stripped galaxy. It is known to have a $\sim 35\Kpc$ long tail of
ionised gas (\citealt{yoshida02,yoshida04}). Additionally, \citet{oosterloo05}
have found a $\sim 120\Kpc$ long HI tail associated with this
galaxy.

NGC~4388 has a high radial velocity of $1400\Kms$ (see \citealt{vollmer03a}
and references therein) with respect to the cluster mean. Given its long tail
in projection, it also must have significant velocity component in the plane
of the sky. A reasonable assumption is that the velocity component in the sky
is comparable to the radial one, which would lead to a total velocity of $\sim
2000\Kms$ in the Virgo cluster rest frame, which is rather high. However,
assuming a much smaller component in the plane of the sky would increase the
true length of the HI tail a lot. Our current assumption already results in a
true length of the HI tail of $\sim 1.5\times 120\Kpc=180\Kpc$.

A reasonable value for the local ICM density at the position of NGC~4388 is a
few $10^{-28}\gccm$. With a likely velocity of about $2000\Kms$, it
experiences a ram pressure of about $10^{-11}\Presunit$. According to our
simulations, this is enough to be stripped heavily, so regarding the degree of
stripping, simulations and observations agree.

Regarding the combination of orbital velocity and local ICM density, in our
simulations, there is no case directly comparable to NGC~4388.  The closest
our simulations get to the velocity-density combination of NGC~4388 is the
peri-centre passage of runs C1-LG-FST-\ldots or at $\sim 800\Myr$ of run
C3-SM-FST-MF. In both cases, the ram pressure is about $10^{-11}\Presunit$,
but the orbital velocity is only slightly below $1500\Kms$. In both cases, the
mass loss per orbital length is about $2.7\cdot 10^{6}M\Sun\Kpc^{-1}$. Given
that NGC~4388 is probably a factor of $\sim 1.4$ faster than the two mentioned
examples, also the mass loss per orbital length should be a factor of $\sim
1.4$ smaller, which would lead to a less dense tail than the ones in the
mentioned runs. As an opposing effect, the line-of-sight and the tail of
NGC~4388 make an angle of about $45\degree$, so that projected densities
appear a factor of $1.4$ higher than they would if the line-of-sight and the
tail were perpendicular to each other. Also the projected gas mass per
orbital length is about $1.4$ times higher. As these two effects cancel each
other, we expect column densities and gas masses per orbital length comparable
to the ones in our simulations.  Thus, our simulations predict that the column
density in the tail near the galaxy should be near $\sim 10^{20}\,\CM^{-2}$,
and decrease gradually with increasing distance to the galaxy to below $\sim
10^{19}\,\CM^{-2}$ at a distance of $\sim 50\Kpc$ or latest $100\Kpc$ behind
the galaxy. At a distance of $\sim 100\Kpc$, the gas mass per orbital length
should correspond to the galaxy's mass loss per orbital length, which we
estimated above to be about $3\cdot 10^{6}M\Sun\Kpc^{-1}$.

However, the tail of NGC~4388 looks different. It is densest at a distance of
$\sim 100\Kpc$ behind the galaxy, here the column density reaches $\sim
10^{20}\,\CM^{-2}$. The spatial width of this dense patch is about $5\Kpc$,
thus this dense patch corresponds to a mass per orbital length of about
$4\cdot 10^{6}M\Sun\Kpc^{-1}$. Towards the galaxy, the column density as well
as the mass per orbital length decrease. In HI, the tail even shows a gap near
the galaxy. In this gap, \citet{yoshida02} observed a plume of ionised
gas. However, the total mass of this plume is only $10^5M\Sun$, which
corresponds to a mass per orbital length to $3\cdot 10^4 M\Sun \Kpc^{-1}$ and
is much less than we expected. Thus, the distribution of stripped gas along
the tail differs from what we expect.

The width of the tail shows some flaring. In our simulations, we do not
observe tail flaring in the sensitivity limit of these simulations. In the
simulations, the tails show flaring only at much lower column densities.  

The ionised part of the tail NGC~4388 shows a filamentary structure. On the
basis of a deep spectroscopic study, \citet{yoshida04} derive the H$\alpha$
velocity field, which turns out to be quite complicated. The authors identify
several kinematic groups among the filaments. However, they also consider 
significant turbulent motions possible. They argue that the kinematic structure
and metallicity of the ionisation are not well-explained by a minor-merger
scenario, but can be explained naturally by ram pressure stripping and an
additional starburst driven superwind in the galactic nucleus. According to our
simulations, turbulent motions in the tail are very likely. However, the mass
of stripped gas near the galaxy is much less than we expected.

The HI tail of NGC~4388 (\citealt{oosterloo05}) remains hard to fit with the
simulations also with respect to the velocity information. The velocity width
of the tail is rather small over the whole length. Additionally, according to
our simulations, the acceleration of the stripped gas towards the ICM velocity
is finished approximately at a distance of $100\Kpc$ behind the galaxy. For
NGC~4388, however, the velocity of at the end of the tail differs from the
galaxy's systemic velocity only by $500\Kms$. According to our simulations, a
value comparable to the galaxy's radial velocity with respect to the cluster
mean, $1400\Kms$, would be expected.

In summary, our simulations have difficulties to explain the details of
NGC~4388's tail, where the most severe points are:
\begin{itemize}
\item The distribution of stripped gas along the tail differs between
  observations and simulations. The simulations predict much more gas near the
  galaxy. The mass per orbital length at a distance of $\sim 100\Kpc$ behind
  the galaxy is approximately at the expected level, however, here the gas is
  confined in a smaller volume than expected and thus shows higher column
  densities than expected.
\item According to our simulations, the stripped gas should be decelerated to
the cluster mean velocity at a distance of $\sim 100\Kpc$. According to the HI
data, for NGC~4388, however, the stripped gas seems to have lost only 1/3 of
the expected velocity change. In contrast to this, the H$\alpha$ tail shows a
velocity gradient as expected from our simulations. Radial velocities at the
end of the H$\alpha$ tail have already reached $300$ to $500\Kms$. Also the
turbulent structure of the H$\alpha$ velocity field is in agreement with our
simulations. We note that here the HI and the H$\alpha$ data seem to be in
disagreement.
\end{itemize}

There are several reasons that could cause the mismatches between the
simulations and the observations. E.g. the ICM may not be smooth but
inhomogeneous, and there may be ambient motions in the ICM, thus leading to a
special ram pressure history of this galaxy. The Virgo spiral NGC~4522 is
another example that suggests ambient motions in the ICM, as its degree of
stripping implies a much stronger ram pressure than expected at its position
in the cluster (\citealt{vollmer06}).  Moreover, NGC~4388 has an active
nucleus and a starburst driven superwind (see \citealt{yoshida04} and
references therein), which will influence the dynamics of the ICM-ISM
interaction.  Additionally, our simulations cannot predict the temperature of
the stripped gas and are thus unable to predict in which wavelengths the
stripped gas is observable. However, the close vicinity of the galaxy has been
searched for ionised and neutral gas and the amounts found are significantly
less than what we expect.

\subsubsection{H$\alpha$ and X-ray tails}
Up to date, only a few galaxies with H$\alpha$ and X-ray tails are
known. There are two irregular galaxies in the cluster A~1367 that have
H$\alpha$ tails of $50\Kpc$ and $75\Kpc$ length and $8\Kpc$ width
(\citealt{gavazzi01}). The tails are straight and show some filamentary
structure along the tail. \citet{yagi07} have observed the Coma galaxy D~100
and found an H$\alpha$ tail that is exceptionally straight and narrow
($60\Kpc$ long, $2\Kpc$ wide). The H$\alpha$ tail of NGC~4388 has been
discussed above in connection with its HI tail.

\citet{sun06} have presented a long and narrow X-ray tail ($70\Kpc$ long,
$7\Kpc$ wide) for the galaxy ESO 137-001 in the Coma-like cluster
A~3627. Additionally, \citet{sun07} have found a $40\Kpc \times 4\Kpc$
H$\alpha$ tail for the same galaxy that coincides spatially with the X-ray
tail. The mass of the X-ray tail is estimated to be $10^9M\Sun$, the H$\alpha$
tail provides another few $10^8M\Sun$. Both, the X-ray and the H$\alpha$ tail
are nearly straight, although some very slight oscillation can be detected. In
both wavebands, the highest emission is found near the galaxy, but there are
some more brighter patches further downstream. If the brightness also traces
the local gas mass, this behaviour would be expected if these tails are caused
by ram pressure stripping. Also the total gas mass of the tails matches the
ram pressure picture.

A characteristic feature of the observed H$\alpha$ and X-ray tails is their
narrowness and straightness.  Given that our simulations cannot estimate the
temperature of the stripped gas, we are unable to predict the brightness
distribution of our tails in H$\alpha$ or in X-rays. However, the projected
density maps presented in this paper generally show much broader tails and
make such narrow and straight tails difficult to understand. Although our
simulations provide an important step towards a realistic description of RPS
by modelling the flight through a cluster, still important physics seems to be
missing.


\section{Summary}
\label{sec:summary}
We have performed 3D hydrodynamical simulations of ram pressure stripping of
disc galaxies on orbits through galaxy clusters. Here we have presented a
detailed study of the wakes associated with these galaxies.

For a sensitivity limit typical for current HI observations
($10^{19}\,\CM^{-2}$), we find that a typical tail length in projected gas
density is $40\Kpc$. In this sensitivity limit, the width of the tails is
similar to the galaxy's width with respect to the tail direction. Our
simulations show that the density and length of the tail is not determined by
the current ram pressure and thus temporal mass loss rate alone, but also by
the galaxy's velocity. These two dependencies can be summarised in the mass
loss per orbital length, which is the crucial parameter that determines the
tail length and density.

If we consider projected gas densities below the sensitivity limit, the tails
show a flaring width, where the flaring ratio depends on the galaxy's
cross-section with respect to its direction of motion. Large cross-sections
lead to strong flaring, while for small cross-sections the flaring is
weak. The flaring is caused by turbulence in the galactic wakes.

The velocity in our wakes shows a significant turbulent component of a few
$100\Kms$. The stripped gas is fully decelerated to the ICM rest frame at a
distance of $\sim 100\Kpc$ behind the galaxy.

We find that the stripped gas does not follow the galaxy for a long way but
remains within $\sim 150\Kpc$ from the position where it is lost. We have
investigated in how far the deceleration of the galaxy by the drag force can
be a heating source for the ICM and found that no significant effect is
expected.

As our simulations neglect thermal conduction, cooling and viscosity, they are
unable to predict the wavelength at which the galactic wakes should be
observable. Nonetheless, we compared our simulations with observations of
galactic wakes.

We can reproduce tails of $40\Kpc$ length at cluster-centric distances between
$0.5$ and $1\Mpc$ in a compact cluster similar to the Virgo cluster. Thus we
conclude that the galactic tails found by \citet{chung07} could indeed be
formed by ram pressure stripping. However, our simulations have difficulties
in reproducing the velocity information of these galaxies. The observations
suggest that RPS proceeds less turbulent than in our simulations. Also, the
structure of the wake of the Virgo spiral NGC~4388 is hard to explain by our
simulations, not only due to shortcomings of our simulations, but also due to
intrinsic discrepancies in this galaxy's data. This HI ``tail'' may not be
caused by RPS at all, or this galaxy may have experienced a complex ram
pressure history.  Given that our simulations cannot predict the correct gas
temperatures, comparisons to the H$\alpha$ and X-ray tails are especially
difficult. However, our simulations do not seem to be able to match the
straightness and narrowness of these tails. Also these observations suggest
that some physics that were neglected in our model,
  e.g. viscosity, cooling, and tidal interactions, may play an important role
  for the detailed structure of ram pressure stripping tails.


\section*{Acknowledgements}
We acknowledge the support by the DFG grant BR 2026/3 within the Priority
Programme ``Witnesses of Cosmic History'' and the supercomputing grants NIC
2195 and 2256 at the John-Neumann Institut at the Forschungszentrum J\"ulich.
The results presented were produced using the FLASH code, a product of the DOE
ASC/Alliances-funded Center for Astrophysical Thermonuclear Flashes at the
University of Chicago. 
We are grateful for helpful discussions with M.~Sun, A.~Chung, J.~van~Gorkom,
T.~Oosterloo, S.~Schindler and W.~Kapferer.

\appendix
\section{Resolution} \label{sec:resolution}
\begin{figure*}
\includegraphics[angle=0,width=0.49\textwidth]{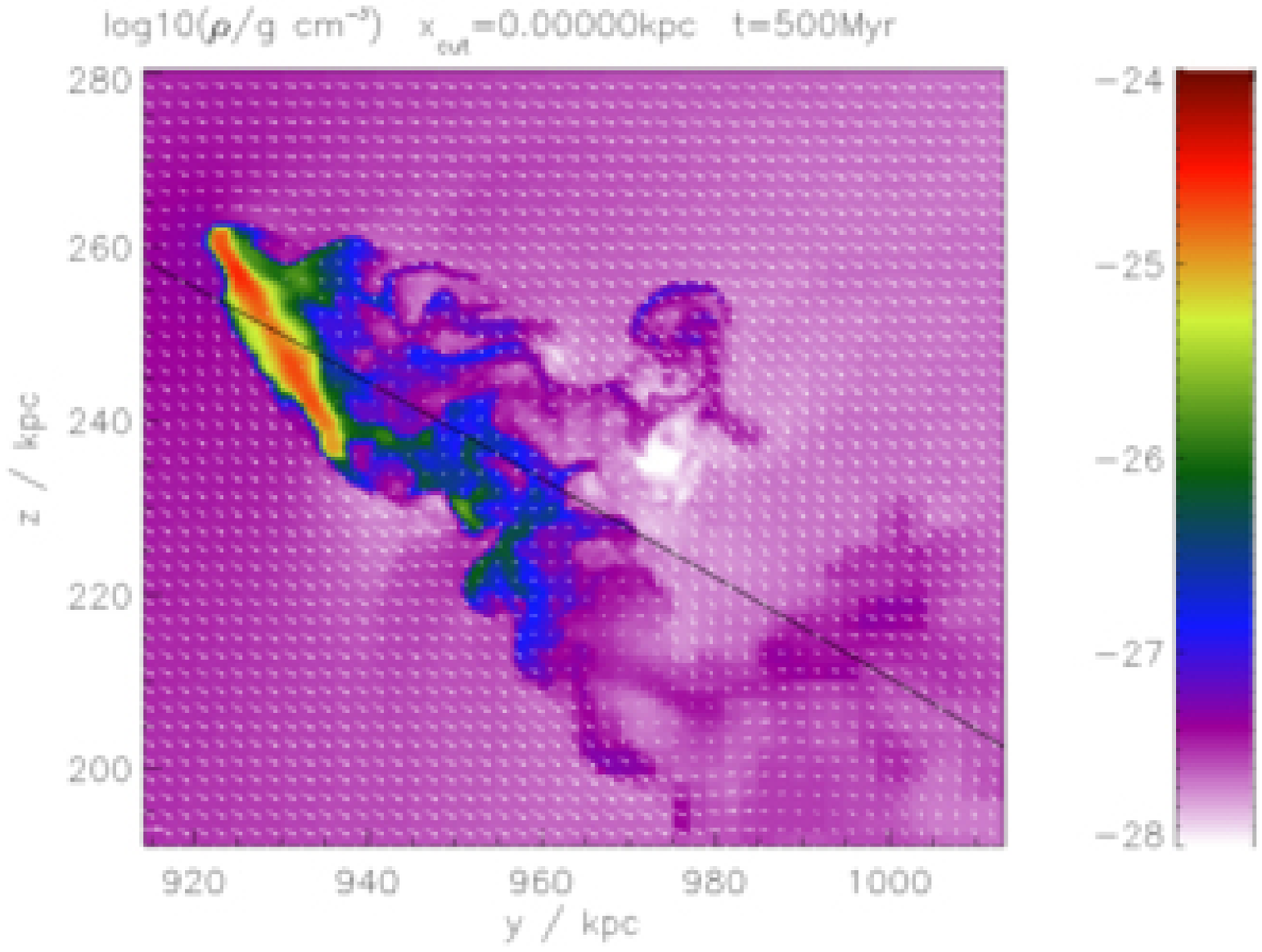}
\includegraphics[angle=0,width=0.49\textwidth]{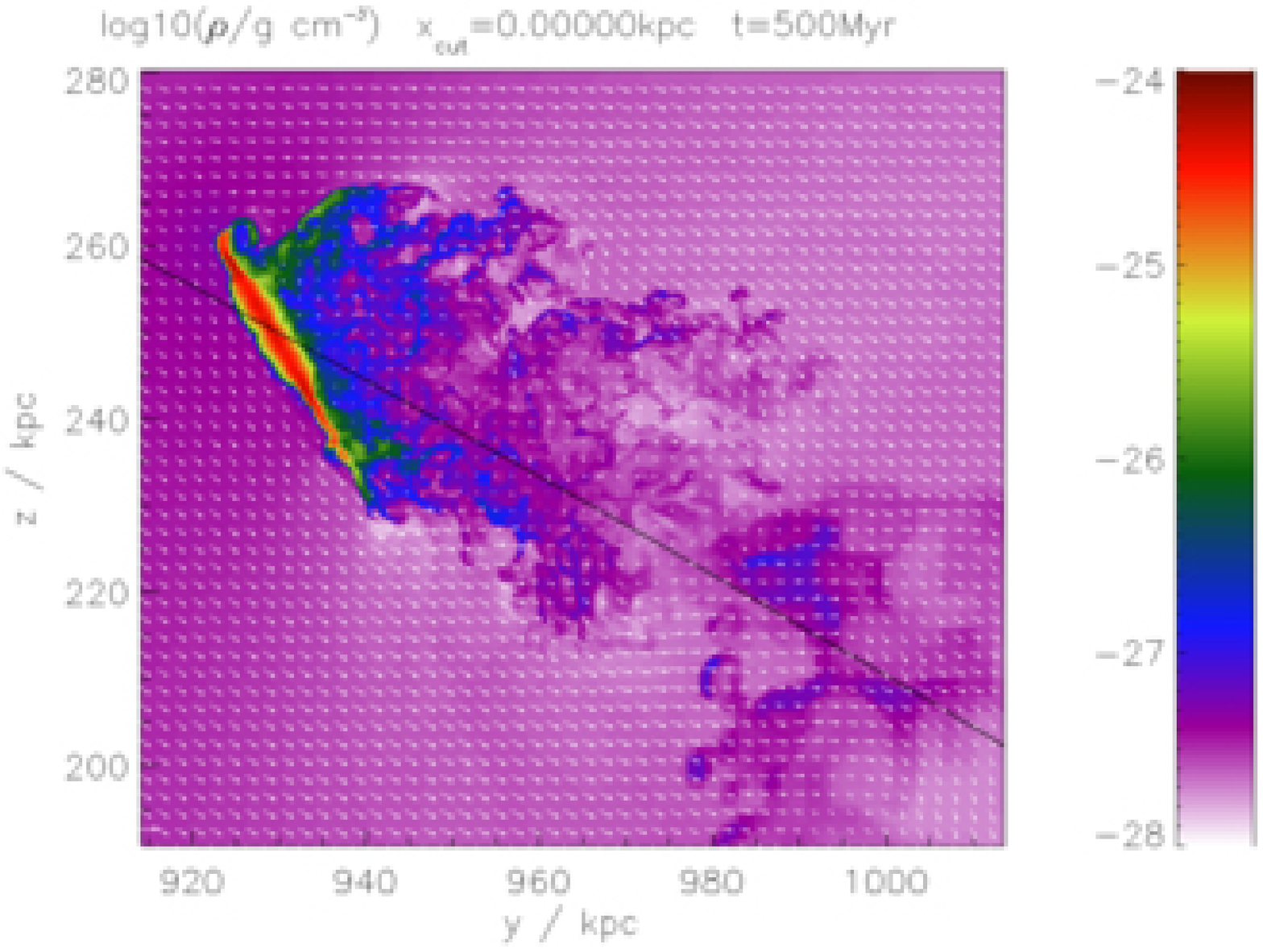}
\includegraphics[angle=0,width=0.49\textwidth]{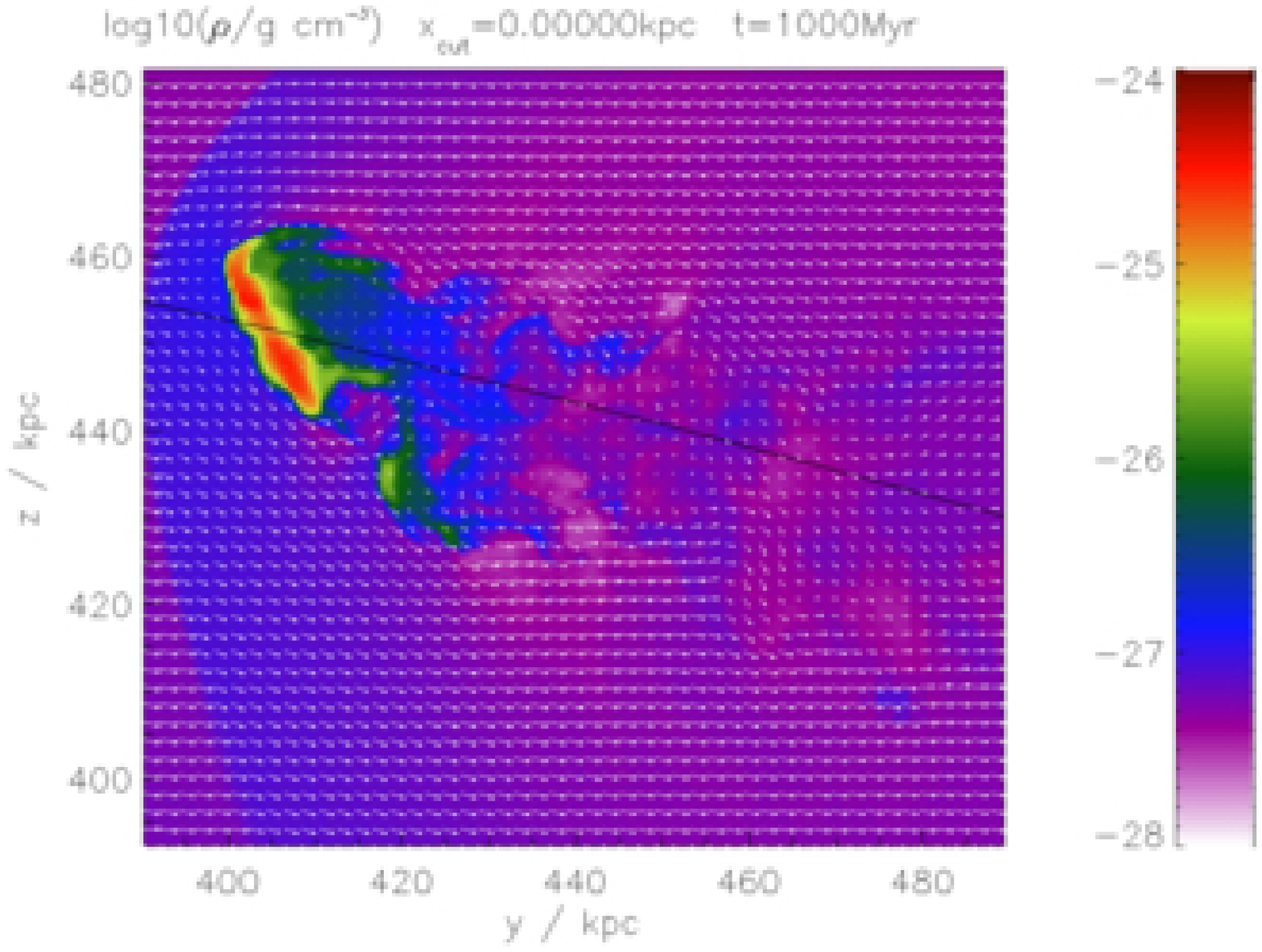}
\includegraphics[angle=0,width=0.49\textwidth]{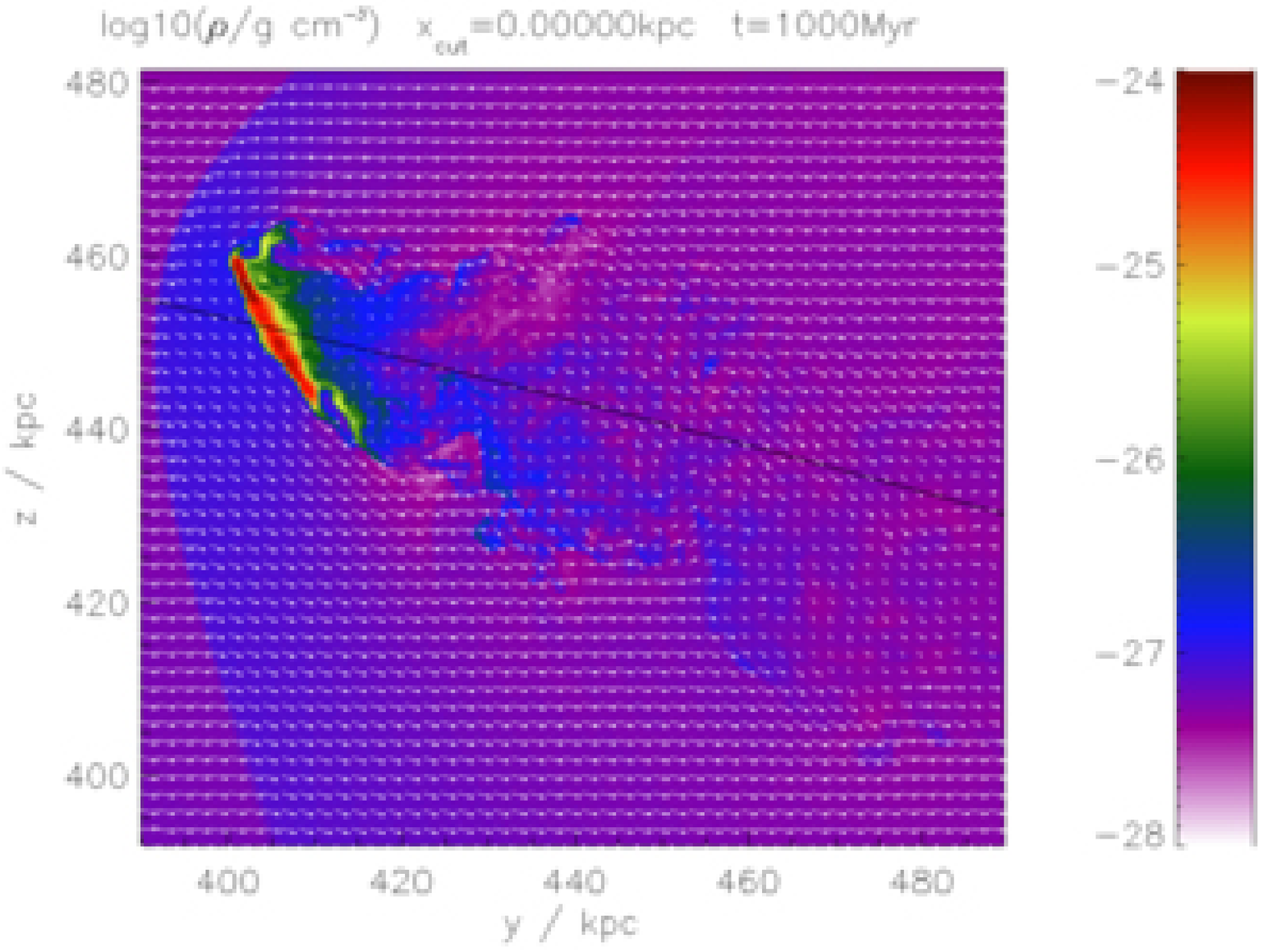}
\includegraphics[angle=0,width=0.49\textwidth]{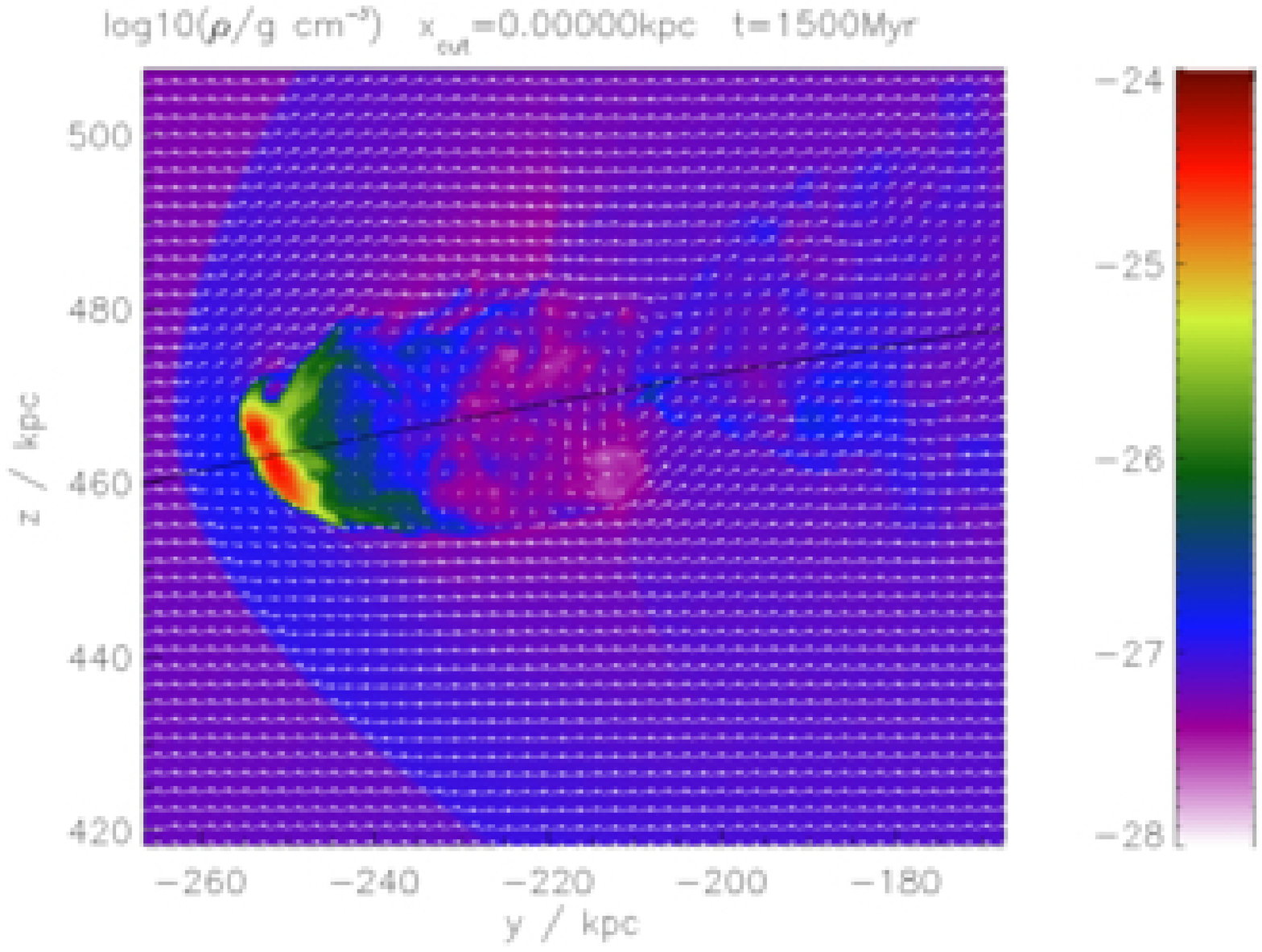}
\includegraphics[angle=0,width=0.49\textwidth]{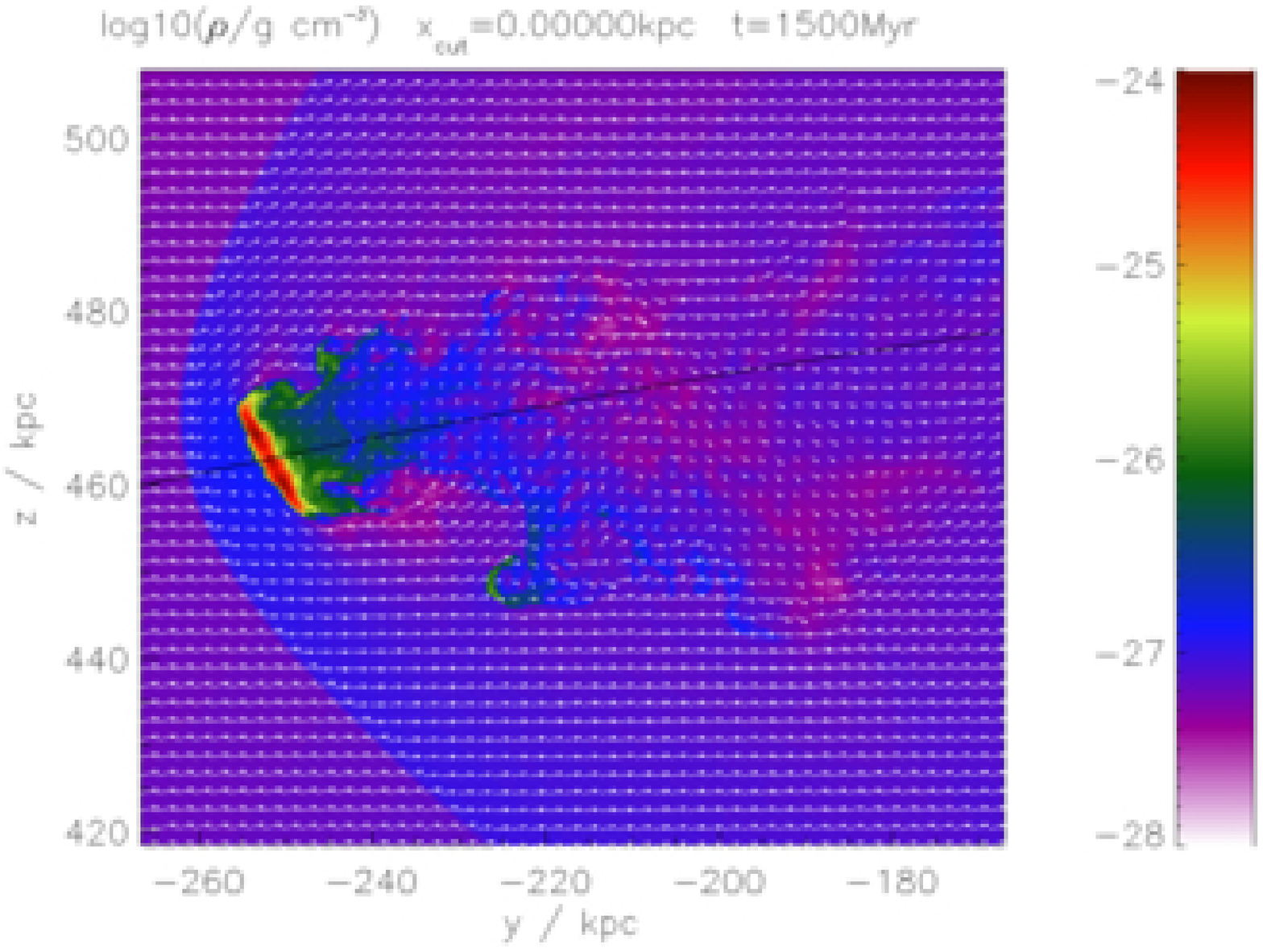}
\caption{Slices in the orbital plane, showing the local gas density
  colour-coded, and the velocity field projected onto this plane by
  arrows, for different time-steps. For run C1-LG-FST-MF. Left column is for resolution as described in
  Sect.~\ref{sec:code}, right column is for resolution improved by a factor 2
  everywhere. The black line marks the galaxy's orbit. The velocities are
  shown in the galactic rest frame. The horizontal arrow in
  the lower left corner shows the length of the magnitude of the ICM wind at
  each time-step. The axis coordinates are in the cluster-centric system.}
\label{fig:res_flow}
\end{figure*}

\begin{figure}
\centering\resizebox{0.49\hsize}{!}{\includegraphics[angle=0,width=0.49\textwidth]{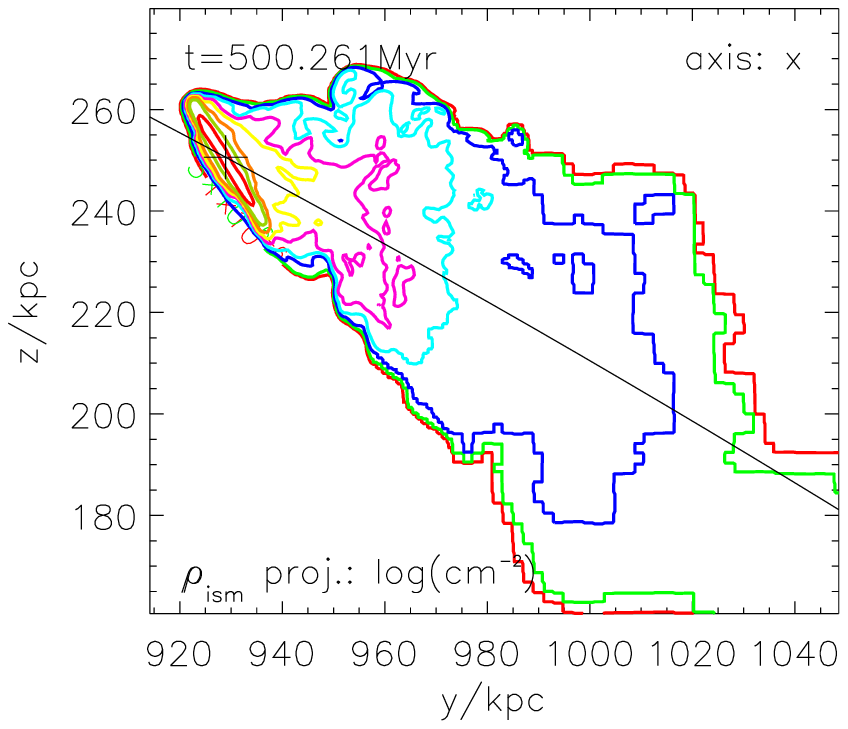}}
\centering\resizebox{0.49\hsize}{!}{\includegraphics[angle=0,width=0.49\textwidth]{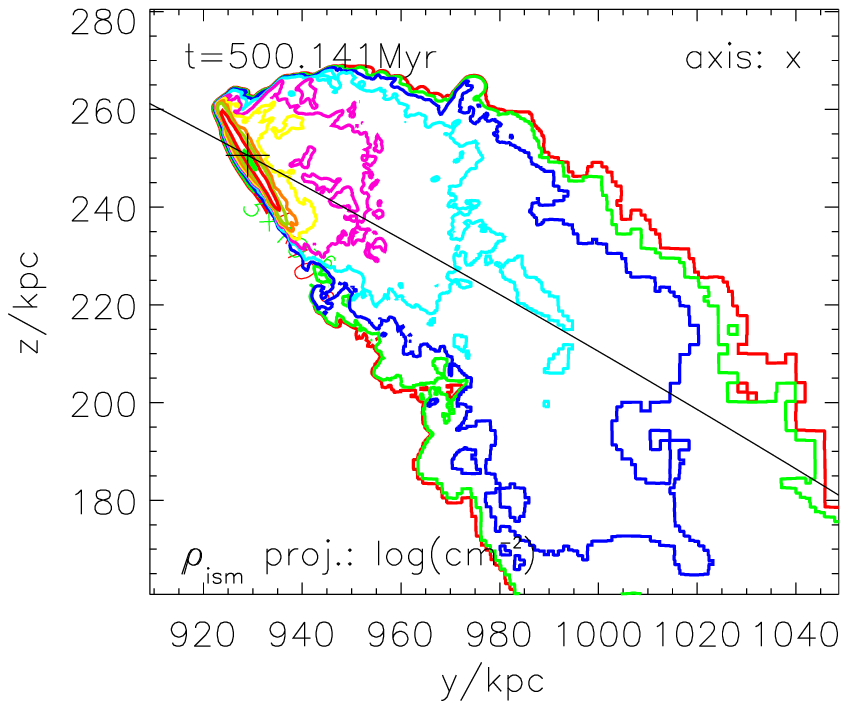}}
\centering\resizebox{0.49\hsize}{!}{\includegraphics[angle=0,width=0.49\textwidth]{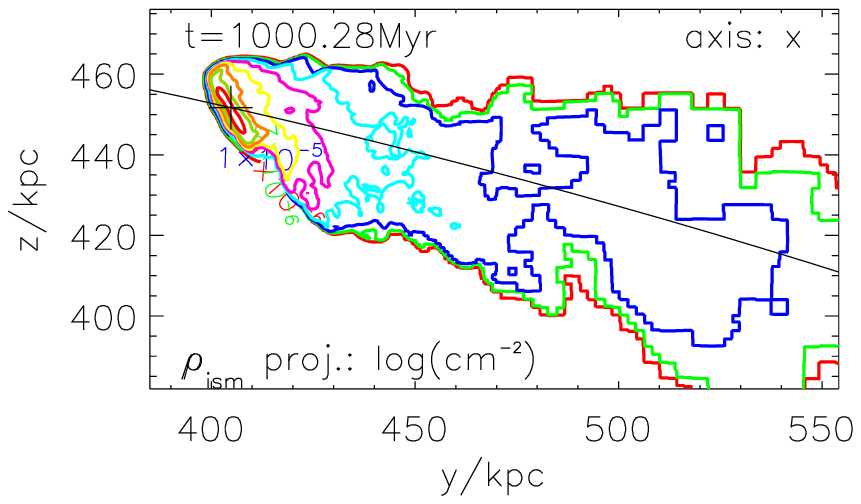}}
\centering\resizebox{0.49\hsize}{!}{\includegraphics[angle=0,width=0.49\textwidth]{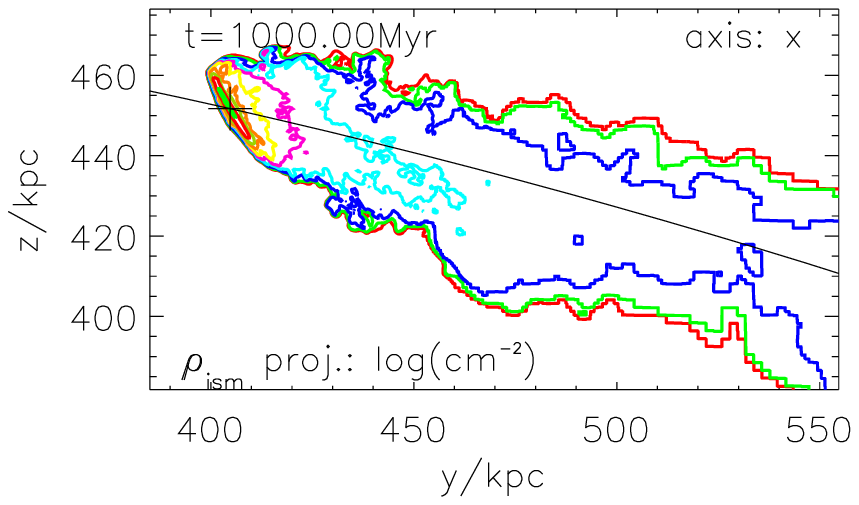}}
\centering\resizebox{0.49\hsize}{!}{\includegraphics[angle=0,width=0.49\textwidth]{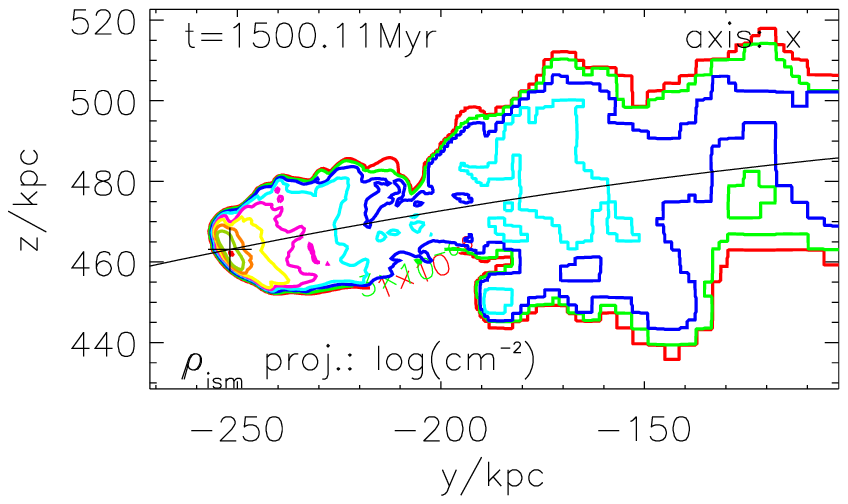}}
\centering\resizebox{0.49\hsize}{!}{\includegraphics[angle=0,width=0.49\textwidth]{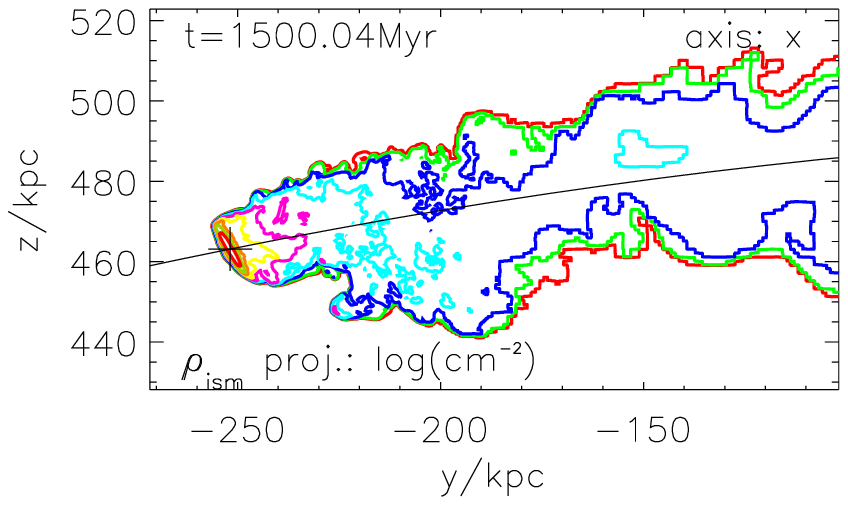}}
\caption{Same as Fig.~\ref{fig:wakes_LG_FST} but for run C1-LG-FST-MF. The lhs
  column is for the resolution described in Sect.~\ref{sec:code}, the rhs
  column for a run where the resolution has been improved by a factor of 2
  everywhere.}
\label{fig:res_proj2}
\end{figure}

\begin{figure*}
\centering\resizebox{0.7\hsize}{!}{\includegraphics[angle=-90,width=\textwidth]{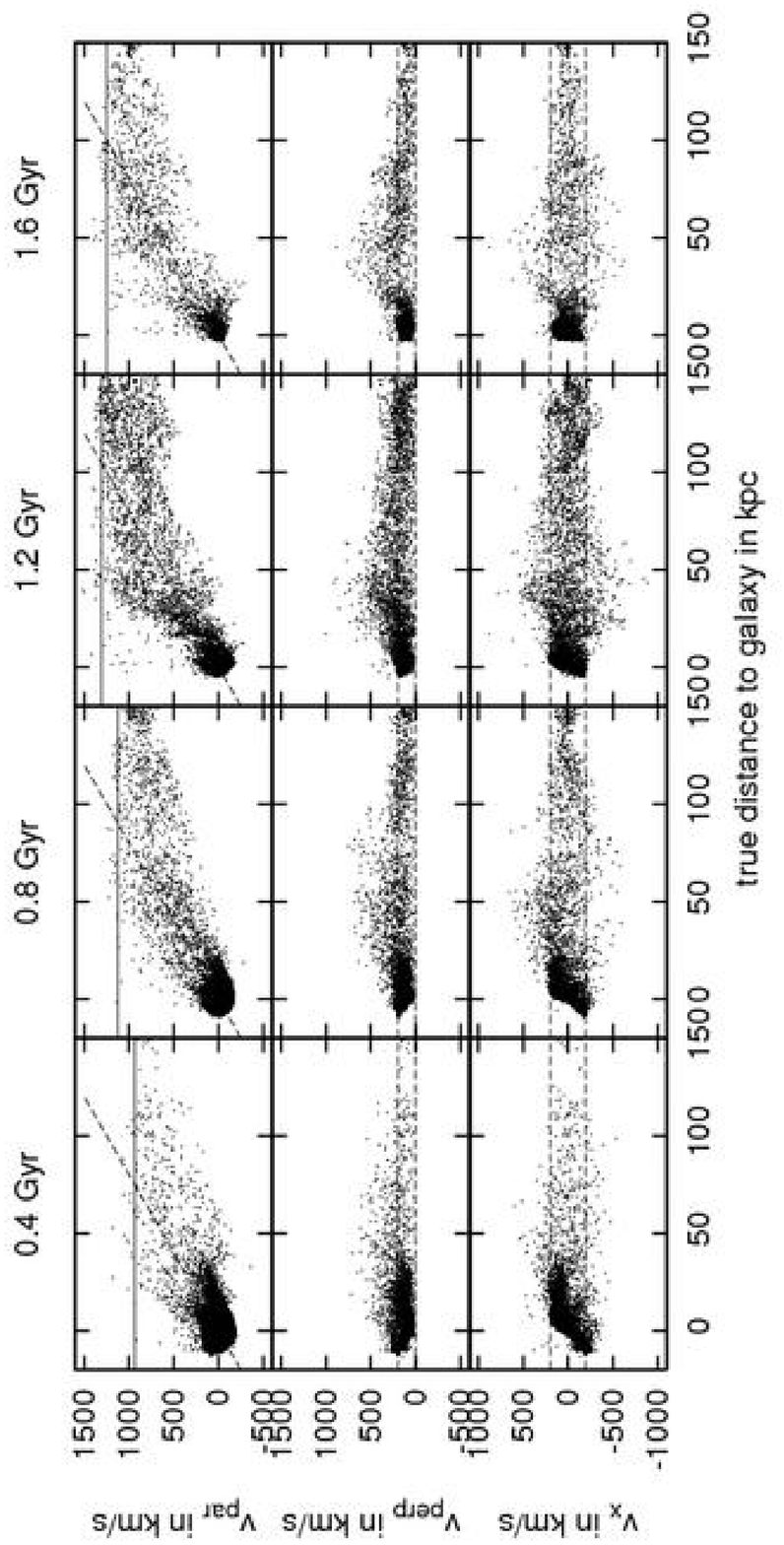}}
\centering\resizebox{0.7\hsize}{!}{\includegraphics[angle=-90,width=\textwidth]{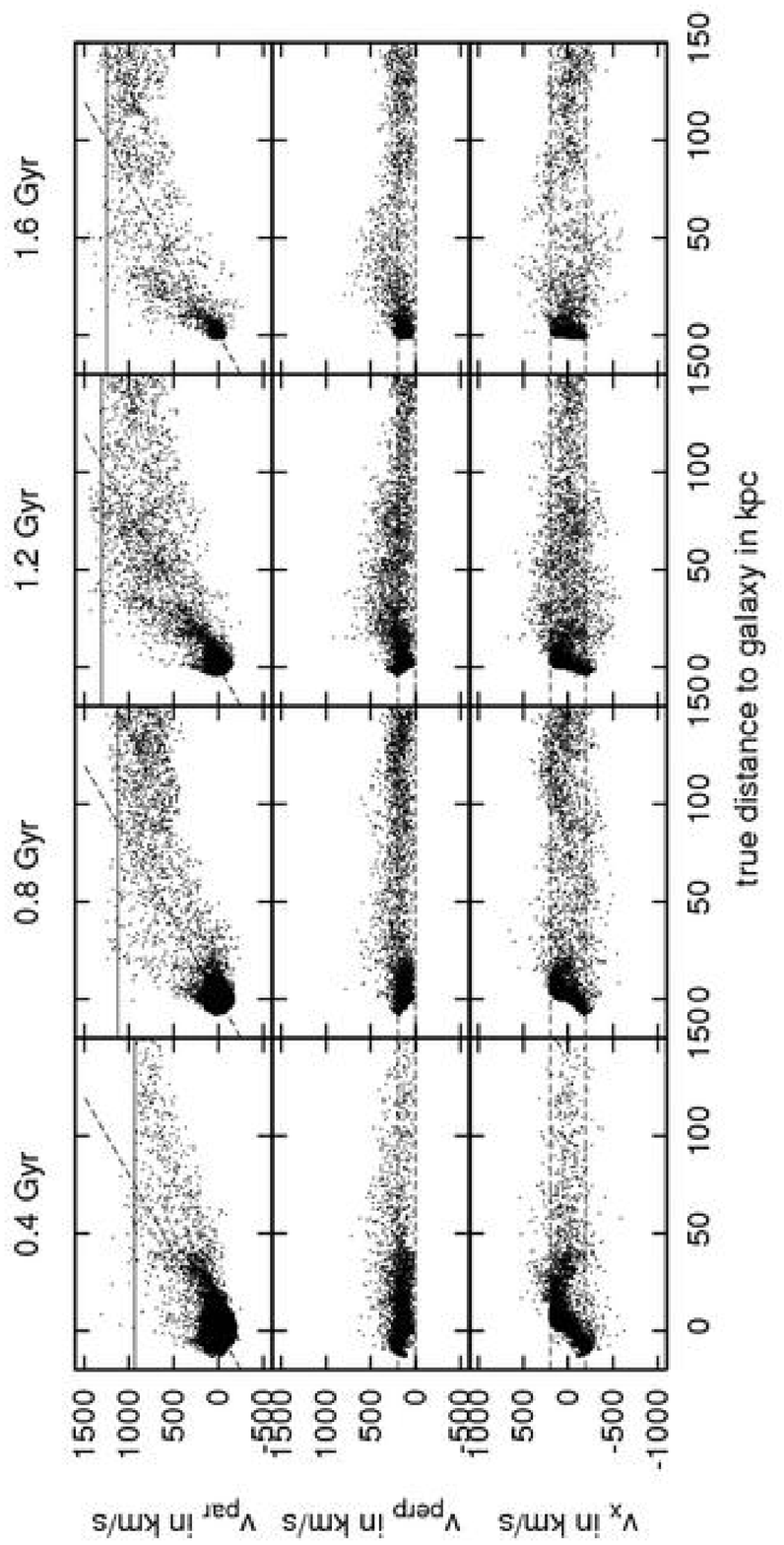}}
\caption{Same as Fig.~\ref{fig:vel_L_EF}, but for run C1-LG-FST-MF. Top plate
  is for resolution as described in Sect.~\ref{sec:code}, bottom plate is for
  run with resolution improved by a factor of 2 everywhere.}
\label{fig:vel_L_MF_HR}
\end{figure*}

\begin{figure}
\centering\resizebox{0.7\hsize}{!}{\includegraphics[angle=-90]{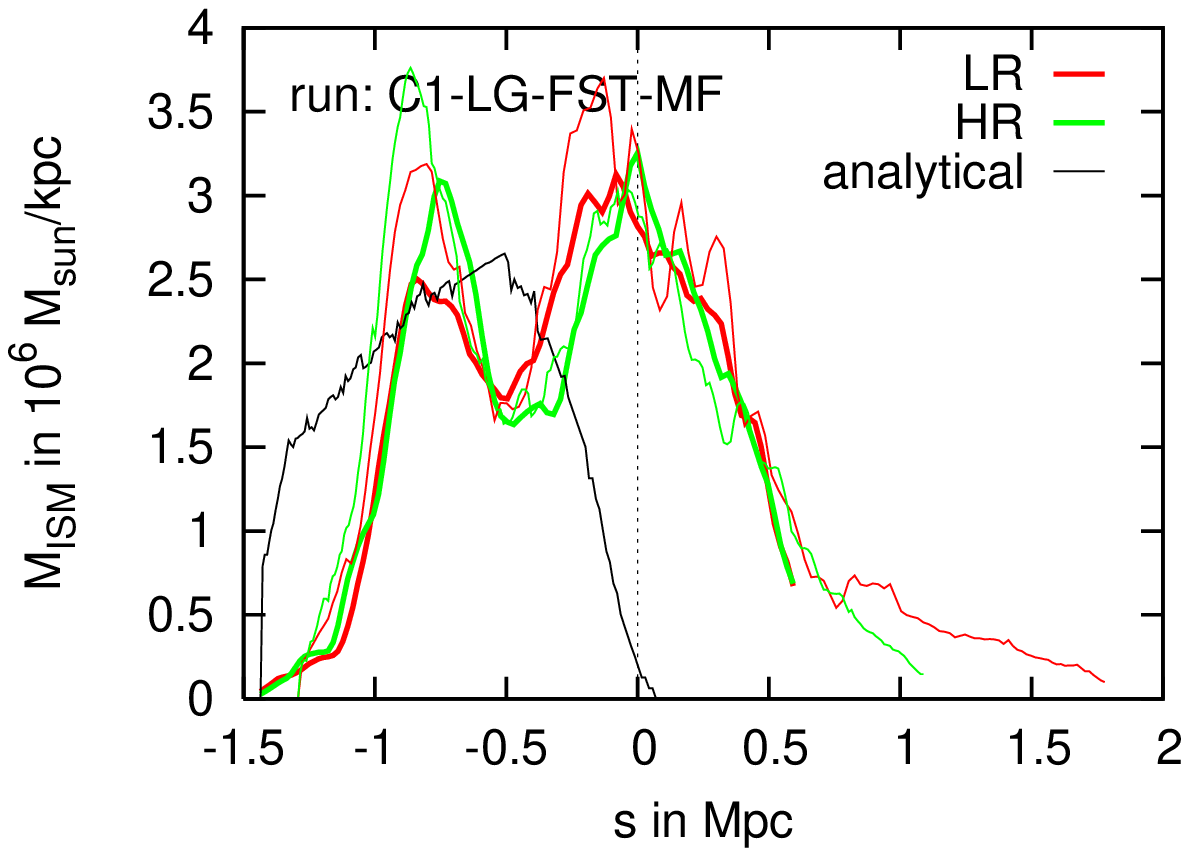}}
\caption{Distribution of stripped gas along orbit (compare to
Figs.~\ref{fig:ism_along_orbit_evol} and \ref{fig:ism_along_orbit}): for run
C1-LG-FST-MF, averaged over $300\Kpc$, for two different resolutions. Colours
code the resolution (LR is low resolution, HR is resolution improved by a
factor of 2 everywhere). The thick lines are distribution as seen in the
simulation. The thin coloured lines are the predictions based on the numerical
result of the bound gas mass as a function of covered distance -- however --
shifted by $150\Kpc$ along the orbit.  The thin black line is the prediction
based on the analytical estimate of the stripping mass (see paper I).  The
zero-point of the $x$-axis is shifted to peri-centre passage.}
\label{fig:ism_along_orbit_res}
\end{figure}

%
In this section, we show slices in the orbital plane that display the local
gas density and the flow structure (Fig.~\ref{fig:res_flow}), projected gas
density maps (Fig.~\ref{fig:res_proj2}) and velocity plots
(Fig.~\ref{fig:vel_L_MF_HR}). These
Figs. are for the run C1-LG-FST-MF, but once for the spatial resolution as
described in Sect.~\ref{sec:code}, and once for a run were the spatial
resolution as been improved by a factor of 2
everywhere. Figure~\ref{fig:ism_along_orbit_res} compares the distribution of
stripped gas along the orbit.

In all plots, the basic structure is same for both resolutions. We conclude
that our conclusions are not influenced by the resolution.

\begin{figure*}
\includegraphics[angle=0,width=0.49\textwidth]{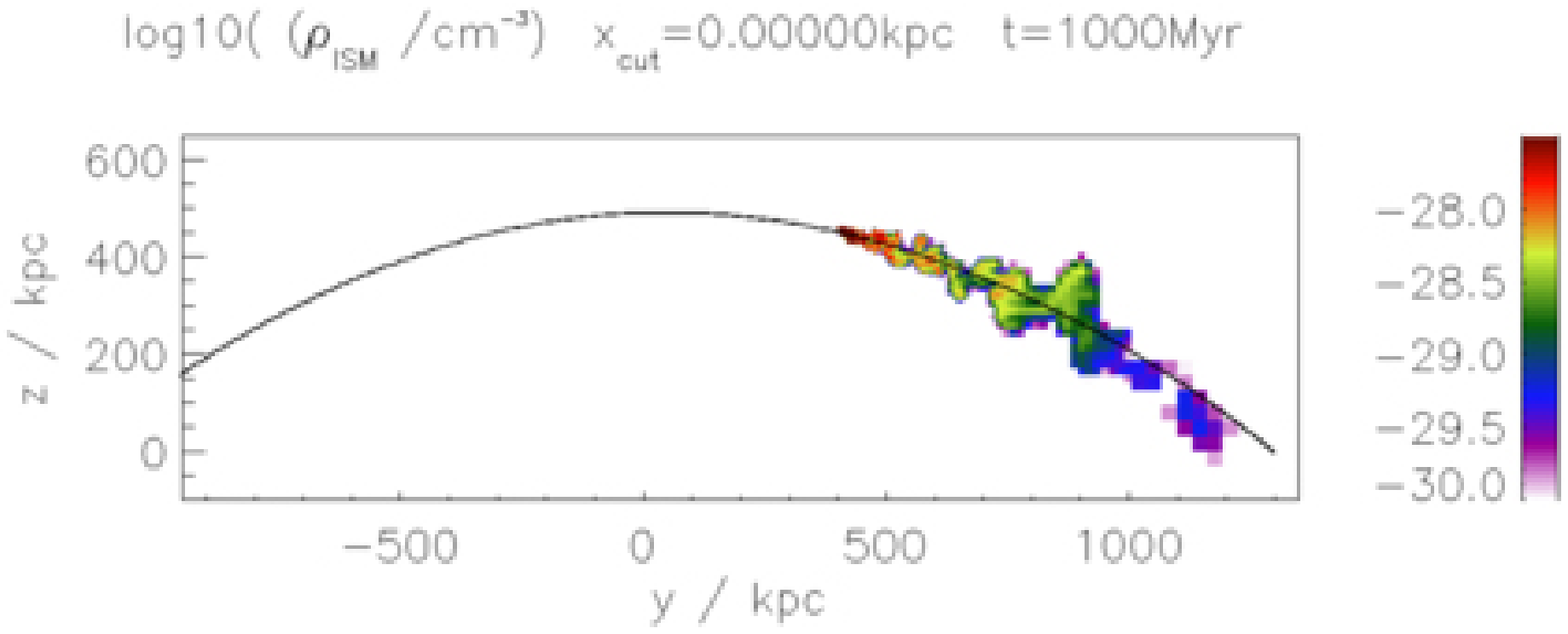}
\includegraphics[angle=0,width=0.49\textwidth]{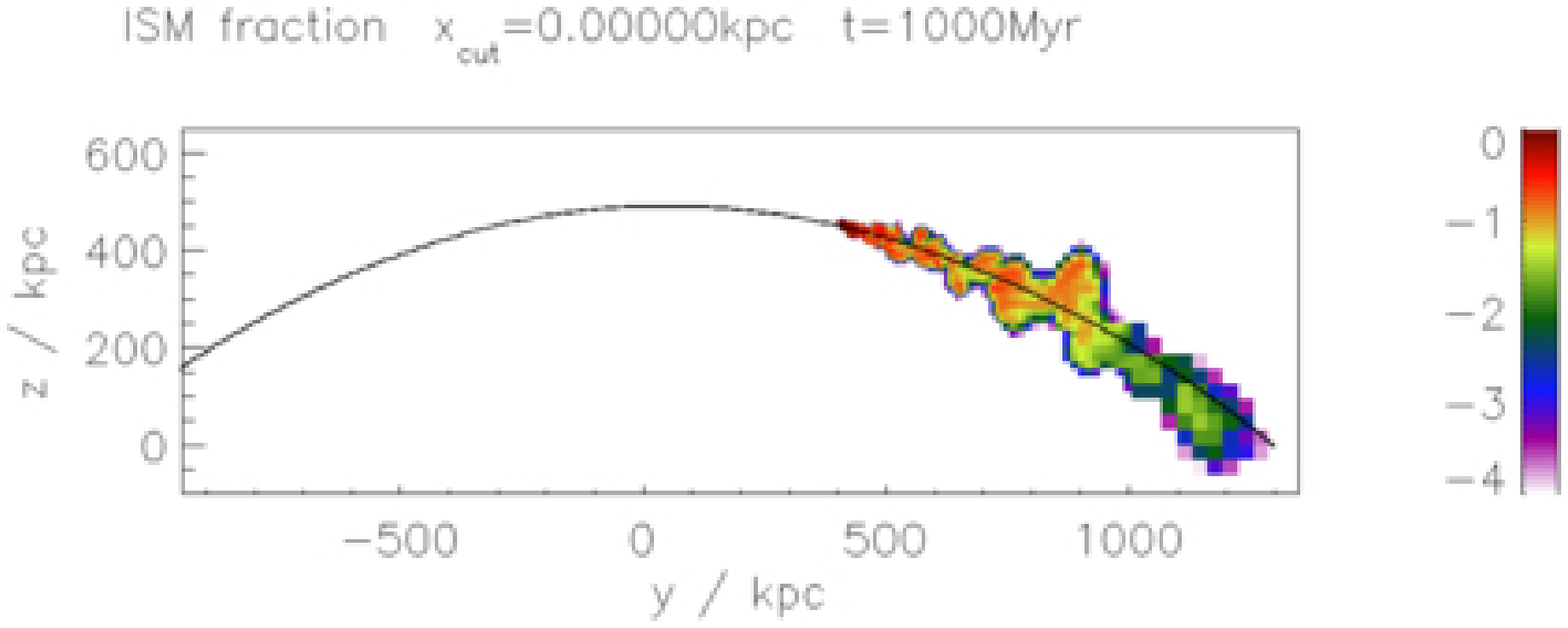}
\includegraphics[angle=0,width=0.49\textwidth]{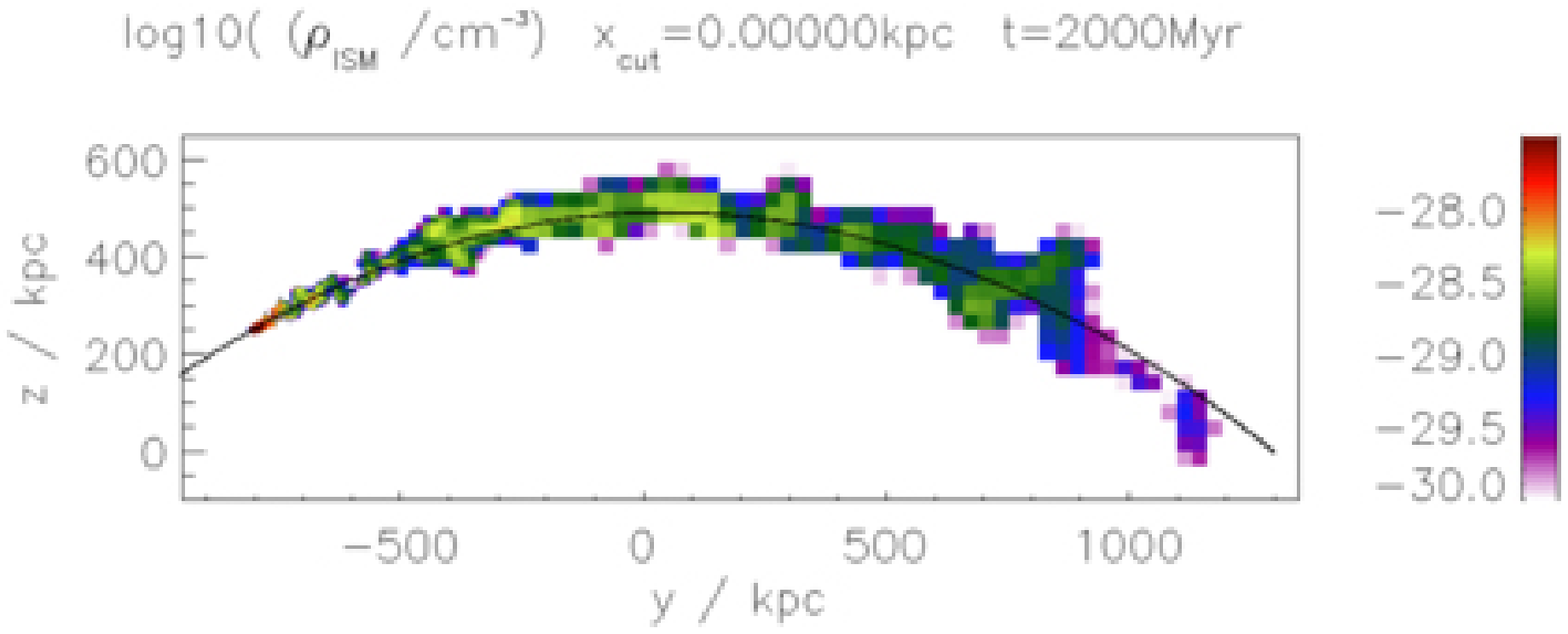}
\includegraphics[angle=0,width=0.49\textwidth]{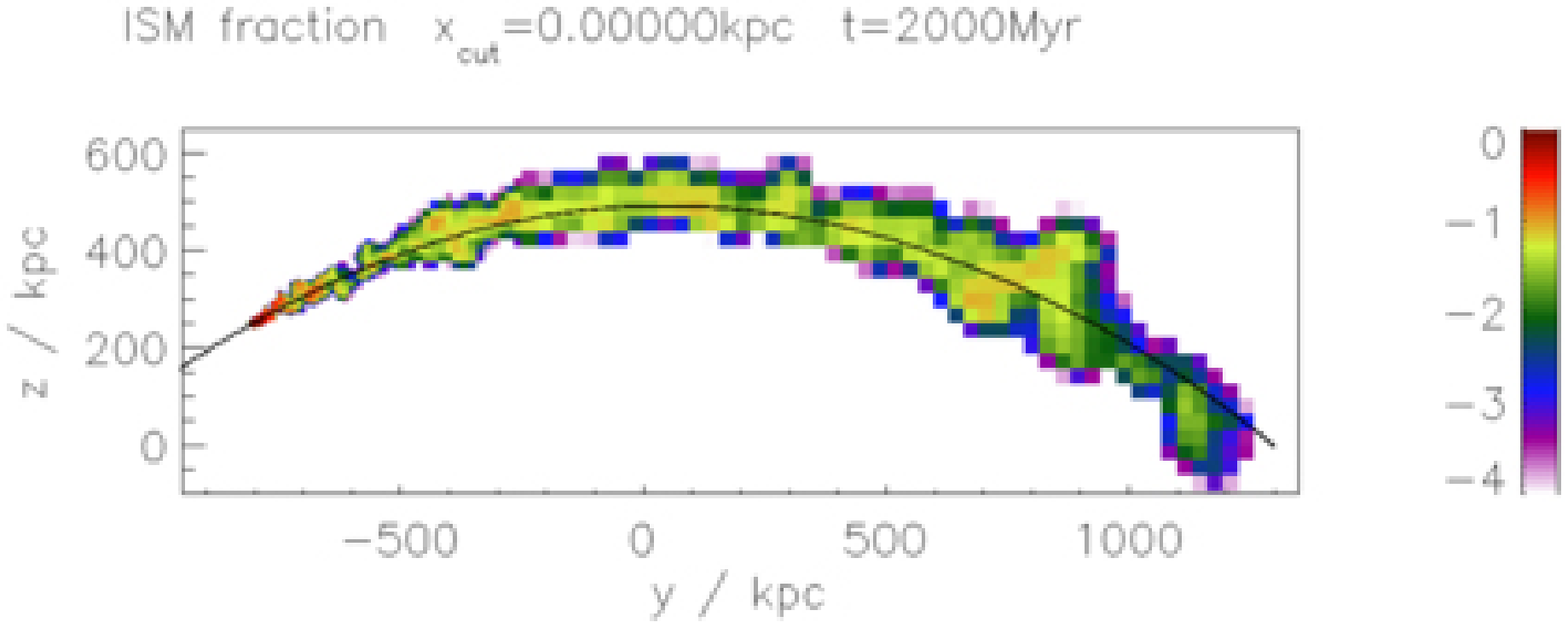}
\vspace*{0.5cm}\rule{\textwidth}{0.2mm}\vspace*{0.5cm}
\includegraphics[angle=0,width=0.49\textwidth]{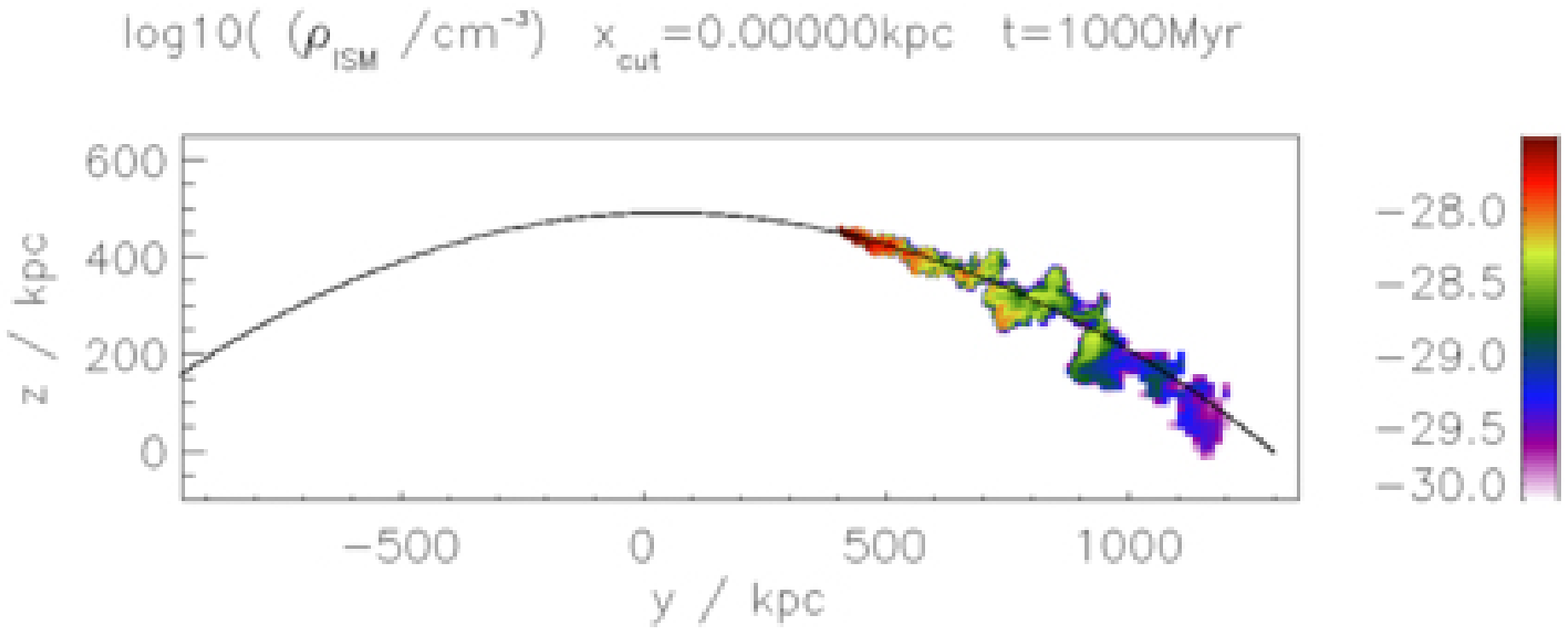}
\includegraphics[angle=0,width=0.49\textwidth]{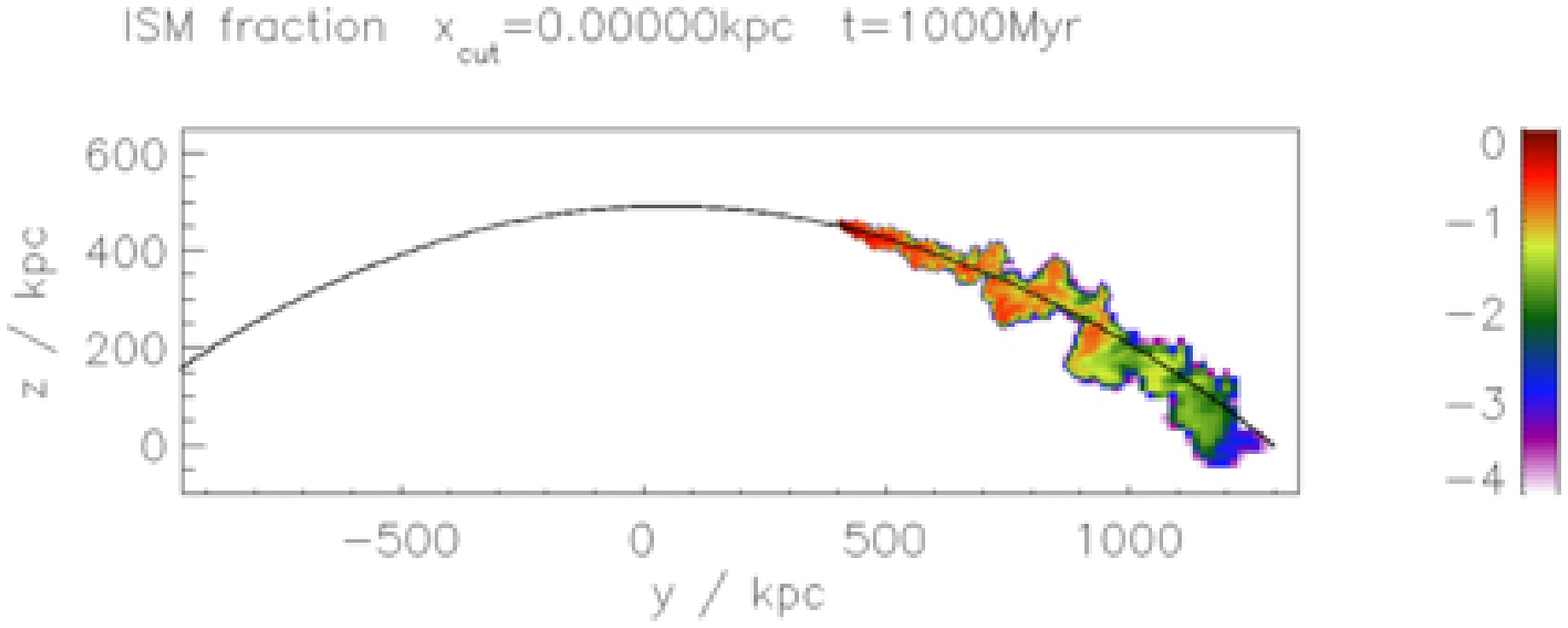}
\includegraphics[angle=0,width=0.49\textwidth]{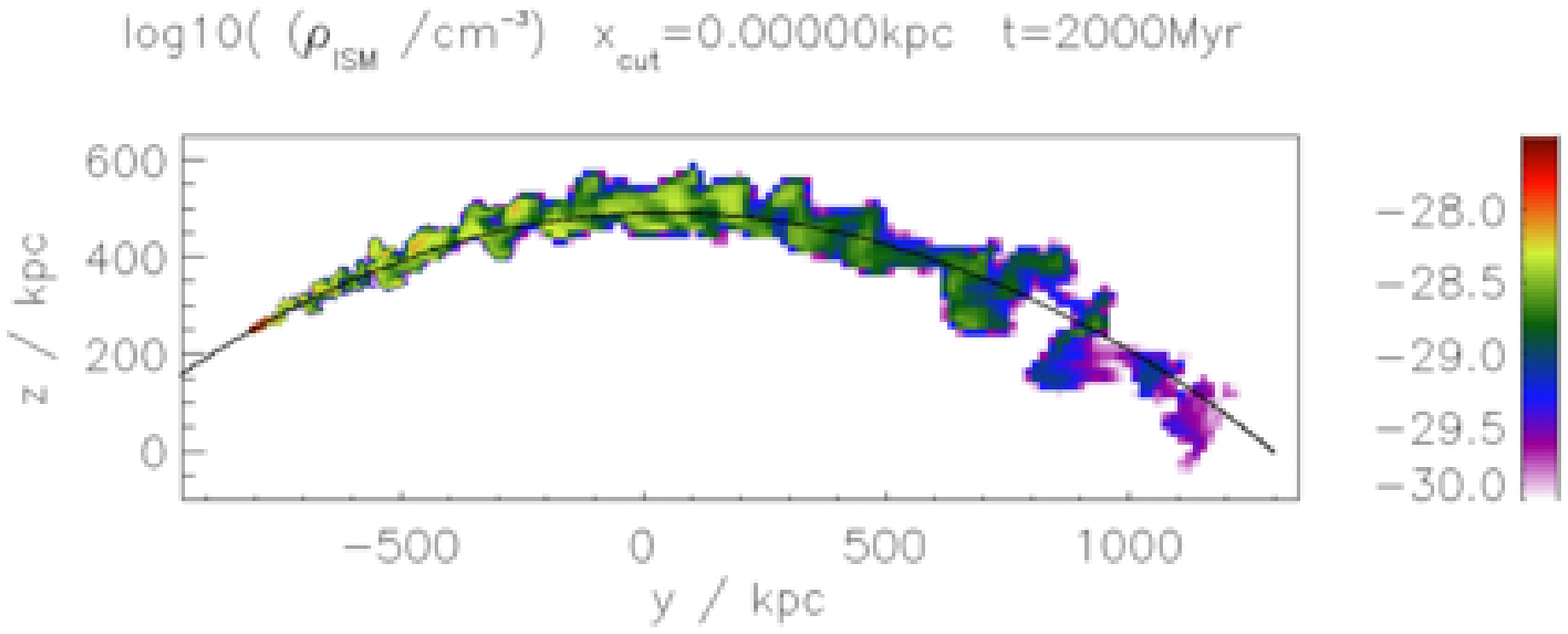}
\includegraphics[angle=0,width=0.49\textwidth]{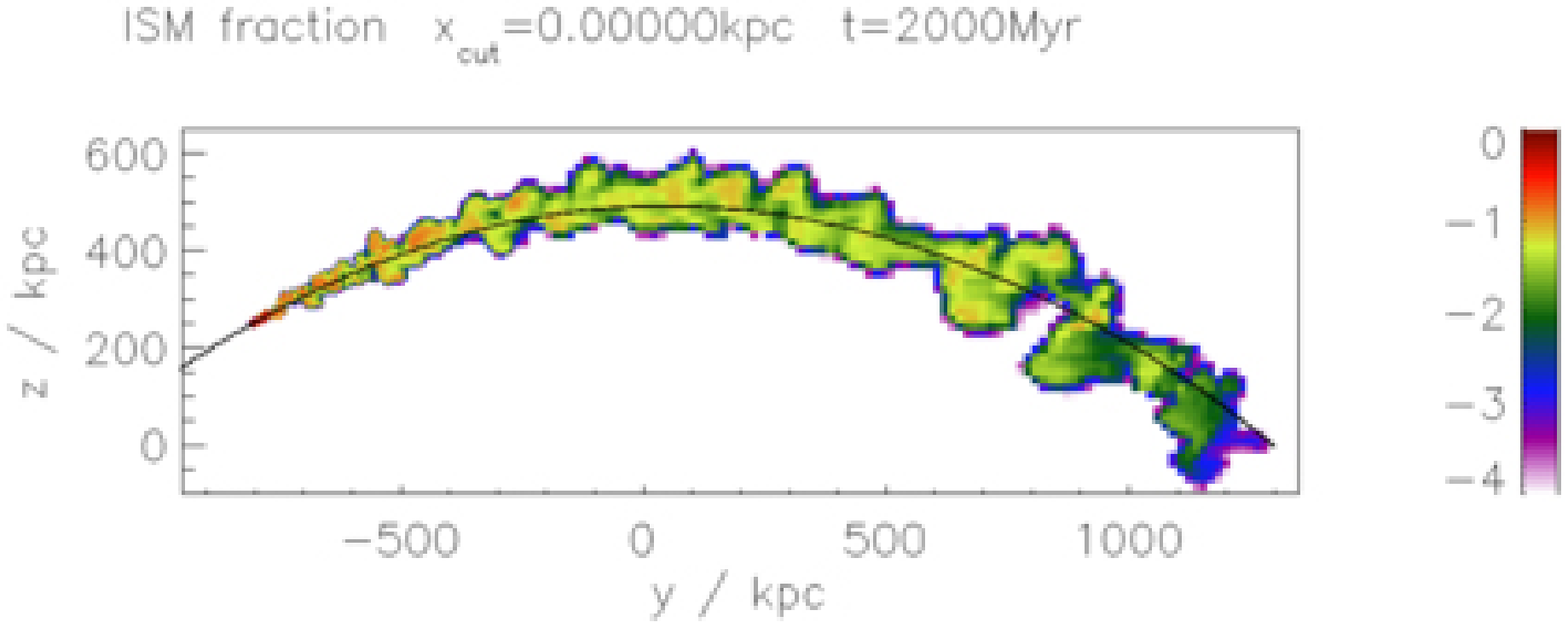}
\caption{Same as Fig.~\ref{fig:slice_ism_L_I45}, but for run C1-LG-FST-MF. The
  two top rows are for a resolution as described in Sect.~\ref{sec:code}, in
  the two bottom rows the resolution is improved by a factor 2 everywhere.}
\label{fig:slice_ism_L_I30_HR}
\end{figure*}

%
\bibliographystyle{mn2e}
\bibliography{%
../../BIBLIOGRAPHY/theory_simulations,%
../../BIBLIOGRAPHY/hydro_processes,%
../../BIBLIOGRAPHY/numerics,%
../../BIBLIOGRAPHY/observations_general,%
../../BIBLIOGRAPHY/observations_clusters,%
../../BIBLIOGRAPHY/observations_galaxies,%
../../BIBLIOGRAPHY/galaxy_model,%
../../BIBLIOGRAPHY/gas_halo,%
../../BIBLIOGRAPHY/icm_conditions,%
../../BIBLIOGRAPHY/clusters,%
../../BIBLIOGRAPHY/agn,%
../../BIBLIOGRAPHY/else}

\bsp

\label{lastpage}

\end{document}